\documentclass[fleqn,usenatbib]{mnras}

\usepackage{newtxtext,newtxmath}

\usepackage[T1]{fontenc}

\DeclareRobustCommand{\VAN}[3]{#2}
\let\VANthebibliography\thebibliography
\def\thebibliography{\DeclareRobustCommand{\VAN}[3]{##3}\VANthebibliography}

\usepackage{graphicx}	
\usepackage{amsmath}	
\usepackage{subcaption}
\usepackage{booktabs}
\usepackage{orcidlink}
\usepackage{xspace} 

\newcommand{\pder}[2]{\frac{\partial#1}{\partial#2}}
\newcommand{\avg}[1]{\langle #1 \rangle}
\newcommand{\BV}{Brunt-V\"{a}is\"{a}l\"{a}\xspace}

\title[Neutrino Heating in 1D, 2D, and 3D]{Neutrino Heating in 1D, 2D, and 3D core-collapse supernovae: characterizing the explosion of high-compactness stars}

\author[L. Boccioli et al.]{
Luca Boccioli,$^{1}$\thanks{E-mail: lbocciol@berkeley.edu} \orcidlink{0000-0002-4819-310X}
David Vartanyan,$^{2}$ \orcidlink{0000-0003-1938-9282}
Evan P. O'Connor$^{3}$ \orcidlink{0000-0002-8228-796X}
and Daniel Kasen$^{4}$ \orcidlink{0000-0002-5981-1022}
\\
$^{1}$Department of Physics, University of California, Berkeley, CA 94720, USA\\
$^{2}$Carnegie Observatories, 813 Santa Barbara St., Pasadena, CA 91101, USA; NASA Hubble Fellow\\
$^{3}$The Oskar Klein Centre, Department of Astronomy, Stockholm University, AlbaNova, SE-106 91 Stockholm, Sweden\\
$^{4}$Department of Astronomy and Theoretical Astrophysics Center, University of California, Berkeley, CA 94720, USA
}

\date{Accepted 2025 June 6. Received 2025 June 6; in original form 2025 January 16}

\pubyear{\the\year{}}

\begin{document}
\label{firstpage}
\pagerange{\pageref{firstpage}--\pageref{lastpage}}
\maketitle

\begin{abstract}
Massive stars can end their lives with a successful supernova explosion (leaving behind a neutron star or, more rarely, a black hole), or a failed explosion that leaves behind a black hole. The density structure of the pre-collapse progenitor star already encodes much of the information regarding the outcome and properties of the explosion. However, the complexity of the collapse and subsequent shock expansion phases prevents drawing a straightforward connection between the pre-collapse and post-explosion properties. In order to derive such a connection several explodability studies have been performed in recent years. However, different studies can predict different explosion outcomes. In this article, we show how compactness, which is related to the average density of the star's core, has an important role in determining the efficiency of neutrino heating, and therefore the outcome of the explosion. Commonly, high-compactness progenitors are assumed to yield failed explosions, due to their large mass accretion rates, preventing the shock from expanding. We show by analyzing $\sim$ 150 2D FLASH and F{\sc{ornax}} simulations and 20 3D F{\sc{ornax}} simulations that this is not the case. Instead, due to the rapid increase of neutrino heating with compactness, high-compactness progenitors lead to successful shock revival. We also show that 1D+ simulations that include $\nu$-driven convection using a mixing-length theory approach correctly reproduce this trend. Finally, we compare 1D+ models, which we show can reproduce some aspects of multi-D simulations with reasonable accuracy, with other widely used 1D models in the literature.
\end{abstract}



\section{Introduction}
The explosion of core-collapse supernovae (CCSNe) is a long-standing problem in astrophysics. The very first simulations of CCSNe were carried out assuming spherical symmetry, simple prescriptions for the equation of state at high densities, and an approximate treatment of neutrino transport and interactions. Despite their simplicity, these pioneering models were the first to establish the role of neutrinos in driving the explosion \citep{Colgate_White1966, Arnett1966, Bethe_Wilson1985}. The tremendous advancements in computational resources in the last few decades allowed for more and more sophisticated simulations to become feasible \citep{Herant1994_first2D, Miller1993_first2D, Fryer2002_first3D, Kuroda2016_GR_nu_transport_code, Summa2018_rot3D_crit_lum, Bruenn2020_CHIMERA_method, Burrows2020_3DFornax, Nakamura2025_LS220_3D_suite}. Moreover, the significant improvement in models for the nuclear equation of state \citep{Lattimer1991_LS,Hempel2010_HS_RMF,Fischer2017_Review_EOS_nu,Raduta2021_EOS_review} and neutrino interactions \citep{Bruenn1985,Hannestad1998_NN_Brem,Horowitz2002,Buras2003_numutau_pair,BRT2006,Bollig2017_muon_creation,Horowitz2017_virial} significantly reduced the uncertainties that enter into the simulations. However, several uncertainties are still affecting the models and their predictions. Therefore, the explosion of core-collapse supernovae remains a very complex phenomenon with many questions that need to be answered.

One particular problem whose solution is currently unclear is the explosion of massive stars with large iron cores or, equivalently, with high compactness, defined as:
\begin{equation}
    \label{eq:compactness}
    \xi_{M} = \dfrac{M/M_\odot}{R(M)/1000\, {\rm km}},
\end{equation}
where $R(M)$ is the radius that encloses a baryonic mass $M$, with $M$ typically being chosen to be between 1.75~$M_\odot$ and 2.5~$M_\odot$. In the vast majority of this paper, we will be using $M=2.0 \,M_\odot$. 

Until recently \citep{Janka2012_review_CCSNe}, it was believed that all stars above $18-20 M_\odot$ (i.e. with large iron cores and high compactness) would not explode, whereas all stars below that would yield a successful supernova. This was in part motivated by the fact that no type II-P SNe were observed for stars above $\gtrsim 20 M_\odot$ \citep{Smartt2009_typeII-P,Smartt2015_observations_CCSNe}, the so-called Red Supergiant Problem. However, this problem is now considered to be much less significant and can mostly be attributed to the large uncertainties in the imaging of the pre-SN progenitors \citep{Davies2018_RSG_IIP_ZAMS_mass,Davies2020_RSGP_upper_lum_IIP,Beasor2025_RSGP}. At the same time, current models have shown that the explodability has a much more complicated dependency on the initial mass of the progenitor \citep{Pejcha2015_explodability, Sukhbold2016_explodability, Muller2016_prog_connection, Ebinger2019_PUSH_II_explodability, Couch2020_STIR, Boccioli2023_explodability, Fryer2022_nu_conv_remnant_masses}. Still, the fate of high-compactness stars is not well understood, and in some studies \citep{Pejcha2015_explodability, Sukhbold2016_explodability, Muller2016_prog_connection, Ebinger2019_PUSH_II_explodability,Fryer2022_nu_conv_remnant_masses}, these stars are predicted (or oftentimes assumed) to yield failed supernovae. In others \citep{Couch2020_STIR,Boccioli2023_explodability,Wang2022_prog_study_ram_pressure}, they are predicted to yield successful explosions instead. Given this tension, part of the motivation behind this study is to analyze the differences among these studies, specifically with respect to the explosion of high-compactness progenitors.

All of the explodability studies mentioned above that, by definition, require hundreds of simulations, have been carried out using spherically symmetric (1D) models. The only exceptions are two sets of hundreds of 2D simulations by \citet{Nakamura2015_2D_explodability} and \citet{Vartanyan2023_nu_100_2D}. The former study involved $\sim 350$ simulations, where, however, the transport of electron-type neutrinos was described using the simple (but computationally efficient) Isotropic Diffusion Source Approximation \citep{Liebendorfer2009_IDSA}, whereas heavy lepton neutrinos were treated using an even more simplified approach following a leakage scheme. Moreover, the ray-by-ray approximation was adopted, i.e. neglecting non-radial effects in the neutrino transport. The latter study is more recent, was carried out for 100 simulations, and uses a state-of-the-art M1 transport scheme to solve for the transport of neutrinos. The explodability of those 100 2D simulations was then analyzed by \citet{Wang2022_prog_study_ram_pressure} and \citet{Tsang2022_ML_explodability}.

All of the 1D models used in the explodability studies mentioned above are characterized by different degrees of approximations. Some rely on simple semi-analytic models and others rely on self-consistent simulations where extra heating has been added. Ultimately, the explosion in all of these models is triggered by neutrinos, as also shown by all modern high-fidelity simulations.  Therefore, to study the explodability of high-compactness progenitors, it is crucial to understand how neutrino heating changes with compactness. 

In the more recent past, there have been several multi-dimensional simulations that showed how high-compactness progenitors can produce successful explosions, often accompanied by the formation of large black holes around $\sim 20-30 M_\odot$ \citep{Muller2018_stripped_SN_to_breakout, Chan2018_BH_SN_40Msol, Powell2021_collapse_PISNe, Sykes2024_2D_fallback_SNe, Burrows2023_BH_supernova_40Msol, Eggenberger2025_BHSNe_EOS, Burrows2024_BH_formation_3D}. This however goes against what has often been assumed by previous studies \citep{OConnor2011_explodability,Fryer2012_remnant_popsynth,Pejcha2015_explodability,Sukhbold2016_explodability,Ebinger2019_PUSH_II_explodability}. This is particularly relevant for population synthesis \citep{Izzard2004_binaryc_popsynth,Belczynski2008_Stratrack_popsynth,Mennekens2014_Brussels_popsynth,Eldridge2017_BPASS_popsynth,Breivik2020_COSMIC_popsynth,Riley2022_COMPAS_popsynth,Kamlah2022_BSE-LEVELC_popsynth,Iorio2023_SEVN_popsynth,Andrews2024_POSYDON2} 
and galactic chemical evolution (GCE) simulations, \citep{Kubryk2015_GCE_Prantzos_code,Cote2017_GCE_OMEGA,Andrews2017_GCE_flexCE,Rybizki2017_GCE_Chempy,Spitoni2020_GCE_Bayesian_MCMC,Johnson2020_GCE_VICE,Gjergo2023_GCE_GalCEM1,Chen2023_GCE_radial_mixing,Diehl2020_GCE_living_review}
both of which adopt prescriptions for the explodability of massive stars where high-compactness progenitors do not explode. Since the overwhelming majority of GCE and population synthesis calculations assume that high-compactness progenitors do not explode (but see \citet{Jost2025_fheat_yields} for an exception), it is not clear how sensitive these models are to this assumption. Therefore, more work is needed to understand the fate of high-compactness stars.

Given the rarity of these events, observational signatures of explosions of high-compactness stars are quite hard to predict. Since high-compactness progenitors tend to be high-mass stars, they are significantly disfavored by the initial mass function (IMF). Additionally, the stars with the largest compactness tend to form at metallicities much lower than solar and are therefore more abundant in distant galaxies, where faint supernovae are harder to detect. If the fallback onto the central black hole is significant, a high-compactness star might produce no explosion at all \citep{Summa2018_rot3D_crit_lum,Kuroda2018_BH_formation_GR,Walk2020_3DBH_neutrinos}, a weak explosion one or two orders of magnitude dimmer than regular type-II CCSNe \citep{Chan2018_BH_SN_40Msol,Eggenberger2025_BHSNe_EOS,Sykes2024_2D_fallback_SNe}, or a quite strong explosion instead \citep{Burrows2023_BH_supernova_40Msol,Burrows2024_BH_formation_3D,Sykes2024_2D_fallback_SNe}. Given the few detailed studies on this subject and the differences among multi-dimensional simulations from different groups, this is still an open problem that requires deeper investigation. Another reason why predicting their observational signatures is challenging is that the uncertainties in the late phases of the explosion are quite large since long-term simulations are computationally very expensive \citep{Vartanyan2025_3D_bounce_to_breakut}.

In the next two years the ground-based Vera C. Rubin Observatory \citep{Ivezic2019_LSST_method_paper} will begin on-sky operations. Similarly, in the next two decades, next-generation gravitational wave observatories such as Cosmic Explorer (CE) \citep{Reitze2019_CE_method_paper1,Evans2021_CE_method_paper2} and Einstein Telescope (ET) \citep{Punturo2010_ET_method_paper1,Maggiore2020_ET_method_paper2} will go online. This will provide an unprecedented number of observed light curves and detections of compact remnants, in particular black holes. Therefore, rarer events such as explosions of high-mass, low-metallicity progenitors, will become more likely to be detected, and better models are needed to understand their properties. Moreover, the much more detailed measurements of the black-hole mass distribution by CE and ET could be used to put constraints on which stars should explode and which ones should collapse into black holes. Specifically relevant for this work, it will become possible to constrain how rare are events where high-compactness stars explode and at the same time leave behind a black hole of $20-30 M_\odot$.

In this paper, we study how neutrino heating develops during the explosion (with a particular focus on high-compactness progenitors) in three different codes: GR1D, FLASH, and F{\sc{ornax}}. We will analyze 1D, 1D+, 2D, and 3D simulations. For the remainder of this paper, we will refer to simulations in spherical symmetry where neutrino-driven convection is included using a parametric model as 1D+ simulations. In Section~\ref{sec:nu_heat}, we describe neutrino heating and neutrino-driven convection in the gain region and describe how convection is modeled in 1D+ simulations. In Section~\ref{sec:comparison_multiD} we compare 1D+, 2D and 3D models from different codes. In Section~\ref{sec:semi_an_qdot} we analyze the interplay between neutrino-driven convection and neutrino heating in 1D+ models as a function of compactness. In Section~\ref{sec:comparison_1D} we compare 1D+ models with other widely-used 1D models in the literature. Finally, in Section~\ref{sec:conclusions} we provide a summary and draw the main conclusions of this work.

\section{Neutrinos in the gain region}
\label{sec:nu_heat}
To understand how the explosion of high-compactness progenitors develops, it is instructive to highlight how neutrinos directly (i.e. by being absorbed by the material close to the shock) and indirectly (by triggering convection) affect the revival of the shock. After the initial shock expansion, the shock stalls during the accretion phase due to a delicate balance between the energy lost by photodissociating the infalling material, and the energy gained due to neutrino absorption. The region where the energy gained by neutrino absorption is larger than the energy lost due to neutrino emission is called the gain region. During this stalled-shock phase, neutrino-driven convection plays a crucial role in further supporting the shock and increasing the time that matter spends in the gain region. In practice, throughout this paper, we define the gain region to be the region where net neutrino heating is positive, density is below $3\times10^{10}\ {\rm g/cm^3}$, and specific entropy per baryon is above $6$. We will come back to the importance of these cuts in density and entropy in Section \ref{sec:comparison_multiD}.

\subsection{Neutrino-driven convection in the gain region}
Neutrino-driven turbulent convection contributes to the overall energy budget in the gain region in two ways: (i) dissipation of turbulent kinetic energy into heat and (ii) transport of hot bubbles of material closer to the shock. These two tightly related effects can be quantitatively understood by analyzing the Reynolds-decomposed Euler equations, which separate each variable into a mean background component and a perturbed (turbulent) one \citep{Murphy2011,Murphy2013_turb_in_CCSNe,Mabanta2018_MLT_turb,Couch2020_STIR}.
\begin{figure*}
\centering
\includegraphics[width=0.99\textwidth]{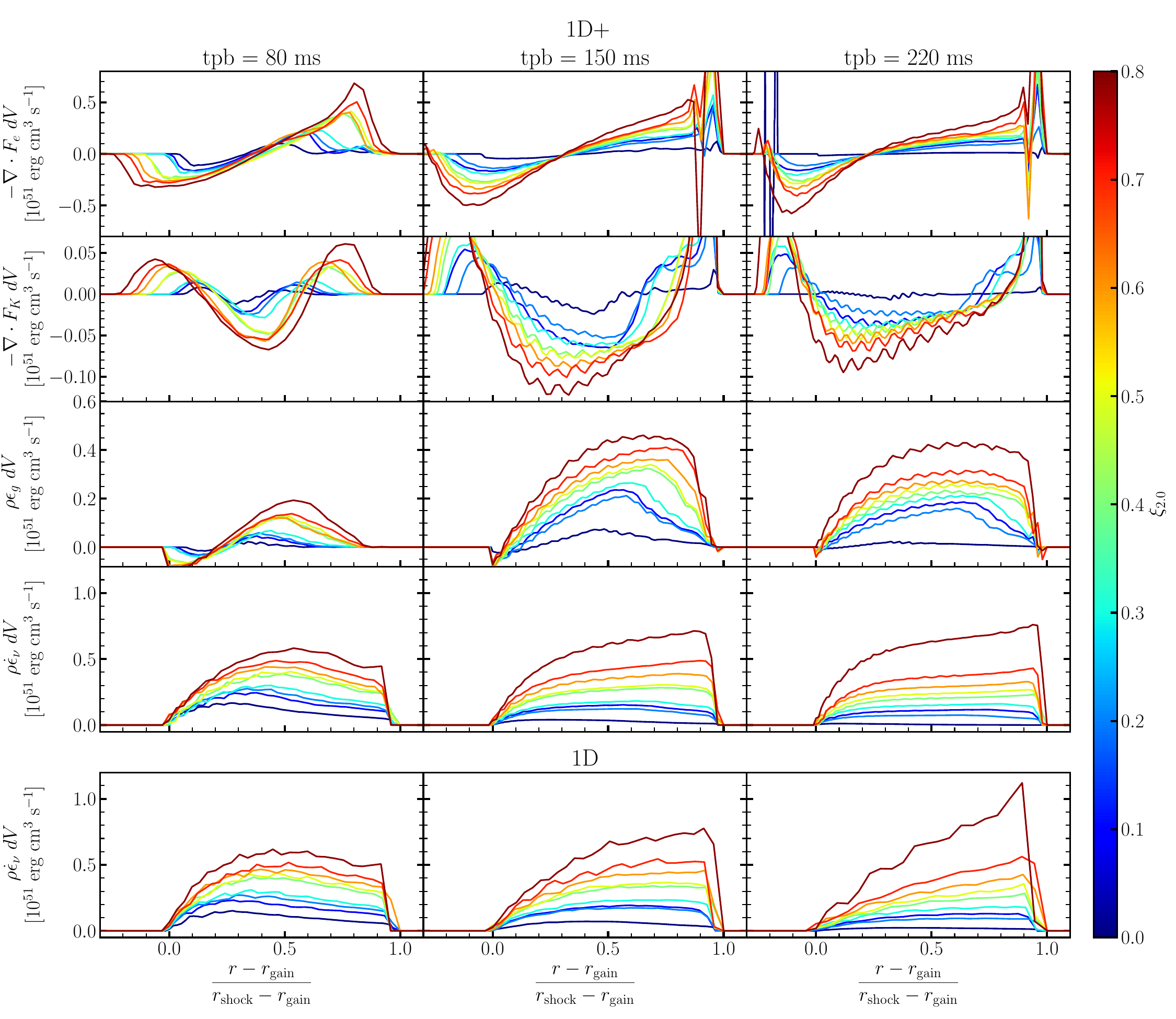}
\caption{The first four rows of this Figure show the 9 1D+ simulations for the progenitors listed in Table~\ref{tab:progs_info} color-coded by compactness. The last row instead show the same 9 progenitors but refers to a simple 1D simulation. Each column shows snapshots at three different times post-bounce. Notice that all quantities have been multiplied by the volume to highlight the contribution to the total net heating at each radius. The x-axis shows the dimensionless radial coordinate within the gain region. This choice is motivated by the need to show different simulations with different shock radii and radial extents of the gain region in the same canvas. The total net neutrino heating rate and rate of turbulent energy generation integrated over the gain region for these same simulations is shown in Section~\ref{sec:semi_an_qdot}. Notice the different ranges for the y-axis. Only the fourth and fifth rows have the same. }
\label{fig:spec_heat}
\end{figure*}

With an appropriate closure relation, the Reynolds decomposed equations can also be used to include the effect of convection into spherically symmetric simulations, also referred to as 1D+ models. The closure is used to determine the rate of turbulent energy dissipation $\epsilon_k$, the Reynolds stress, and the internal energy flux. In this paper, we consider the STIR model, originally developed by \citet{Couch2020_STIR} and generalized to GR hydrodynamics by \citet{Boccioli2021_STIR_GR}, although similar models have also been developed \citep{Mabanta2018_MLT_turb,Sasaki2024_STIR_diffusion}. The closure utilized by \citet{Couch2020_STIR} is based on mixing length theory, i.e. the internal energy flux is treated as a diffusion term, the Reynolds stress is expressed in terms of the convective velocity, and the turbulent energy dissipation is expressed in terms of the mixing length. To properly account for all of the extra terms due to turbulence, the Euler equations should be decomposed in a background and perturbed component, and more details on this can be found in the original papers \citep{Murphy2011,Murphy2013_turb_in_CCSNe,Mabanta2018_MLT_turb,Couch2020_STIR}. Here, we simply show the evolution equation for the turbulent energy in the STIR model:
\begingroup
\allowdisplaybreaks
\begin{align}
    \label{eq:vturb_eq}
    &\pder{\rho v_{\rm turb}^2}{ t} + \frac{1}{r^2}\pder{}{r} [r^2 v_r(\rho v_{\rm turb}^2)] + \nabla \cdot F_K = \rho \epsilon_g - \rho \epsilon_k, \\
    \label{eq:epsk_eq}
    &\rho \epsilon_k = \rho \frac{v_{\rm turb}^3}{\Lambda_{\rm MLT}}, \\
    \label{eq:epsk_g}
    &\rho \epsilon_g = -\rho v_{\rm turb}^2 \pder{v_r}{r} + \rho v_{\rm turb} \omega_{\rm BV}^2 \Lambda_{\rm MLT} = W_s + W_b, \\
    \label{eq:lambda_eq}
    &\Lambda_{\rm MLT} = \alpha_{\rm MLT} \frac{P}{\rho g}, \\
    \label{eq:FK_eq}
    &F_K = -\alpha_{\rm MLT} D_K \rho \nabla v_{\rm turb}^2 = -\alpha_{{\rm D}_K} v_{\rm turb} \Lambda_{\rm MLT} \rho \nabla v_{\rm turb}^2.
\end{align}
\endgroup
In the above equations, $r$ is the radius, $\rho$ is the density, $v_{\rm turb}$ is the turbulent (or convective) velocity, $v_r$ is the radial velocity, $\Lambda_{\rm MLT}$ is the mixing length, $\omega_{\rm BV}$ is the \BV/ frequency, $P$ is the pressure. We also define the quantities $\rho \epsilon_g$ and $\rho \epsilon_k$ to be the rates of generation and dissipation of turbulent kinetic energy per volume, respectively. The term $F_K$ indicates the turbulent energy flux, and in this model, it is treated as a diffusion term, where $D_K$ is its respective diffusion coefficient. 

The most important parameter in this model is $\alpha_{\rm MLT}$, a parameter of $\mathcal{O}(1)$ that effectively controls the strength of convection. The other parameter that appears in the above equations is $\alpha_{{\rm D}_K}$, which controls how fast the diffusion of turbulent energy is. Similar terms are present in the equations for internal energy, electron fraction, and neutrino energy density, each with their own free parameter analogous to $\alpha_{{\rm D}_K}$. In accordance with several other studies in the literature \citep{Muller2016_Oxburning,Couch2020_STIR}, all of these parameters are kept constant at a value of $1/6$. However, see \citet{Sasaki2024_STIR_diffusion} for a study where the strength of diffusion is changed independently for different quantities.

As seen in Eq.~\eqref{eq:epsk_g}, the turbulent kinetic energy can be generated via shear ($W_s$) or buoyancy  ($W_b$). In the full set of hydrodynamic equations, the term $\epsilon_k$ is added as a positive source term to the RHS of the evolution equation for the internal energy. However, as pointed out by \citet{Gogilashvili2024_FEC+}, it is the term $W_b$ (i.e. the term responsible for generating the convection) that enters the equation for the total energy. 

It should be highlighted that in the literature, the shear term is usually neglected since it is quite small, but is included in STIR. Therefore, it is the term $\rho \epsilon_g = W_s + W_b$ that is responsible for generating turbulent energy. This can be more clearly seen if one considers the equation for the total energy, which includes the contribution from internal, kinetic, and turbulent energy:
\begin{equation}
\begin{split}
\label{eq:STIR_tot_ene}
\pder{\rho e_{\rm tot}}{ t} &+ \frac{1}{r^2}\pder{}{r} [r^2 v_r(\rho e_{\rm tot} + P + P_{\rm turb})] + \nabla \cdot F_e + \nabla \cdot F_K\\ 
&= -\rho v_r g + \dot{q}_\nu + \rho \epsilon_g,
\end{split}
\end{equation}
where $F_e$ is a term analogous to $F_K$, where instead of the gradient of $v_{\rm turb}^2$ one has the gradient of internal energy, $g$ the local gravitational acceleration, and $\dot{q}_\nu = \rho \dot{\epsilon}_\nu$ is the net neutrino heating rate. As mentioned above, the main effects of convection are the generation and dissipation of turbulent kinetic energy into heat (i.e. $\rho \epsilon_g$ and $\rho \epsilon_k$) and transport of energy within the gain region (i.e. $\nabla \cdot (F_e + F_K)$). In addition to the turbulent pressure $P_{\rm turb}$, which is quite small, these are the most important extra terms added to the spherically symmetric Euler equations. 

\subsection{Neutrino heating in 1D+ models}
\label{sec:nu_heat_1D+}
The main "extra" 1D+ contribution to the spherically symmetric Euler equations is $\rho \epsilon_g - \nabla \cdot F_e - \nabla \cdot F_K$, as can be seen from Eq.~\eqref{eq:STIR_tot_ene}. To quantify how much neutrino-driven convection changes the overall heating in the gain region, we selected 9 different 1D and 1D+ simulations from \citet{Boccioli2024_remnant} for 9 different values of compactness $\xi_{2.0}$. The 1D+ and 1D simulations only differ for the inclusion of the STIR turbulent model, with everything else being equal. The properties of the selected progenitors are shown in Table~\ref{tab:progs_info}. These simulations were run with GR1D \citep{OConnor2010}, an open-source spherically symmetric, general relativistic code with an M1 neutrino transport and neutrino opacities from the open-source library NuLib, described in \citet{OConnor2015}. The publicly available code\footnote{https://github.com/evanoconnor/GR1D} has been modified with the addition of STIR \citep{Couch2020_STIR,Boccioli2021_STIR_GR} described in the previous section. More details on the numerical setup adopted for GR1D can be found in \citet{Boccioli2021_STIR_GR} and \citet{Boccioli2024_remnant}.

To illustrate the effects of convection and neutrino heating in the STIR model we show in Figure~\ref{fig:spec_heat} the time and radial dependence of several key quantities for these 9 1D+ and 1D simulations. Notice that, to better show how much each quantity contributes to the overall heating, each quantity is multiplied by the volume of each zone. The first two rows show the (negative) divergence of the internal and turbulent energy convective fluxes (i.e. $-\nabla \cdot F_e$ and  $-\nabla \cdot F_K$) in the 1D+ simulations. As mentioned at the start of this section, these can be thought of as additional source terms in Eq.~\eqref{eq:STIR_tot_ene}. The term $-\nabla \cdot F_e$ represents the net change in internal energy resulting from the transport of hot bubbles closer to the shock. One would expect that below the gain region, i.e. where the entropy gradient vanishes, $-\nabla \cdot F_e$ would vanish. However, this is not the case and it has been pointed out by \citet{Muller2019_STIR} that this is indeed a potential issue of the STIR model. This relatively small inconsistency is caused by an inadequate expression for the convective flux $F_e$, which instead of being expressed in term of $\nabla \epsilon$, as done in STIR, should instead depend on $\nabla \epsilon + P \nabla (1/\rho)$. 

In the context of this paper, we allow for this inconsistency that, as can be inferred by the upper panels of Figure~\ref{fig:spec_heat}, causes an overall positive net heating when $-\nabla \cdot F_e$ is integrated over the gain region. Importantly, even if $-\nabla \cdot F_e$ would vanish at the bottom of the gain region, causing vanishing net heating within the gain region, the material behind the shock would still experience positive net heating, which is what is most important for shock revival. As shown in the second row, $-\nabla \cdot F_K$ is instead negative throughout the gain region, which indicates that turbulent energy is taken from the center of the region (i.e. the minimum of $-\nabla \cdot F_K$, which corresponds to the maximum of $v_{\rm turb}$) and transported away from it. Turbulent energy is therefore transported toward the shock or toward the bottom of the gain region, depending on whether it is located above or below the radial coordinate of the maximum of $v_{\rm turb}$ (i.e. the minimum of $-\nabla \cdot F_K$). At the edges of the convective region $-\nabla \cdot F_K$ is instead positive, since $v_{\rm turb}$ vanishes. As expected, $-\nabla \cdot F_K$ is roughly zero when integrated over the gain region. 

The third row of Figure~\ref{fig:spec_heat} shows $\rho \epsilon_g$, which has a similar radial dependence of the neutrino heating shown on the fourth row. This is consistent with the fact that neutrino heating determines the entropy profile in the gain region, which in turn determines the strength of convection. So far, the quantities in the first four rows were discussed for the 1D+ simulations. The fifth row shows instead the neutrino heating for the respective 1D simulations. As one can see, the radial dependence and magnitude of $\rho \epsilon_\nu$ are the same in 1D and 1D+ models. However, as we will show in detail in Section~\ref{sec:semi_an_qdot} the crucial difference is the shock radius and hence the size of the gain region, which is much larger in 1D+ models. Therefore, the total net heating in the gain region is much larger in the 1D+ simulations because the size of the gain region is larger. We will come back to this later in the manuscript. 

One exception is the last snapshot at $\sim 220$ ms, which shows a different radial dependence in the 1D simulation, where $\rho \epsilon_\nu\, dV$ quickly increases behind the shock. When the shock recedes, as is the case for the last snapshot, the dependence of the net neutrino heating becomes inversely proportional to radius. Since $dV \propto r^2$, overall, one can see a linear increase in $\rho \dot{\epsilon}_\nu dV$. The higher compactness progenitors will have the same linear dependence on radius, although steeper, since luminosities and neutrino energies are larger. The reason why the net neutrino heating is inversely proportional to radius can be understood by the fact that, in general, neutrino heating is proportional to $1/r^2$, whereas neutrino cooling is proportional to $1/r^6$ in a radiation-dominated region. When the shock recedes, the entire gain region will be located at small radii, where the cooling is larger, hence softening the overall dependence of net neutrino heating from $1/r^2$ to $\sim 1/r$.


\begin{table}
\centering 
\begin{tabular}{ccccc}
\toprule
$\xi_{2.0}$ & M$_{\rm ZAMS}$ & Metallicity & Explosion & Reference \\
\midrule
$5\times 10^{-5}$ & 9.0 & $z_\odot$ & yes & \citet{Sukhbold2016_explodability} \\
0.1 & 13.0 & 0 & no & \citet{Woosley2002_KEPLER_models} \\
0.2 & 18.0 & $10^{-4} z_\odot$ & yes & \citet{Woosley2002_KEPLER_models} \\
0.3 & 15.4 & $10^{-4} z_\odot$ & no & \citet{Woosley2002_KEPLER_models} \\
0.4 & 25.4 & $10^{-4} z_\odot$ & no & \citet{Woosley2002_KEPLER_models} \\
0.5 & 27.4 & $10^{-4} z_\odot$ & yes & \citet{Woosley2002_KEPLER_models} \\
0.6 & 24.7 & $z_\odot$ & yes & \citet{Sukhbold2016_explodability} \\
0.7 & 22.6 & $z_\odot$ & yes & \citet{Sukhbold2016_explodability} \\
0.8 & 39.0 & $10^{-4} z_\odot$ & yes & \citet{Woosley2002_KEPLER_models} \\

\bottomrule
\end{tabular}

\caption{Summary of the properties of the progenitors used in Section~\ref{sec:semi_an_qdot}. The simulations were performed in \citet{Boccioli2024_remnant}. \label{tab:progs_info}}
\end{table}

Progenitors with higher compactness are characterized by shallower density profiles, and therefore larger mass accretion rates. As a consequence, they generate larger neutrino heating, as can be seen in Figure~\ref{fig:spec_heat}, while at the same time, they also generate larger ram pressure onto the shock. In the remainder of this paper, we will show that high-compactness progenitors, in the presence of $\nu$-driven convection, are able to generate large neutrino heating very efficiently, which has a direct impact on explodability.

The energy associated with neutrinos can be described by:
\begin{equation}
    \label{eq:Qdot}
    \dot{Q}_\nu = \int_{r_{\rm gain}}^{r_{\rm shock}} \dot{q}_\nu dV = \int_{r_{\rm gain}}^{r_{\rm shock}} \rho \dot{\epsilon_\nu} dV
\end{equation}
which is the net neutrino heating rate in the gain region. The energy associated with neutrino-driven convection can instead be described by:
\begin{align}
    \label{eq:Eg}
    E_g &= \int_{r_{\rm gain}}^{r_{\rm shock}} \rho \epsilon_g dV \\
    \label{eq:Ek}
    E_k &= \int_{r_{\rm gain}}^{r_{\rm shock}} \rho \epsilon_k dV \\
\end{align}
which represent the rates of turbulent generation and dissipation, respectively. Notice that, in principle, one expects $E_g \approx E_k$, since the overall turbulent energy generated should balance the turbulent energy dissipated, as shown by \citet{Murphy2013_turb_in_CCSNe}. This is however not necessarily true in STIR, and instead, $E_g$ is about $10 \%$ larger than $E_k$. The origin of this difference lies in the fact that when one integrates Eq.~\eqref{eq:vturb_eq} over the gain region, $\nabla \cdot F_e$ is non-zero as mentioned above. Instead, the first and second terms of Eq.~\eqref{eq:vturb_eq} are both roughly zero when integrated over the gain region, a sign that during the accretion phase, the gain region is in a quasi-steady-state.

It should be highlighted that $\epsilon_g$ and $\epsilon_k$ are rates of turbulent dissipation and generation, and therefore $\epsilon_{g,k}$ and $E_{g,k}$ are dimensionally equivalent to $\dot{\epsilon}_\nu$ and $\dot{Q}_\nu$, respectively. This makes the notation slightly ambiguous since one might expect those quantities to be defined as $\dot{\epsilon}_{g,k}$ and $\dot{E}_{g,k}$. However, to be consistent with previous literature \citep{Murphy2013_turb_in_CCSNe,Mabanta2018_MLT_turb,Couch2020_STIR}, we decided to omit the dot.


\section{Comparison between STIR and multi-dimensional simulations}
\label{sec:comparison_multiD}
\begin{figure*}
\centering
\includegraphics[width=\textwidth]{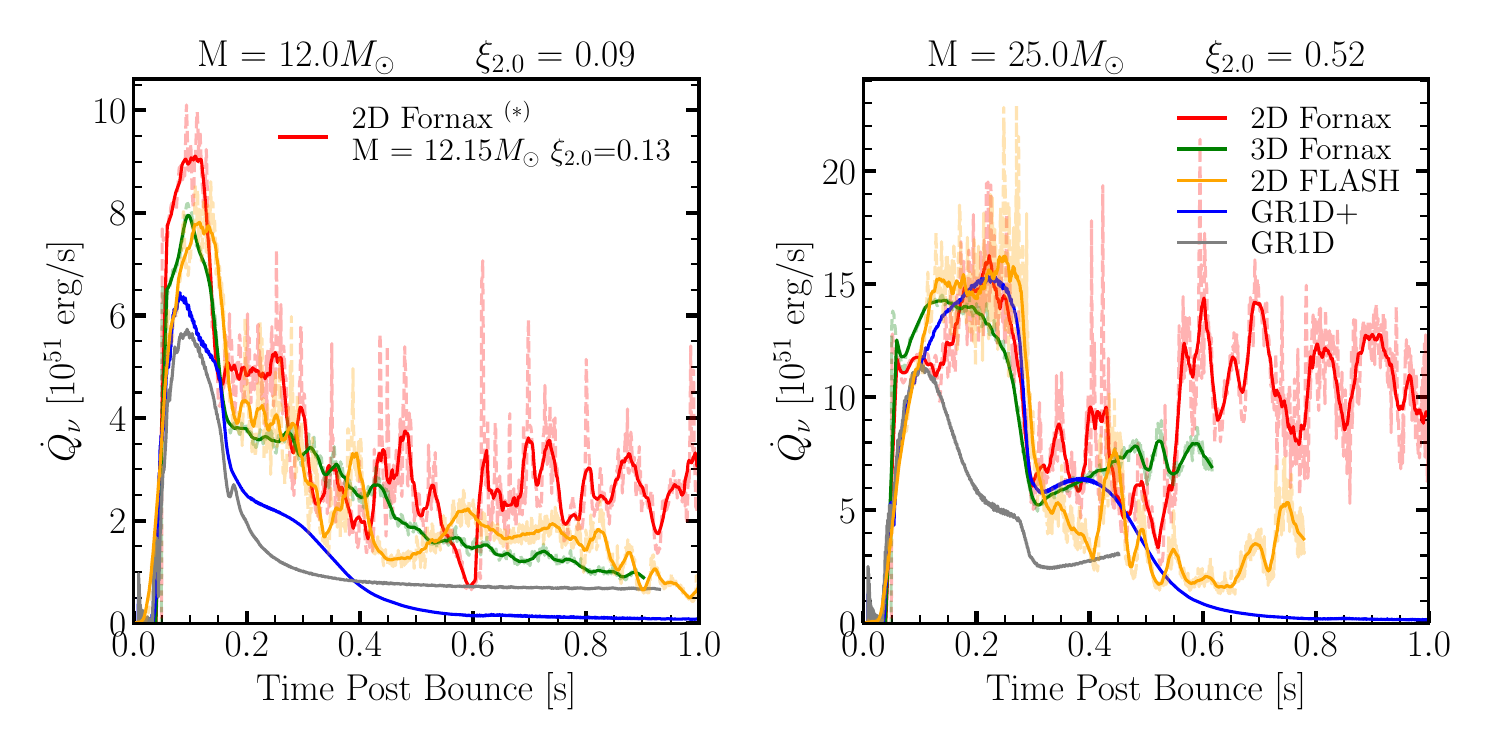}

\caption{Comparison of neutrino heating in the gain region in 1D, 1D+, 2D and 3D simulations. For the multi-dimensional simulations, we show both the raw data as a faded dashed line and a smoothing over a window of 20 ms as a solid line. The left panel shows the comparison for a 12 $M_\odot$ progenitor from \citet{Sukhbold2016_explodability}. However, notice that the 2D F{\sc{ornax}} simulation was done for a 12.15 $M_\odot$ $M_\odot$ progenitor from \citet{Sukhbold2018_preSN_KEPLER_bug}, which has a very similar compactness $\xi_{2.0}$. The right panel shows the comparison for a 25 $M_\odot$ progenitor from \citet{Sukhbold2016_explodability}, which has instead a relatively high compactness. There are some discrepancies between the two 2D simulations in the left panel, where the 1D+ simulations also struggle to produce significant neutrino heating. All simulations in the right panel agree quite well except for the 1D simulation, as expected, where no convection is present.}
\label{fig:compare_1D_2D_3D}
\end{figure*}

To illustrate how the STIR model compares to multi-dimensional simulations, we compare neutrino heating in STIR with neutrino heating in state-of-the-art 2D and 3D F{\sc{ornax}} and 2D FLASH simulations. Specifically, we will analyze 98 2D and 20 3D F{\sc{ornax}} simulations and 55 2D FLASH simulations. The 2D F{\sc{ornax}} simulations \citep{Vartanyan2023_nu_100_2D} were performed on solar metallicity progenitors from \citet{Sukhbold2016_explodability} with initial masses between 9 and 12 $M_\odot$, and on solar metallicity progenitors from \citet{Sukhbold2018_preSN_KEPLER_bug} with initial masses between 12 and 26.99 $M_\odot$. The 3D F{\sc{ornax}} simulations \citep{Burrows2020_3DFornax} were performed on solar metallicity progenitors from \citet{Sukhbold2016_explodability}, with masses ranging from 12 to 60 $M_\odot$, as well as a $25 M_\odot$ progenitor from \citet{Sukhbold2018_preSN_KEPLER_bug}. All of the F{\sc{ornax}} simulations adopted the SFHo equation of state. The 2D FLASH simulations were performed on different progenitors with masses from $12-60 M_\odot$ from \citet{Sukhbold2016_explodability} as well as a few from \citet{WH07} and some with artificially reduced mass loss from \citet{Sukhbold2018_preSN_KEPLER_bug}. Of these progenitors, 37 were simulated using the SFHo EOS, whereas the rest were simulated using other EOSs from \citet{Schneider2019}. The FLASH simulations are collected from various studies but also from publications in preparation \citep{Eggenberger2021_EOS_dependence_GW_2D,Eggenberger2025_BHSNe_EOS,Li2024_2D_EOS_GW_thesis, Andresen2024_GreyM1}.

Several groups have performed comparisons between different codes, although usually it has been done for simulations in either 1D \citep{OConnor2018_comparison}, 2D \citep{Pan2019_nu_transport_2D_comparison}, or 3D \citep{Cabezon2018_3Dcomparison,Glas2019_comparison_nu_transport_AA}, and more rarely have 1D simulations been thoroughly compared to multi-dimensional ones \citep{Richers2017_nu_transport_comparison,Just2018_1D2D_Vertex_vs_Alcar,Nagakura2020_PNS_convection}. The latter is a particularly difficult comparison to perform given that 1D simulations do not explode, whereas multi-dimensional ones often do, and therefore the most significant differences between 1D and multi-D results can be attributed to the different outcome rather than specific multi-dimensional effects.

In this section, we focus on comparing neutrino heating in 1D, 1D+, 2D, and 3D simulations. We adopt simulations already performed with other studies in mind, which means that neutrino transport, opacities, and overall simulation setups are not necessarily the same across codes. These differences prevent a detailed and thorough comparison study, which is however beyond the scope of this paper. Therefore, some of the discrepancies between 1D+ and multi-D, as well as discrepancies between the 2D FLASH and F{\sc{ornax}} simulations, are to be expected. The largest differences are most likely due to neutrino opacities and transport methods, and we refer to the original papers listed at the beginning of this section for details regarding the specific neutrino treatments in the different codes.

In Figure~\ref{fig:compare_1D_2D_3D} we show the net neutrino heating in the gain region as a function of time for a low-compactness (left panel) and a high-compactness (right panel) progenitor. For the low-compactness progenitor, we compare simulations of the $12 M_\odot$ progenitor from \citet{Sukhbold2016_explodability}. The only exception is the 2D F{\sc{ornax}} simulations, for which we selected the $12.15 M_\odot$ from \citet{Sukhbold2018_preSN_KEPLER_bug}. The reason is that, among the progenitors used in the 2D suite of simulations, it is the one with the closest compactness $\xi_{2.0}$ to the $12 M_\odot$ progenitor from \citet{Sukhbold2016_explodability}. For the high-compactness progenitor, we compare simulations of the $25 M_\odot$ progenitor from \citet{Sukhbold2016_explodability}. All simulations lead to a successful explosion, except for the 1D case, for which we only expect a successful explosion for very few edge cases.

In general, neutrino heating (but also several other quantities) in 2D simulations tends to be a lot noisier than in 3D, even though the neutrino luminosities of 2D and 3D F{\sc{ornax}} simulations agree very well at early times $\lesssim 1$s \citep{Vartanyan2023_nu_100_2D}. Because of the imposed axial symmetry, 2D simulations are characterized by much larger structures than their 3D counterparts. The average in the phi direction in 3D simulations is what smooths out the large oscillations. Indeed, in a 3D simulation, the neutrino heating calculated in different wedges at different values of phi, show large variations in neutrino heating up to $\sim 20 \%$, which are then smoothed out by the phi-average. This is the main reason why neutrino heating in 2D simulations does not average out as much compared to 3D.

For the low-compactness progenitor, the 2D F{\sc{ornax}} simulations produce a larger neutrino heating at early times, compared to both the 2D FLASH and 3D F{\sc{ornax}} simulations, which instead seem to agree relatively well. This is seen for all of the lower-compactness progenitors in the 2D F{\sc{ornax}} models compared to 3D F{\sc{ornax}} and 2D FLASH models (see Section~\ref{sec:comparison_multiD} for more details). The larger neutrino heating in the 2D F{\sc{ornax}} simulations can be caused by a number of factors, such as the different geometry, grid resolution, and neutrino opacities and transport schemes. Here we simply describe these differences and argue what their potential impact might be, but a thorough analysis of this discrepancy is beyond the scope of this work. 

FLASH and F{\sc{ornax}} are both Newtonian codes, that include general relativistic corrections to the monopole gravitational potential using Case A of \citet{Marek2006}. The grid resolutions for all the multi-D simulations are overall quite similar, with the smallest radial zone being $\sim 500$ m. The hydrodynamics solver is however slightly different. FLASH simulations adopt an unsplit hydrodynamic solver, Spark \citep{Couch2021_Spark}, which makes use of an adaptive mesh refinement grid and uses fifth order WENO reconstruction. F{\sc{ornax}} instead is a finite-volume code \citep{Skinner2019_Fornax_methods_paper} adopting a ``dendritic grid", i.e. a static-mesh refinement near the center and the poles, and adopts a custom reconstruction algorithm described in \citet{Burrows2020_3DFornax}. Both codes use an HLLC Riemann solver that reduces to HLLE near shocks. The neutrino transport scheme is also very similar, and both codes adopt a multi-dimensional, multi-energy transport algorithm that solves the Boltzmann equations using an M1-scheme and the maximum-entropy Minerbo closure \citep{Minerbo1978_closure}. F{\sc{ornax}} solves the transport equations in the co-moving frame, whereas FLASH solves them in the laboratory frame. The opacities adopted are, for the most part, the same, but we refer to the method papers cited above for more details. The most significant difference between 2D F{\sc{ornax}} models and 2D FLASH models is perhaps the adoption in F{\sc{ornax}} of a Kompaneets approach for inelastic neutrino-nucleon scattering \citep{Wang2020_Kompaneets_NNS}, whereas the FLASH simulations only consider the elastic version of this process. This is also the main difference (other than, of course, geometry) between the 2D and 3D F{\sc{ornax}} simulations. That said, it seems however unlikely that this could cause such large differences in neutrino heating at early times, as seen for the 12.0 and 12.15 $M_\odot$ progenitors, although more comparison work in this direction is needed. Notably, no thorough comparisons among state-of-the-art multi-dimensional codes regarding the impact of different neutrino opacities have been carried out yet, due to their complexity.

An important detail worth mentioning is that, depending on how the gain region is defined, certain computational zones might be included or excluded from the integral in Eq.~\eqref{eq:Qdot}. Therefore, it is important to define a cut in density (i.e., exclude zones with densities above $3\times10^{10}\ {\rm g/cm^3}$) to exclude zones inside the PNS which might experience large numerical fluctuations in net heating. However, this is not enough since some zones just above the PNS might still be included in the integral, even though they do not influence the neutrino heating in the vicinity of the shock, which ultimately is what Eq.~\eqref{eq:Qdot} should reflect. Requiring the specific entropy per baryon to be above 6 ensures that only material in the shocked region is not included in the integral. Not including this extra condition was seen to increase the neutrino heating in the 2D F{\sc{ornax}} simulations by up to $\sim 10 \%$, and it is therefore important to take it into account.

For the high compactness progenitor, the 2D and 3D simulations produce comparable neutrino heating, with the 3D simulation decreasing a bit earlier compared to the 2D simulations, which instead keep increasing for a few tens of milliseconds before dropping. Both of the F{\sc{ornax}} simulations seem to show an increase in neutrino heating at late times, which is also seen for other progenitors, but it does not occur in the 2D FLASH simulation. This can be caused by more pronounced asymmetries and late-time accretion, but a more detailed analysis goes beyond the scope of this work.

For the low-compactness progenitor, the 1D+ simulations underestimate the neutrino heating at early times by 10--15~$\%$ compared to the 3D F{\sc{ornax}} and 2D FLASH simulations, and by 25--30~$\%$ compared to the 2D F{\sc{ornax}} simulation. As one can see by comparing the 2D simulations for the low and high-compactness progenitors, the neutrino heating for the low-compactness progenitor increases very rapidly, and it starts deviating from the 1D case at around 70--80~ms. For the high-compactness progenitor, this increase is instead more gradual, and neutrino heating starts deviating from the 1D case at around $100$ ms. Therefore, a possible explanation for this could be prompt convection, which is not properly captured by STIR. Moreover, the lower neutrino heating found in STIR explains the tendency of this 1D+ model to yield failed explosions for the low compactness progenitors. Although 2D simulations also show several failed explosions for these lower-compactness progenitors \citep{OConnor2018_2D_M1,Vartanyan2023_nu_100_2D}, typically with initial masses in the range $12 M_\odot < M_{\rm ZAMS} < 15 M_\odot$, 1D+ models generally produce even less explosions. 

For the high-compactness progenitor, neutrino heating in the 1D+ simulation is overestimated by $\sim 5 \%$ compared to the 3D F{\sc{ornax}} simulation, whereas it agrees remarkably well with the 2D F{\sc{ornax}} and FLASH simulations. Considering the difference in neutrino transport, opacities, and geometry, one can consider the 1D+ model to be in extremely good agreement with all of the multi-dimensional simulations. 

Overall, the $25 M_\odot$ produces a much larger neutrino heating than the $12 M_\odot$ and $12.15 M_\odot$ progenitors, due to its higher compactness. This is not surprising, since neutrino heating is larger when mass accretion rates (and therefore compactness) are larger.

\subsection{Compactness dependence of neutrino heating}
In this section, we will focus on several suites of simulations in 1D, 1D+, 2D and 3D, to study how neutrino heating depends on the progenitor compactness. For that purpose, we will use all of the 2D and 3D simulations described above. In addition to that, we selected several suites of both 1D and 1D+ simulations ran with \texttt{GR1D}: 309 FRANEC progenitors \citep{Chieffi2021_C12_compactness} from \citet{Boccioli2023_explodability}; 341 KEPLER progenitors \citep{Woosley2002_KEPLER_models,Sukhbold2016_explodability} from \citet{Boccioli2024_remnant}; 88 (only 1D+ simulations) KEPLER progenitors \citep{Sukhbold2018_preSN_KEPLER_bug}, which are the same progenitors as the 2D F{\sc{ornax}} simulations by \citet{Vartanyan2023_nu_100_2D}. Finally, we also selected 148 1D FLASH simulations \citep{Segerlund2021_distance_SN_nns} of KEPLER progenitors from \citet{Sukhbold2016_explodability} with masses between 9 and 120~$M_\odot$, as well as 40 1D+ FLASH simulations \citep{Couch2020_STIR} with masses from 9 to $30 M_\odot$. In total, we therefore have 173 multi-D simulations, 778 1D+ simulations, and 798 1D simulations, for a total of $\sim 1500$ simulations.

To quantitatively analyze neutrino heating during the explosion of stars with different compactness, we chose to use $\dot{Q}_\nu^{\rm max}$, i.e. the maximum value that $\dot{Q}_\nu$ reaches during the simulation, as a proxy of the strength of neutrino heating. An alternative could be to use the average of $\dot{Q}_\nu$ during the accretion phase, or more in general during some relevant time window. However, as seen in Figure~\ref{fig:Qdot} and discussed later in the paper, $\dot{Q}_\nu$ increases more rapidly in time for progenitors with higher compactness, for which the accretion phase also tends to be longer. By averaging in time one would therefore be less sensitive to the fact that longer accretion phases cause larger neutrino heating and therefore favor explosions. We ultimately decided to adopt the much simpler $\dot{Q}_\nu^{\rm max}$ as a proxy for the strength of neutrino heating during the simulations, which also makes the interpretation of the results much more straightforward (see for example the derivation in Appendix~\ref{sec:appendix_Qdot_comp}).

Figure~\ref{fig:Qdot_vs_comp_multiD} shows $\dot{Q}_\nu^{\rm max}$ as a function of the pre-supernova compactness for the $\sim 1500$ simulations described above. The bottom panel is color-coded based on explosion outcome, and it shows that high-compactness progenitors above $\xi_{2.0} \sim 0.5$ lead to successful explosions both in multi-D, and 1D+ simulations, with the only exception of four progenitors simulated with $\texttt{GR1D+}$ and one progenitor simulated in 2D with FLASH. The former are a 25 $M_\odot$ progenitor at zero metallicity and the 23.2, 23.8, and 24 $M_\odot$ at a metallicity of $10^{-4} z_\odot$ from \citet{Woosley2002_KEPLER_models}, where $z_\odot$ is solar metallicity. The latter is a 35 $M_\odot$ progenitor from \citet{WH07} simulated using a very stiff equation of state, i.e. the Skyrme-type EOS with effective nucleon mass $m^*$ = 0.55 from \citet{Eggenberger2021_EOS_dependence_GW_2D}, and therefore it is not surprising that it does not explode. 

Notice that if one were to use a stiffer equation of state, the overall curve $\dot{Q}_\nu^{\rm max}$ vs. $\xi_{2.0}$ would be shifted down, since stiff equations of state lead to colder neutrinospheres, and therefore lower neutrino heating \citep{Schneider2020_EOS_dependence_BH,Eggenberger2021_EOS_dependence_GW_2D}. The few 2D FLASH simulations performed with stiffer equations of state (in particular, with lower effective nucleon masses) are compatible with this finding.

From Figure~\ref{fig:Qdot_vs_comp_multiD} we can confirm that modern multi-dimensional simulations clearly show that high-compactness progenitors explode. It is however important to define what "explosion" means in this case. All of the simulations analyzed in this paper only simulate up to 1-5 seconds after bounce, depending on the model. In all simulations of high-compactness progenitors (except for very stiff EOSs, as discussed above), the shock was successfully revived and expanded to radii of thousands of kilometers. Whether the shock will successfully break out of the star and, more importantly, what energy that explosion will be, is however not obvious. Several studies \citep{Chan2018_BH_SN_40Msol,Powell2021_collapse_PISNe,Burrows2023_BH_supernova_40Msol,Eggenberger2025_BHSNe_EOS,Sykes2024_2D_fallback_SNe,Burrows2024_BH_formation_3D} have shown that it is indeed possible to achieve an explosion (with energies between one-hundredth of a Bethe to a few Bethe) of these high-compactness progenitors. In particular, \citet{Sykes2024_2D_fallback_SNe} have shown that as long as a black hole forms after the shock has propagated past the sonic point within the envelope, the shock will break out of the star. With all of that considered, we can assume that simulations, where the shock has been successfully revived, yield an explosion. Therefore, we assume that all of the 2D and 3D simulations with compactness $\xi_{2.0} \gtrsim 0.5$ explode.

We can also conclude from Figure~\ref{fig:Qdot_vs_comp_multiD} that significant neutrino heating is generated for progenitors with high compactness, and 1D+ simulations can reliably reproduce this trend. The difference with the 1D simulations starts growing at compactnesses $\xi_{2.0} \sim 0.3-0.4$ and then experiences an even more rapid increase at compactnesses $\xi_{2.0} \gtrsim 0.7$. The two trends shown in Figures~\ref{fig:Qdot_vs_comp_multiD} can be derived from simple fits based on a semi-analytical expression of neutrino heating (see Section~\ref{sec:semi_an_qdot} and Appendix~\ref{sec:appendix_Qdot_comp}). For 1D simulations, one has:
\begin{equation}
    \label{eq:Qdot_fit_1D}
    \dot{Q}_\nu^{\rm max} = 5.18 \times 10^{51}\ \frac{\rm erg}{\rm s} \left(0.82 + 0.37\, \xi_{2.0}^{1/3} + 1.82\, \xi_{2.0}^{5/3} + 0.81\, \xi_{2.0}^{2} \right),
\end{equation}
For multi-D and 1D+ simulations one instead has, for progenitors with $\xi_{2.0} < 0.72$: 
\begin{equation}
    \label{eq:Qdot_fit_STIR1}
    \dot{Q}_\nu^{\rm max} = 5.18 \times 10^{51}\ \frac{\rm erg}{\rm s} \left(0.66 + 0.30\, \xi_{2.0}^{4/3} + 1.45\, \xi_{2.0}^{5/3} + 0.65\, \xi_{2.0}^{3} \right),
\end{equation}
and for progenitors with $\xi_{2.0} \geq 0.72$:
\begin{equation}
    \label{eq:Qdot_fit_STIR2}
    \dot{Q}_\nu^{\rm max} = 5.18 \times 10^{51}\ \frac{\rm erg}{\rm s} \left(-27.7 -66.6 \, \xi_{2.0}^{1/3} + 36.0 \, \xi_{2.0}^{2/3} + 87.0 \, \xi_{2.0}\right).
\end{equation}
We will quantitatively explain this trend in the next section and in more detail in Appendix \ref{sec:appendix_Qdot_comp}. 

Another remarkable conclusion that can be drawn from Figure~\ref{fig:Qdot_vs_comp_multiD} is that, for a given nuclear EOS, the trend of $\dot{Q}_{\nu}^{\rm max}$ as a function of compactness is very robust. This is confirmed by the fact that all of the 1D+ simulations lead to very similar $\dot{Q}_\nu^{\rm max}$, despite the simulation setups and progenitors adopted being quite different. Therefore, one can conclude that details of stellar evolution do not enter the dependence of $\dot{Q}_{\nu}^{\rm max}$ on compactness, although it should be noted that at compactnesses $\xi_{2.0} \sim 0.6$ the FRANEC progenitors seems to generate a bit more heating than the KEPLER ones. 

It is also instructive to run two identical 1D+ models implemented in different codes \citep{Couch2020_STIR,Boccioli2021_STIR_GR}. The 1D+ FLASH simulations of the KEPLER progenitors from \citet{Sukhbold2016_explodability} follow the same trend of the \texttt{GR1D+} simulations, with a systematically slightly lower neutrino heating. The reason is that 1D+ models can be quite sensitive to the mixing-length parameter $\alpha_{\rm MLT}$. Moreover, FLASH adopts modified Newtonian gravity, whereas \texttt{GR1D+} is a general relativistic code. Therefore, the expression for the \BV frequency is different, which results in different calibration of the mixing length parameter $\alpha_{\rm MLT}$, which is $\sim 1.23-1.27$ in FLASH \citep{Couch2020_STIR} and $\sim 1.48-1.52$ in GR1D \citep{Boccioli2021_STIR_GR}. The fact that the heating is consistently lower in FLASH for all progenitors indicates that if one were to use a slightly higher value of $\alpha_{\rm MLT}$ in FLASH, or slightly lower in \texttt{GR1D+}, within the ranges mentioned above, one would get an even better agreement between the two 1D+ models.

More details on the comparison between the 2D and 1D+ suites of simulations are included in Appendix~\ref{sec:appendix_Qdot_1D_2D}.

\begin{figure}
\centering
\includegraphics[width=\columnwidth]{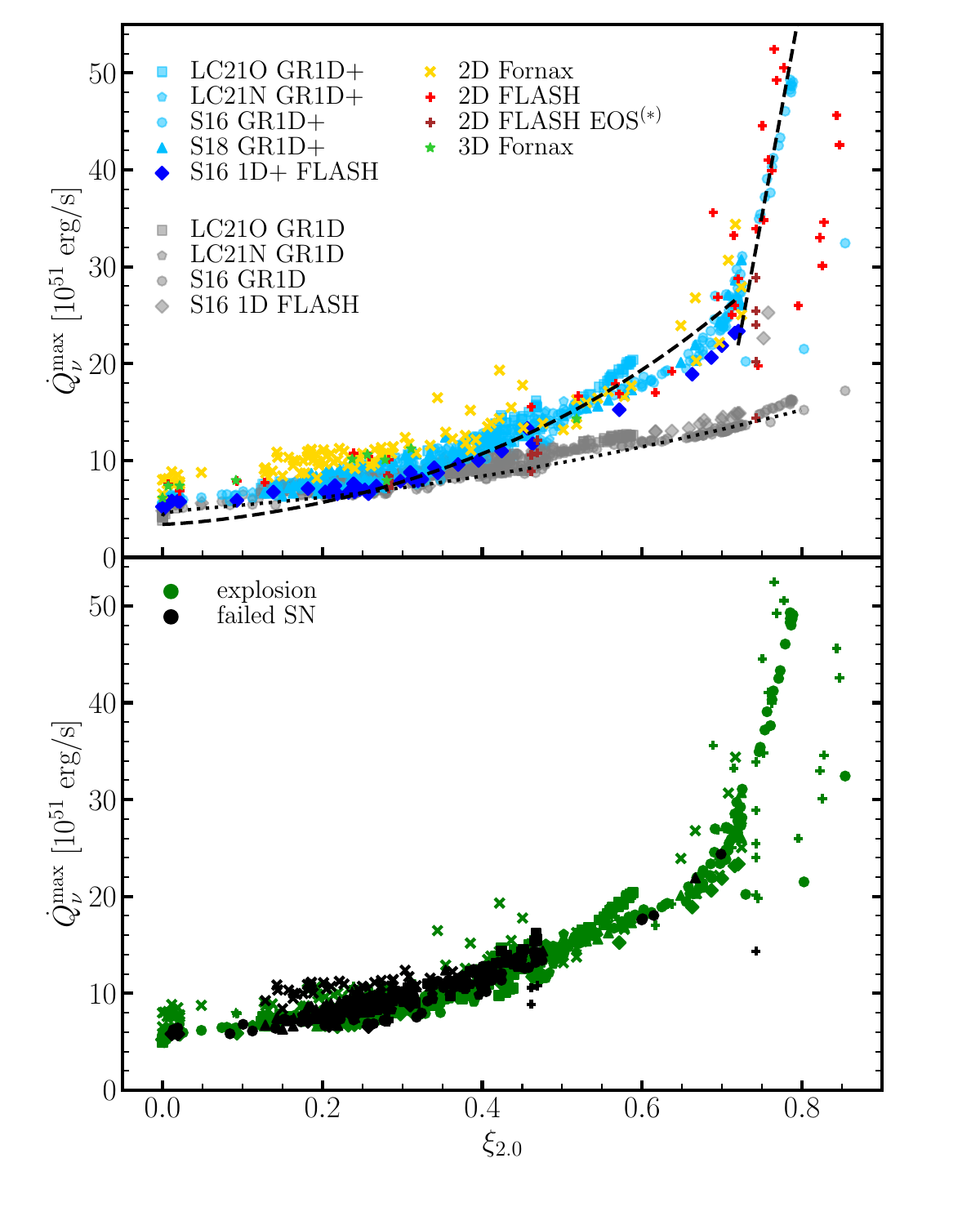}

\caption{Maximum of the neutrino heating as a function of compactness. The bottom panel shows only 1D+, 2D, and 3D simulations, color-coded based on explosion outcome. Notice that the failed explosions have been plotted on top of the successful ones, and since several points overlap, it looks like the number of failed explosions is much higher than it is in reality. All simulations (except for four 1D+ and one 2D simulation discussed in the text)} above $\xi_{2.0} \sim 0.5$ lead to successful explosions. In the top panel, we also include 1D simulations (in grey), and different symbols and colors represent different sets of simulations. All simulations adopt the SFHo EOS (except for the dark brown pluses(*), which indicate 2D FLASH simulations run with a few different EOSs). Typically, the stiffer the EOS the smaller the neutrino heating. The agreement between 1D+, 2D, and 3D simulations is remarkable. Three general trends can be seen in the Figure: the one followed by 1D simulations, shown in Eq.~\eqref{eq:Qdot_fit_1D}, the one followed by multi-D and 1D+ simulations for compactnesses below $\xi_{2.0} \sim 0.72$, shown in Eq.~\eqref{eq:Qdot_fit_STIR1}, and the even steeper one for compactnesses above $\xi_{2.0} \sim 0.72$, shown in Eq.~\eqref{eq:Qdot_fit_STIR2}. Different trends are illustrated by the two dashed lines and the dotted black line in the upper panel. Notice that these are not direct fits to $\dot{Q}_\nu^{\rm max}$ vs. $\xi_{2.0}$, and the details of how they were derived are given in Appendix~\ref{sec:appendix_Qdot_comp}.
\label{fig:Qdot_vs_comp_multiD}
\end{figure}

\section{Interplay between neutrino heating and convection}
\label{sec:semi_an_qdot}
To illustrate the physical mechanisms that cause high-compactness progenitors to explode, it is useful to introduce a semi-analytical expression for the total net neutrino heating in the gain region \citep{Janka2012_review_CCSNe}:
\begin{equation}
\label{eq:Qdot_semi_an}
\begin{split}
    \dot{Q}_\nu = &\ 5.18 \times 10^{51}\ \dfrac{\rm erg}{\rm s} \times  \left(\frac{M_{\rm g}}{0.01\ M_\odot}\right) \left(\frac{\bar{R}_{\rm g}}{100\ \rm km}\right)^{-2} \\
    &\frac{1}{2}\sum_{\nu \in \{\nu_e, \bar{\nu}_e \}}\frac{\avg{\epsilon^2_\nu}}{\left(18\ \rm MeV\right)^2} \left(\frac{L_\nu (\bar{R}_{\rm g})}{3 \times 10^{52}\ \rm erg/s}\right).
\end{split}
\end{equation}
The equation above differs slightly from the expression derived by \citet{Janka2012_review_CCSNe}. The two main differences are that, instead of the neutrinosphere temperature $T_\nu$, we use directly the neutrino root mean squared energy $\langle \epsilon_\nu^2 \rangle$, since the cross section explicitly depends on energy. The other difference is that we do not use the neutrino luminosity at the gain radius $R_{\rm g}$, but rather at a slightly larger radius:
\begin{equation}
    \label{eq:Rgbar}
    \bar{R}_{\rm g} = R_{\rm g} + 0.15 \times (1 + \xi_{2.0}) (R_{\rm s} - R_{\rm g}),
\end{equation}
where $R_{\rm s}$ is the shock radius. Therefore, $\bar{R}_{\rm g}$ is located at roughly 15\% and 30\% of the size of the gain region for the lowest and highest compactness progenitors, respectively. This was done to better reproduce the simulation results, especially in the 1D+ case.  Notice that $\langle \epsilon_\nu^2 \rangle$ is also calculated at $\bar{R}_{\rm g}$, and both the luminosity and the neutrino energy are calculated in the fluid frame. If instead one uses luminosities and energies at 500 km in the lab frame (as it is common in the field), one would obtain a similar expression, with the overall factor being $\sim 40 \%$ larger. More details on how and why this expression differs from the one from \citet{Janka2012_review_CCSNe} are given in Appendix \ref{sec:appendix_diff_Janka}.  The remaining term in Eq.~\eqref{eq:Qdot_semi_an} is the total mass in the gain region $M_{\rm g}$.
\begin{figure*}
\centering
\includegraphics[width=\textwidth]{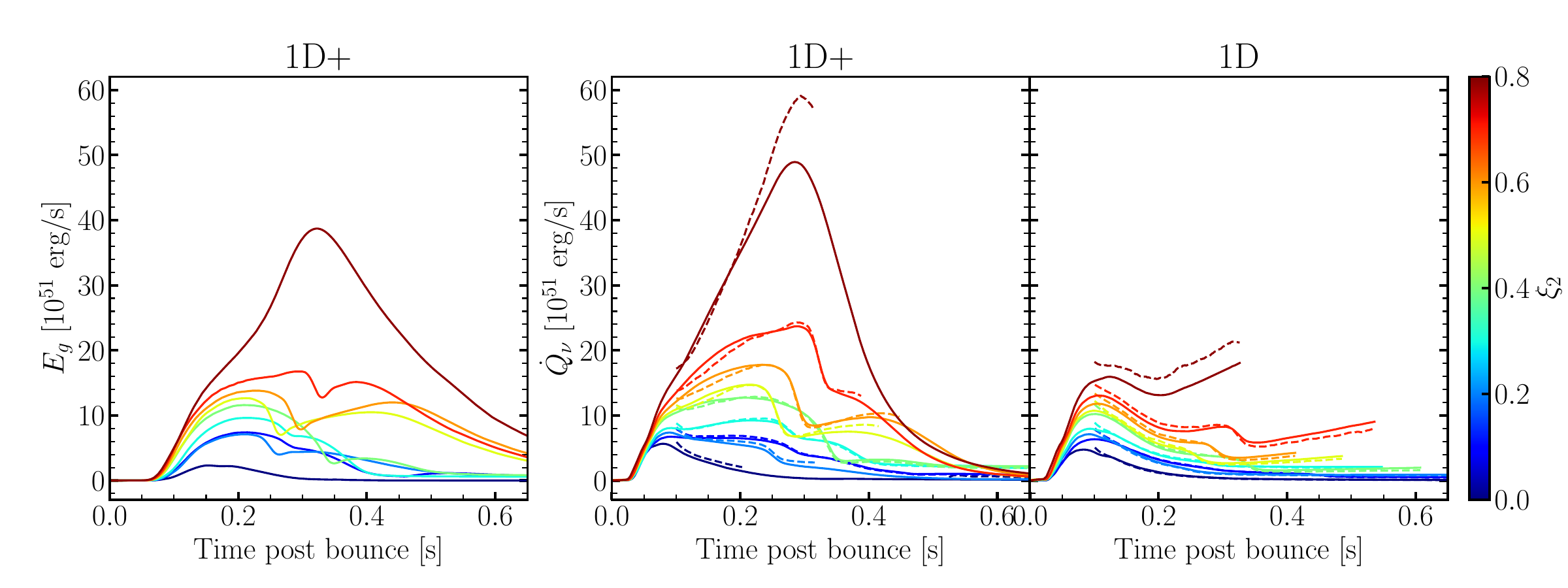}

\caption{Integrated turbulent energy generation rate in the gain region (left panel) and neutrino heating in the gain region (middle panel) for 1D+ and 1D simulations of the progenitors listed in Table~\ref{tab:progs_info}. The integrated turbulent energy is roughly half of the neutrino heating. The dashed lines in the two rightmost panels show the semi-analytical expression in Eq.~\eqref{eq:Qdot_semi_an}, which is only valid during the stalled-shock phase, and therefore we only compute it starting at 100 ms, until either the shock crosses 500 km or the simulation ends. The excellent agreement between dashed and solid lines shows that Eq.~\eqref{eq:Qdot_semi_an} is a good description of neutrino heating in both 1D+ and 1D simulations, i.e. with or without convection.}
\label{fig:Qdot}
\end{figure*}

To carry out this part of the analysis, we compare the same nine 1D and 1D+ simulations from \citet{Boccioli2024_remnant} already discussed in Section~\ref{sec:nu_heat_1D+}. The agreement between the semi-analytical expression in Eq.~\eqref{eq:Qdot_semi_an} and the neutrino heating calculated directly from the simulation is excellent, for both 1D and 1D+ simulations, as shown in Figure~\ref{fig:Qdot} by the solid and dashed lines. It should be highlighted that the difference between the two rightmost panels of Figure~\ref{fig:Qdot} is only the inclusion of $\nu$-driven convection, with everything else in the simulation being kept the same. For completeness, we show in the left panel the rate of turbulent energy generation in the gain region $E_{\rm g}$, defined in Eq.~\eqref{eq:Eg}, which is roughly half of the neutrino heating, although the exact value depends on compactness. If one considers neutrino heating plus turbulent energy generation, this is compatible with the findings of \citep{Gogilashvili2024_FEC+} that convection decreases the explosion condition by $\sim$ 30 \%.


\begin{figure}
\centering
\includegraphics[width=\columnwidth]{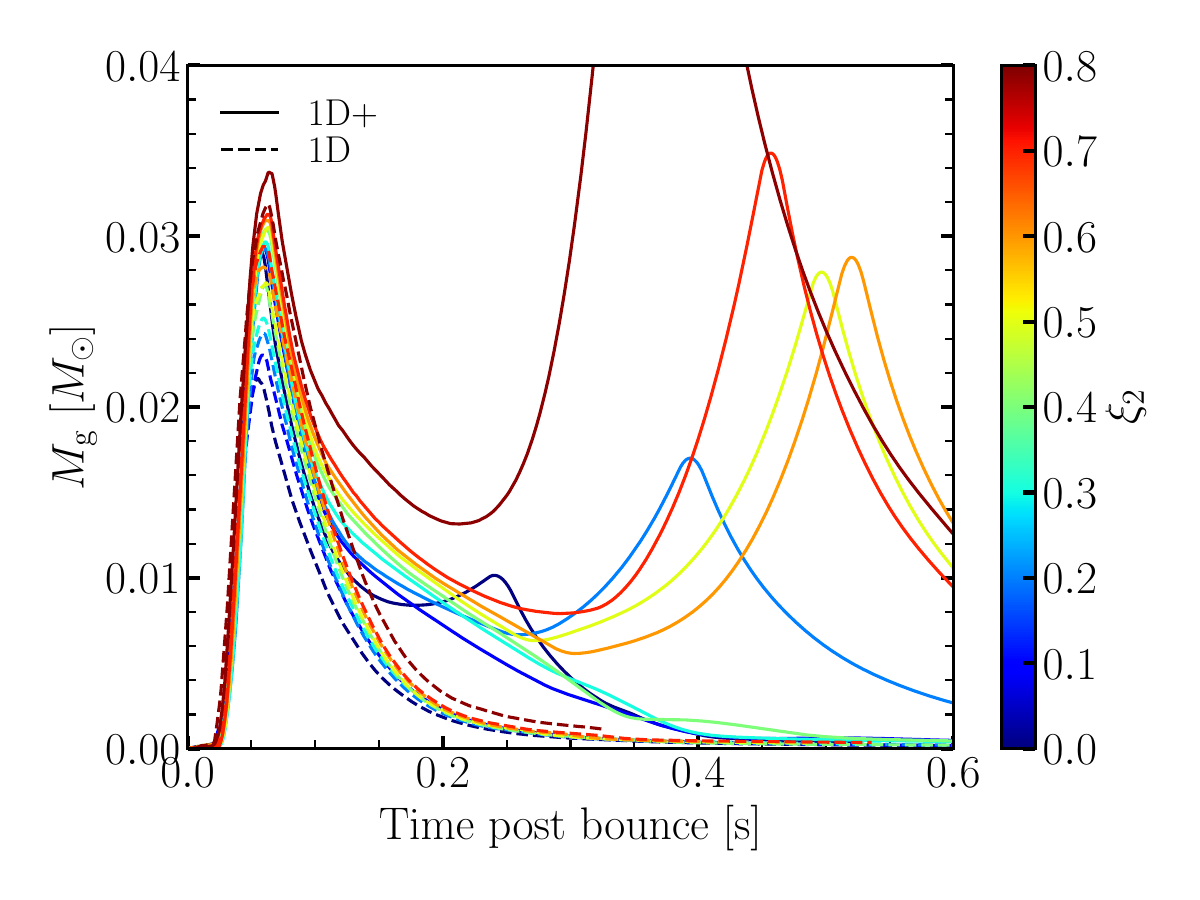}

\caption{Total mass in the gain region for 1D (dashed lines) and 1D+ (solid lines) simulations of the progenitors listed in Table~\ref{tab:progs_info}. Because of the quick decrease in shock radius (see Figure~\ref{fig:radii_1D_1D+}), the mass in the gain region in 1D simulations quickly decreases, leading to a quick decrease in $\dot{Q}_\nu$, as shown in Eq.~\ref{eq:Qdot_semi_an}. On the contrary, in 1D+ simulations the shock radius stalls at larger radii for a long time, and therefore the mass in the gain region does not drop as quickly, avoiding the quick decrease in $\dot{Q}_\nu$. Notice that the fast increase in $M_{\rm g}$ for some 1D+ simulations is a sign that the explosion has been triggered, and therefore the shock is quickly expanding and Eq.~\eqref{eq:Qdot_semi_an} is not valid anymore. Later, matter has been swept by the shock and therefore $M_{\rm g}$ drops again.}
\label{fig:mgain}
\end{figure}

\begin{figure*}
\centering
\includegraphics[width=\textwidth]{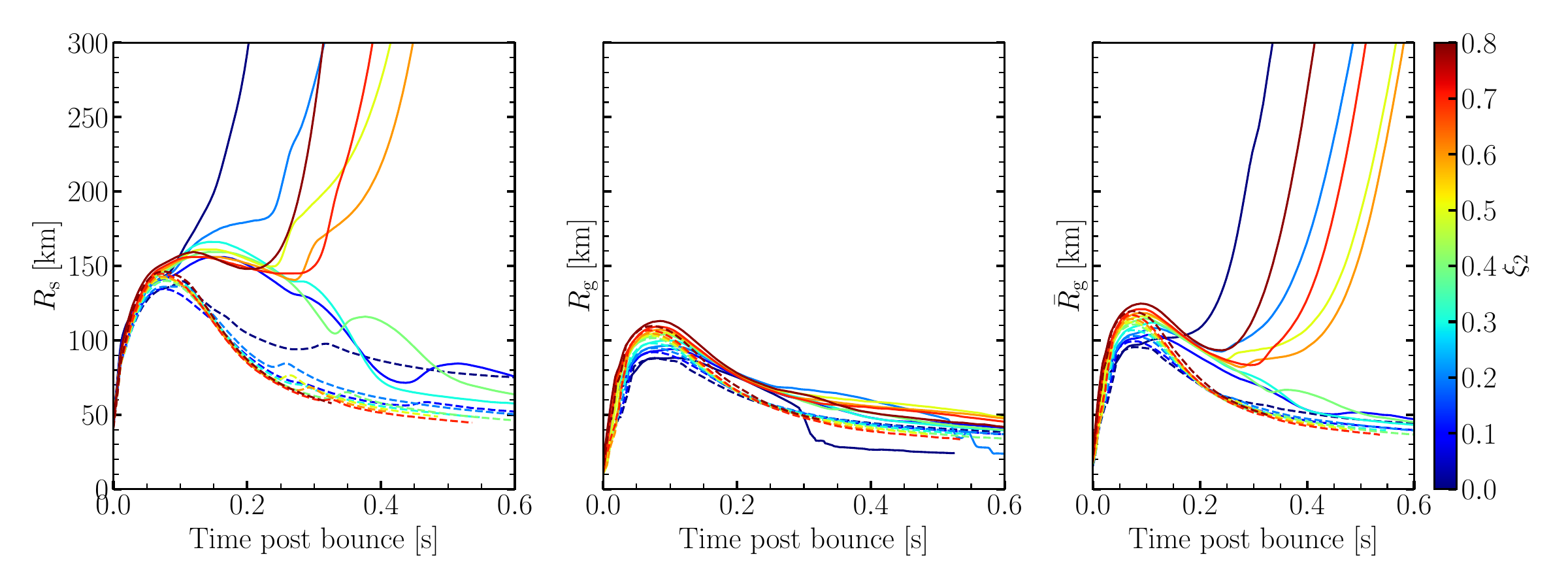}

\caption{The left and middle panels show the evolution of the shock and gain radius, respectively, for 1D (dashed lines) and 1D+ simulations (solid lines) of the progenitors listed in Table~\ref{tab:progs_info}. The right panel shows the radius defined in Eq.\eqref{eq:Rgbar}, which is located between 15\% and 30\% of the gain region, depending on compactness. The presence of convection in 1D+ simulations allows the shock to stall at larger radii for a longer time, hence increasing the size of the gain region, and therefore the total neutrino heating, as discussed in Section~\eqref{sec:semi_an_qdot}.}
\label{fig:radii_1D_1D+}
\end{figure*}

\begin{figure}
\centering
\includegraphics[width=\columnwidth]{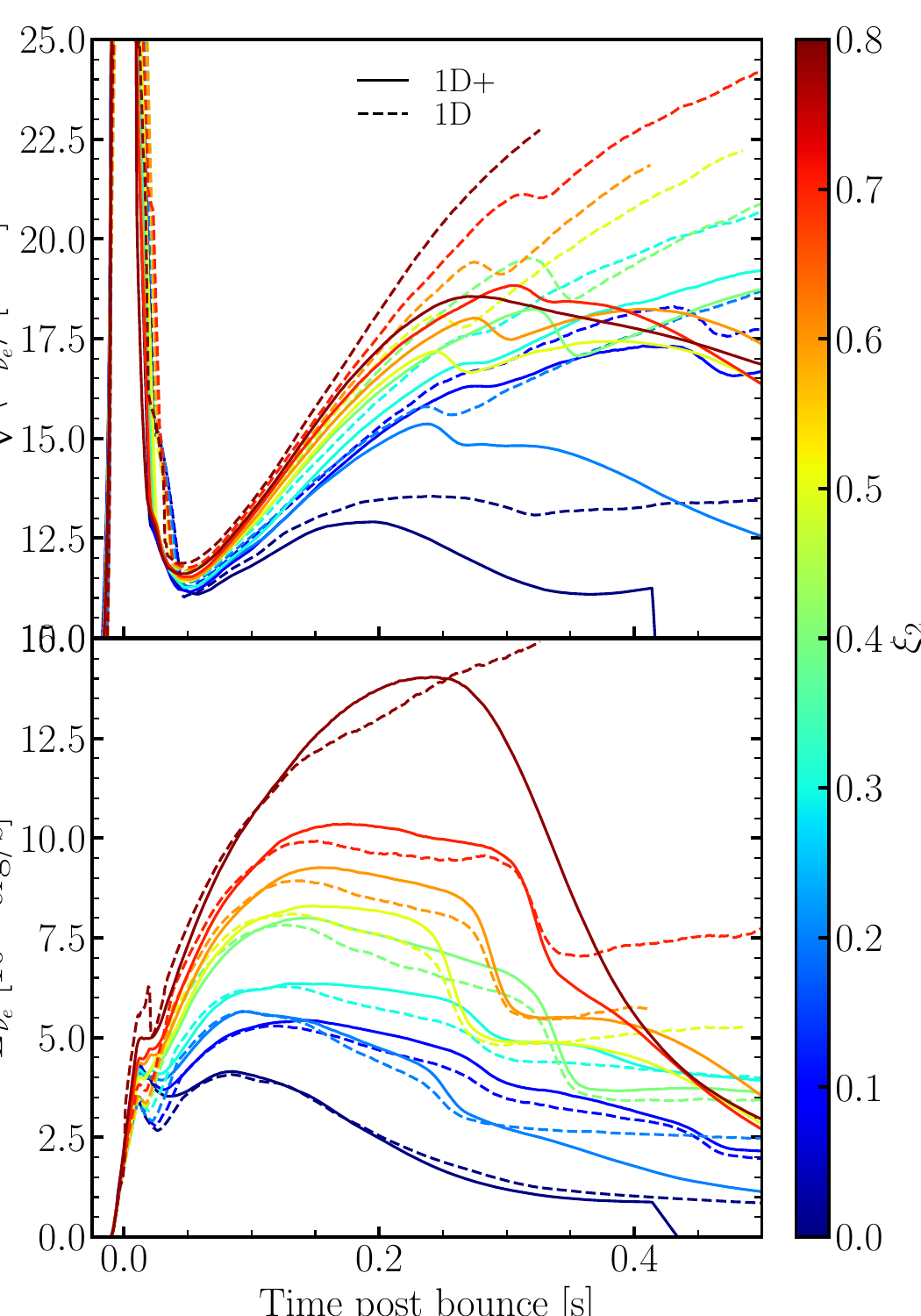}

\caption{Neutrino root mean squared energies (upper panel) and neutrino luminosities calculated at {$R=\avg{R}_g$} (lower panel) for electron-type neutrinos. These are 1D (dashed lines) and 1D+ (solid lines) simulations for the progenitors listed in Table~\ref{tab:progs_info}. The luminosities are very similar since they mostly depend on the accretion history, which are the same in 1D and 1D+ (but also, to very good approximation, multi-D) simulations. Deviations only appear after the explosion sets in, when the mass accretion rate and temperature of the neutrinosphere significantly drop. A similar argument holds for neutrino energies. Neutrino energies for electron antineutrinos have a very similar time dependence and are $\sim 10 \%$ larger. Luminosities, on the other hand, are quite similar in the accretion phase, i.e. later than $\sim 100$~ms after bounce. }
\label{fig:Enue_Lnue_vs_time}
\end{figure}

The main difference between $\dot{Q}_\nu$ in 1D and 1D+ is that in 1D $\dot{Q}_{\nu}$ decreases immediately after peaking at around 100 ms. In 1D+, instead, $\dot{Q}_{\nu}$ keeps increasing, typically until the Si/O interface is accreted. At that point, either an explosion ensues or, in the case of a failed SN, the heating drops, and then the shock slowly recedes until a black hole is formed at later times. In the remainder of this section, we will use Eq.~\eqref{eq:Qdot_semi_an} to analyze the cause of the difference between the behavior of neutrino heating in 1D and 1D+ simulations. 

The largest difference between 1D and 1D+ is $M_{\rm g}$, which after $\sim 100 $ ms drops significantly in 1D, whereas in 1D+ it does not, as shown in Figure~\ref{fig:mgain}. This is because in 1D+ simulations the shock radius is sustained by the extra-heating from $\nu$-driven convection and therefore the volume of the gain region is roughly constant during the accretion phase, whereas it decreases in 1D simulations. This can be seen from the large differences in shock radii in 1D and 1D+ simulations shown in Figure~\ref{fig:radii_1D_1D+}, whereas the gain radius is $\sim 10-20 \%$ larger in 1D+ simulations. It can indeed be shown that the reason for the increase in the gain region mass is due to the increase in volume, and not an increase in average density, which on the contrary is lower in 1D+ simulations since the shock stalls at larger radii and therefore when the infalling matter is accreted onto the shock it has not reached densities as large as in the 1D case.

The time evolution of $\avg{\epsilon^2_{\nu_e}}$ and $L_{\nu_e}$ are shown in Figure~\ref{fig:Enue_Lnue_vs_time}. They are both quite similar between the 1D and 1D+ cases since at early times the mass accretion rate onto the PNS is the same, and they deviate only after the explosion sets in and the accretion stops. During the stalled-shock phase, $L_{\nu_e}$ is roughly constant (and the same is true for $L_{\bar{\nu}_e}$), whereas $\avg{\epsilon_\nu^2}$ steadily increases. The reason is that as mass is accreted onto the PNS, its radius shrinks and temperature increases, leading to larger neutrinosphere temperatures. Consequently, $\avg{\epsilon_\nu^2}$ grows more rapidly for stars with larger mass accretion rates and, therefore, compactness. Notice however that for the progenitor with the largest compactness, the luminosity is instead increasing during the accretion phase (see bottom panel of Figure~\ref{fig:Enue_Lnue_vs_time}). This causes an even more rapid increase in neutrino heating compared to the other progenitors, and more details on this are given in Appendix~\ref{sec:appendix_Qdot_comp}.

To summarize, $L_{\nu_i}$ is roughly constant, $\bar{R}_{\rm g}$ slowly decreases and then increases close to explosion, whereas $M_{\rm g}$ and $\avg{\epsilon^2_{\nu_i}}$ decrease and increase, respectively. In 1D simulations, the rapid decrease of $M_{\rm g}$ causes neutrino heating to immediately decrease after peaking at $\sim 100$~ms at the beginning of the accretion phase. However,  in 1D+ simulations the decrease of $M_{\rm g}$ is much tamer. One can show that $L_{\nu_i} \avg{\epsilon^2_{\nu_i}} / \bar{R}_{\rm g}^2$ is roughly constant during the accretion phase (for a more complete discussion see Appendix~\ref{sec:appendix_Qdot_comp}). Therefore, it is the neutrino energy (or equivalently neutrinosphere temperature) that determines the increase of $\dot{Q}_\nu$ shown in the middle panel of Figure~\ref{fig:Qdot}. Neutrino heating will then significantly drop at the end of the accretion phase, which typically coincides with the accretion of the Si/O interface or, for the highest compactness progenitors,  with the onset of explosion. The interplay among these timescales is more thoroughly analyzed in Section~\ref{sec:timescales_correlations}.

In 1D+ simulations, after the end of the stalled-shock phase, the mass accretion rate drops significantly causing $\avg{\epsilon^2_{\nu_e}}$, and therefore $\dot{Q}_\nu$, to also drop. For failed explosions, the situation is not too different. At some point (see Section~\ref{sec:timescales_correlations} for more details regarding the timescale) the mass accretion rate drops and the stalled-shock phase ends. The difference with the previous case is that, after $\dot{Q}_\nu^{\rm max}$ is reached, instead of a rapid shock expansion leading to an explosion, the shock quickly recedes until a black hole is formed.

It is also instructive to analyze what happens in 1D simulations. From Figure~\ref{fig:radii_1D_1D+} one can see that the stalled-shock phase starts at $\sim 100$~ms but is not very long. Rather than a longer phase where the shock stalls at roughly a constant radius for hundreds of milliseconds, the shock slowly recedes through a series of quasi-steady state phases at progressively smaller radii. As can be seen from Figure~\ref{fig:Qdot}, neutrino heating in 1D peaks at around 100 ms for all progenitors, i.e. when the stalled-shock phase starts, and then it inevitably decreases following the slow decrease in shock radius. This means that the maximum of $\dot{Q}_\nu^{\rm max}$ is always reached at the beginning of the stalled-shock phase, and therefore there is no late-time increase due to the increase in PNS mass and neutrino energy. This is the reason why in Figure~\ref{fig:Qdot_vs_comp_multiD} 1D simulations follow a linear trend, whereas 1D+ and multi-D simulations show a rapid increase of $\dot{Q}_\nu^{\rm max}$ with compactness (see Appendix~\ref{sec:appendix_Qdot_comp} for a more quantitative proof).

To summarize, the main difference between 1D and 1D+ models is that 1D+ models are able to sustain longer stalled-shock phases during which the gain region mass does not decrease very rapidly, allowing the hot neutrinos coming from the PNS to successfully revive the shock for high-compactness progenitors, contrary to what has been commonly assumed. The good agreement between neutrino heating in 1D+ and multi-dimensional models (at least for high-compactness progenitors) shows that indeed 1D+ models can model the stalled-shock phase very well, because of how the extra heating is introduced in the model (i.e. using $\nu$-driven convection). The way the extra-heating in 1D models is introduced can indeed lead to very different post-bounce dynamics, and therefore explodabilities, and we will discuss this in detail in Section~\ref{sec:comparison_1D}.

\begin{figure*}
\centering
\includegraphics[width=\textwidth]{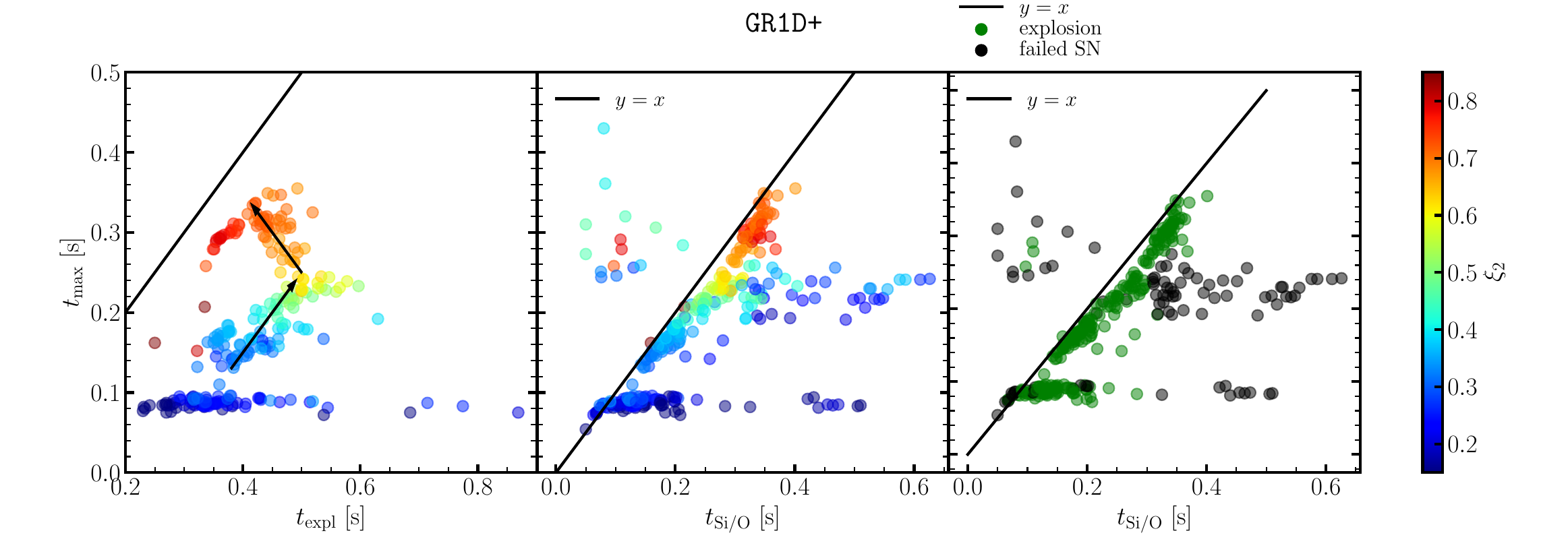}

\caption{The vertical axis shows the time at which $\dot{Q}_\nu^{\rm max}$ is reached. For the left panel, the horizontal axis shows the explosion time, which is defined as the time when the shock crosses 500 km. The right panels have on the horizontal axis the time when the Si/O interface is accreted through the shock. The middle one is color-coded according to compactness, whereas the right panel is color-coded according to explosion outcome (i.e. successful or failed). Solid black lines in all panels show the bisectors, i.e. $t_{\rm max} = t_{\rm Si/O}$, and the arrows in the first panels have sole purpose of guiding the eye. Exploding progenitors show a correlation between $t_{\rm Si/O}$ and $t_{\rm max}$, . Outliers and failed SNe are discussed in the text. There is instead no apparent correlation between $t_{\rm expl}$ and $t_{\rm max}$. This is because high-compactness progenitors have very large explosion energies, and therefore very high shock velocities, as discussed in the text. }
\label{fig:tmax_vs_taccr_vs_texpl}
\end{figure*}

\subsection{Interplay among relevant timescales of the problem}
\label{sec:timescales_correlations}
It is instructive to analyze the interplay between relevant timescales of the problem, namely the time at which the maximum of $\dot{Q}_\nu$ is reached ($t_{\rm max}$), the time when the explosion sets in ($t_{\rm expl}$, defined as the time when shock reaches 500 km), and finally the time when the Si/O interface is accreted ($t_{\rm Si/O}$). The latter is important since it plays an important role in triggering the explosion \citep{Lentz2015_3D,Summa2016_prog_dependence_vertex,Vartanyan2018_2D_simulations_revival_fittest} and can be used to predict the outcome of the explosion \citep{Ertl2016_explodability,Wang2022_prog_study_ram_pressure,Tsang2022_ML_explodability,Boccioli2023_explodability}. For this part of the analysis, we will adopt the same 341 1D+ models used by \citet{Boccioli2024_remnant}.

The maximum of $\dot{Q}_\nu$ for successful explosions is reached, in most cases, when the Si/O interface is accreted. This is shown in the right panel of Figure~\ref{fig:tmax_vs_taccr_vs_texpl}. The left panel instead shows how there is no correlation between $t_{\rm expl}$ (defined as the time when the shock crosses 500 km) and $t_{\rm max}$, as one could naively expect. Notice that, even though it is not immediately clear from Figure~\ref{fig:tmax_vs_taccr_vs_texpl}, $t_{\rm Si/O}$ also positively correlates with compactness, which is not surprising since $t_{\rm Si/O}$ depends on the mass location of the Si/O interface $M_{\rm Si/O}$, and $M_{\rm Si/O}$ positively correlates with compactness, simply because high-compactness progenitors have larger iron cores.

It can therefore be concluded that the maximum of $\dot{Q}_\nu$ for successful explosions is reached when the Si/O interface is accreted (i..e when the shock is revived), except for a few outliers discussed later in the section. After that, the shock moves to lower densities where neutrino heating significantly decreases, and then the explosion is launched. 

More subtly, the lack of correlation between $t_{\rm max}$ and $t_{\rm expl}$ is due to the much faster shock propagation in high-compactness progenitors after the shock revival is initiated. In other words, if shock revival is initiated roughly at the same time for a higher-compactness and a lower-compactness progenitor, the shock in the high-compactness progenitor will reach 500 km much faster, and therefore despite them having a similar $t_{\rm max}$, the high-compactness progenitor will have a smaller $t_{\rm expl}$. This is exactly what is seen in the left panel of Figure~\ref{fig:tmax_vs_taccr_vs_texpl}. For progenitors with $\xi_{2.0} \lesssim 0.5$, there is a positive correlation between $t_{\rm max}$ and $t_{\rm expl}$. However, for progenitors with $\xi_{2.0} \gtrsim 0.5$ this trend is reversed, because of the high initial shock velocities at the time of shock revival for high-compactness progenitors. These progenitors are also the ones for which the line $\dot{Q}_\nu^{\rm max}$ vs. $\xi_{2.0}$ in Figure~\ref{fig:Qdot_vs_comp_multiD} changes slope and grows like a power law, and they will also be the ones with the largest explosion energies \citep{Burrows2024_Phys_correlations}.

\begin{figure}
\centering
\includegraphics[width=\columnwidth]{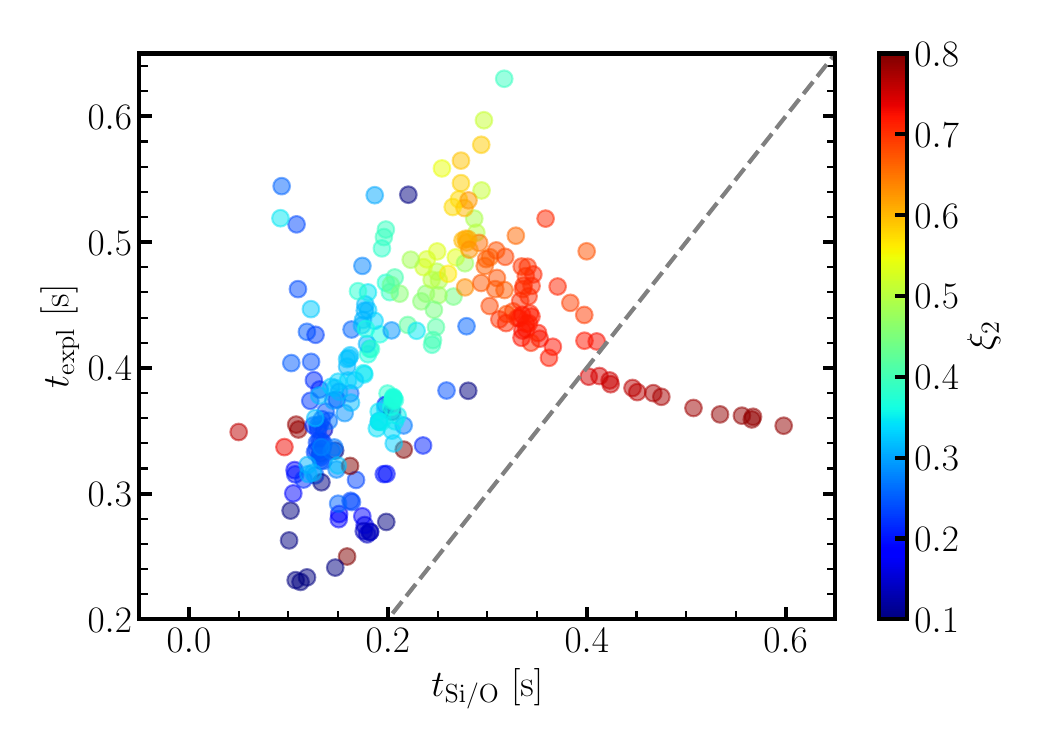}

\caption{Only successful explosions of the 341 1D+ simulations from \citet{Boccioli2024_remnant} are shown in this Figure. The dashed line indicates where $t_{\rm Si/O} = t_{\rm expl}$. Therefore, points to the right of that refer to simulations for which the Si/O interface is accreted after the explosion sets in. Therefore, despite the negative correlation seen for those points, the accretion of the Si/O interface does not actually influence the explosion. As discussed in the text, only in the region where $t_{\rm expl}$ positively correlates with $t_{\rm Si/O}$ (i.e. left of the dashed line) the accretion of the Si/O interface significantly affects the explosion.}
\label{fig:texpl_vs_taccr}
\end{figure}

It is also instructive to analyze the correlation between $t_{\rm expl}$ and $t_{\rm Si/O}$ shown in Figure~\ref{fig:texpl_vs_taccr}. For $\xi_{2.0} \lesssim 0.5$, $t_{\rm expl}$ positively correlates with $t_{\rm Si/O}$. For $\xi_{2.0} \gtrsim 0.5$, however, there is no correlation anymore, indicating that the accretion of the Si/O interface is not important for the explosion, in line with the findings of \citet{Boccioli2024_FEC+_SiO}. The density drop at the Si/O interface is important to determine the explodability of progenitors with $\xi_{2.0} < 0.5$, but it has no impact on the explodability of progenitors with $\xi_{2.0} > 0.5$ that instead all explode, as highlighted by \citet{Boccioli2024_remnant} and shown again in this work. 

It is also important to highlight that from Figure~\ref{fig:texpl_vs_taccr} one might be tempted to conclude that there is a negative correlation between $t_{\rm expl}$ and $t_{\rm Si/O}$ for high-compactness progenitors. However, this is simply a consequence of the fact that the Si/O interface is located at larger masses for high-compactness progenitors, i.e. $t_{\rm Si/O}$ is larger for larger $\xi_{2.0}$. At the same time, high-compactness progenitors develop very large neutrino heating and therefore explode with very high shock velocities, and reach 500 km in a shorter time, causing $t_{\rm expl}$ to be smaller. Therefore, these opposite dependencies on compactness cause the negative correlation between $t_{\rm expl}$ and $t_{\rm Si/O}$ seen in Figure~\ref{fig:texpl_vs_taccr}. This does not mean that the Si/O interface causes the explosion, whereas it does for progenitors with $\xi_{2.0} < 0.5$. 

The case of failed 1D+ simulations is slightly different. Most of them have small to moderate density drops in correspondence with the Si/O interface. In those cases, neutrino heating resembles the case of a successful explosion, with the only difference being that, after the accretion of the Si/O interface, the shock expansion to lower densities is only temporary. This is the cause for the initial drop in $\dot{Q}_\nu$ (see for example the $\xi_{2.0} \sim 0.4$ progenitor in the middle panel of Figure~\ref{fig:Qdot} at $\sim 350$ ms after bounce). After that, the shock falls back and the behavior resembles the case of a pure 1D simulation, with the gain region shrinking and therefore neutrino heating decreasing. 

It is also important to highlight that the failed SNe from the 1D+ simulations experience a longer stalled-shock phase compared to their 1D counterparts. However, this might not significantly affect the time of black-hole formation, and since we did not follow the failed SNe until black-hole formation, due to our numerical setup, we could not quantify this. Both in the 1D and 1D+ cases, one expects the failed SN to eventually experience a sharp increase in neutrino heating before black hole formation, which is likely even larger than the maximum reached at $\sim 100 $ ms. This can be seen in the rightmost panel of Figure~\ref{fig:Qdot}, where there is a rapid increase in neutrino heating before the simulation stops close to (although not exactly at) black hole formation. One can also appreciate that the high compactness progenitors experience this increase earlier, and therefore they also form black holes much sooner after bounce, as shown by \citet{OConnor2011_explodability}.

There are some outliers in the right panel of Figure~\ref{fig:tmax_vs_taccr_vs_texpl} worth analyzing. In the upper left corner, there are a few failed SNe that accrete the Si/O interface before $\dot{Q}_\nu^{\rm max}$ is reached. These are progenitors that have relatively small density drops at the Si/O interface, and therefore the accretion of the interface does not impact the shock, which keeps stalling at a relatively constant radius for a few hundred milliseconds (i.e. until $t=t_{\rm max}$, depending on the progenitor), before falling back. The four successful explosions are instead high-compactness progenitors with small Si/O interfaces. Therefore, the accretion of the Si/O interface does not affect the shock dynamics (and hence neutrino heating). However, due to their high compactness and consequently large neutrino heating, these progenitors can explode even without a strong interface. Not surprisingly, in their case $t_{\rm max}$ and $t_{\rm expl}$ coincide.

The majority of the remaining failed SNe are instead characterized by $t_{\rm Si/O} > t_{\rm max}$. Most of these have $t_{\rm max} \approx 300$~ms and $t_{\rm Si/O} > 300$~ms which, as one can see from Figure~\ref{fig:radii_1D_1D+}, is roughly the time after which the shock falls back onto the NS if the explosion is not triggered. These are cases where the accretion of the Si/O interface is not important since the fallback of the shock has already been initiated, and neutrino heating has already started decreasing significantly. In other words, the Si/O interface for these progenitors is located at very large masses, since their iron core is more massive. An example of this is again given by the progenitor with $\xi_{2.0} \sim 0.4$ shown in \hbox{Figures \ref{fig:Qdot} through \ref{fig:Enue_Lnue_vs_time}}. 

Finally, there are roughly $\sim 80$ progenitors with a wide range of $t_{\rm Si/O}$ and $t_{\rm expl}$, which however all have $t_{\rm max} \approx 100 $ ms. As can be seen from the middle panel of Figure~\ref{fig:tmax_vs_taccr_vs_texpl}, these are all low-compactness progenitors, which, as pointed out in Section~\ref{sec:comparison_multiD}, do not experience the typical increase in $\dot{Q}_\nu$ seen in for high-compactness progenitors. Instead, after peaking at $\approx 100 $ ms, neutrino heating does not increase significantly during the stalled-shock phase because the temperature of the neutrinosphere does not increase significantly). Therefore, one expects the neutrino heating for these low-compactness progenitors to peak at around 100 ms, which is also confirmed by the 20 3D F{\sc{ornax}} simulations analyzed in this paper. Then, some of them will explode after the accretion of a significant density drop at the Si/O interface, and others will not. A more detailed analysis of the interplay between the explosion condition and the accretion of the Si/O interface is given in \citet{Boccioli2024_FEC+_SiO}.


\section{Comparison between STIR and other 1D explosion models}
\label{sec:comparison_1D}

Historically, extensive explodability studies involving hundreds of simulations have been mostly carried out using 1D simulations, with possibly only one exception of a very large 2D study by \citet{Nakamura2015_2D_explodability}. Notably, in that work, the authors carried out hundreds of 2D simulations and found that most high-compactness progenitors did not explode. The reason for the disagreement in explodability with the 2D simulations presented here is not easy to assess. Possible explanations could be that, in the study of \citet{Nakamura2015_2D_explodability}, the simulations were run with the IDSA approximate neutrino transport from \citet{Liebendorfer2009_IDSA} and a simple leakage scheme for heavy-lepton neutrinos. Moreover, they adopted the LS220 nuclear EOS from \citet{Lattimer1991_LS}, which typically leads to a faster contraction of the PNS compared to the SFHo EOS adopted for most simulations analyzed in this work. The faster PNS contraction typically leads to earlier shock revival \citep{Boccioli2022_EOS_effect,Janka2023_EOS_and_dynamics}, but at the same time, it also causes a faster collapse into a black-hole \citep{Schneider2020_EOS_dependence_BH}. Therefore, this could potentially inhibit the successful explosion of high-compactness progenitors, which tend to explode later, as shown in Figure~\ref{fig:tmax_vs_taccr_vs_texpl}. It should also be noted that, in their simulations, \citet{Nakamura2015_2D_explodability} adopt fully Newtonian gravity (i.e. without any general relativistic effective potential), which is known to delay even further (i.e. later than 1 second post bounce) the explosion of high-compactness progenitors \citep{OConnor2018_2D_M1}. Therefore, without any general relativistic corrections, high-compactness progenitors will be harder to explode, which might explain the lack of exploding high-compactness progenitors in the study of \citet{Nakamura2015_2D_explodability}. As we have shown in previous sections, the time dependence of neutrinosphere temperatures (i.e. neutrino energies) is very important in order to obtain successful shock revival of high-compactness progenitors. Therefore, less accurate neutrino transport schemes such as the IDSA, lack of general relativistic corrections and, more importantly, the simpler leakage scheme for heavy lepton neutrinos, could significantly alter the cooling of the PNS.

Only very recently state-of-the-art 2D studies \citep{Vartanyan2023_nu_100_2D} have become feasible, and in the near future 3D simulations will also become more affordable, although carrying out extensive studies requiring hundreds of simulations is still an extremely challenging computational task. Moreover, 1D simulations are much easier to analyze, given their simplicity, and therefore showcase an attractive framework for large parameter studies given both their low computational cost and simplicity in analyzing the results. In the rest of this section, we will provide a brief overview of the most popular 1D explodability studies, and highlight the differences with the 1D+ models based on STIR adopted in this work, in particular regarding the explosion (or lack thereof) of high-compactness progenitors.

\subsection{O'Connor 2011}
\label{sec:Oconnor}
An often-used explodability criterion, because of its simplicity, is the one developed by \citet{OConnor2011_explodability}. It is based on the compactness $\xi_{2.5}$ calculated at bounce (the difference with the pre-supernova compactness is however rather small), and they found that progenitors with $\xi_{2.5} \lesssim 0.45$ are more likely to explode, whereas the high compactness progenitors are more likely to collapse. Their study has the benefit of exploring a larger set of progenitor models generated with different codes, i.e. both KEPLER \citep{Woosley2002_KEPLER_models} and FRANEC \citep{Limongi2006_preSN_models} progenitors, as well as different equations of state. However, it was carried out with a simpler version of the code \texttt{GR1D} adopted in this work, with a simple leakage scheme to describe neutrino heating, rather than solving for neutrino transport, as it is done in the more recent version of the code adopted in this work through an M1 scheme. Moreover, the explosion was achieved by multiplying the neutrino heating term in the leakage scheme by a factor $f_{\rm heat}$ of order 1.

In their simplified 1D simulations, they noticed that the critical value of $f_{\rm heat}$ needed to explode a star had a minimum at $\xi_{2.5} \sim 0.45$, and was larger for stars of both larger and smaller compactness. However, the critical value for the heating efficiency (i.e. $\eta = \dot{Q}_\nu / (L_{\nu_e} + L_{\bar{\nu}_e})$) was instead roughly the same for low-compactness progenitors, whereas it was larger for progenitors with compactnesses above $\sim 0.45$. With the assumption that multi-D simulations could achieve a fixed heating efficiency independent of compactness, they interpreted their results as an indication that high-compactness progenitors do not explode, whereas low-compactness progenitors do. 

Interestingly, in a more recent study, \citet{Jost2025_fheat_yields} found instead that for a given value of $f_{\rm heat} \sim 2.5$, high-compactness progenitors can easily explode, similarly to what is seen in 1D+ models. Compared to \citet{OConnor2011_explodability}, they adopted a more sophisticated neutrino transport and a different equation of state, which indicates that perhaps the simple leakage scheme in combination with $f_{\rm heat}$ might artificially prevent explosions of high-compactness progenitors, whereas with more accurate neutrino transport adopting an artificial heating based on $f_{\rm heat}$ might give similar results to 1D+ models.

Another well-known problem with artificially enhancing heating in 1D is that unphysical oscillations of the shock radius by hundreds of kilometers occur \citep{Gogilashvili2023_FEC_GR1D}. Moreover, it is not clear whether the same value of $f_{\rm heat}$ should be applied to all progenitors. As mentioned above, \citet{OConnor2011_explodability} assumed that multi-D simulations could achieve a fixed heating efficiency independent of compactness, which however is not guaranteed. 

As shown in Section~\ref{sec:semi_an_qdot}, the amount of extra heating in a 1D+ simulation, which as shown in Section~\ref{sec:comparison_multiD} well reproduces the neutrino heating found in multi-D simulations, varies with compactness. The effects of convection in increasing the volume of the gain region in order for the hot neutrinos emitted in the PNS are crucial to reliably model the explosion, and cannot be reproduced by simply multiplying $\dot{Q}_\nu$ by an arbitrary factor. Notice that, even if one allows $f_{\rm heat}$ to vary for different progenitors, it is not guaranteed that the correct time-dependence of $\dot{Q}_\nu$, which is highly impacted by time-dependent $\nu$-driven convection, would be retrieved. Nonetheless, \citet{Jost2025_fheat_yields} showed that multiplying the neutrino heating by an arbitrary factor might qualitatively yield similar results to a 1D+ model, although it is affected by larger uncertainties and it might lead to unphysical shock expansions.

\subsection{PUSH}
\label{sec:PUSH}
Another method to trigger explosions in 1D, developed by the North Carolina State group, is called PUSH \citep{Perego2015_PUSH1, Ebinger2019_PUSH_II_explodability}.
This method consists on adding an energy source to the hydrodynamic equations given by \citep[Eqs. 4, 5, 8]{Perego2015_PUSH1}:
\begin{equation}
    \label{eq:QPUSH}
    Q^{+}_{\rm push}(t,r) = \frac{4 \mathcal{G}(t) \sigma_0}{4 m_b} \int_{0}^{\infty} \left( \frac{E}{m_e c^2} \right)^2 \frac{1}{4 \pi r^2} \left( \frac{d L_{\nu_x}}{d E} \right) \mathcal{F}(r,E) \,dE,
\end{equation}
where $\mathcal{G}(t)$ and $\mathcal{F}(r,E)$ control the time and spatial dependence of the extra energy injection, respectively. $\mathcal{F}(r,E)$ is equal to $\exp{\left(-\tau_{\nu_e} (r,E) \right)}$, with $\tau_{\nu_e}$ being the opacity, in regions where neutrino heating is positive and zero everywhere else. Notice that early versions of PUSH also required the entropy gradient to be negative (i.e. the gain region) in order for $\mathcal{F}(r,E)$ to be nonzero, which is relevant for the discussion that follows. 

$\mathcal{G}(t)$ is initially zero, and then at a time $t_{\rm on}$ gradually increases until it reaches its maximum value $k_{\rm push}$, and eventually gradually decreases back to zero after the explosion has been launched. As can be seen from Eq.~\eqref{eq:QPUSH},
the extra heating in PUSH depends on the luminosity and energy of the heavy lepton neutrinos, which are evolved with the simplified IDSA formalism \citep{Liebendorfer2009_IDSA}, that is known to not properly capture the time evolution of the neutrino energy. However, the energy dissipation due to $\nu$-driven convection can be directly related to the luminosity of the electron flavor neutrinos \citep{Murphy2013_turb_in_CCSNe} or, equivalently, to the neutrino heating in the gain region \citep{Gogilashvili2024_FEC+}. This is the first major difference between PUSH and STIR. 

The biggest difference between STIR (but also the 2D and 3D simulations considered here) and PUSH is however the calibration method. The factor $\mathcal{G}(t)$ in Eq.~\eqref{eq:QPUSH} mainly depends on two parameters: $t_{\rm rise}$ and $k_{\rm push}$. Two more parameters are present in $\mathcal{G}(t)$ (i.e. $t_{\rm on}$ and $t_{\rm off}$), but their impact on the explosion dynamics is quite small since they are constrained quite well, and we will therefore not discuss them. We refer the reader to the original paper \citep{Perego2015_PUSH1} for more details. The first parameter, $t_{\rm rise}$, controls how fast the extra heating in PUSH reaches the maximum value. This can be thought of as the growth time scale of the largest multi-dimensional perturbations. The second parameter, $k_{\rm push}$, is the one responsible for setting the amount of extra heating that will be injected into the model.

The calibration of $k_{\rm push}$ was performed by \citet{Perego2015_PUSH1} and \citet{Ebinger2019_PUSH_II_explodability}, with a parabolic dependence of $k_{\rm push}$ on compactness (see Figure 8 of \citet{Ebinger2019_PUSH_II_explodability}). The three points used to constrain the parabolic dependence of $k_{\rm push}$ are: $k_{\rm push} = 4.3$ at $\xi_{2.0} = 0.245$, which represents the best fit for SN 1987A \citep{Sonneborn1987A}, and is also very close to the overall maximum of $k_{\rm push}$. Then, $k_{\rm push} = 2.5$ at $\xi_{2.0} = 0.0$, justified by the fact that low-compactness stars lead to weaker explosions and therefore $k_{\rm push}$ needs to be smaller. 

Most important for the current discussion is the value of $k_{\rm push}$ at large compactnesses. \citet{Ebinger2019_PUSH_II_explodability} constrain $k_{\rm push}$ to be zero for $\xi_{2.0} \geqslant 0.7$. This was done because the convective neutrino-driven mechanism was believed to be less efficient in these high-compactness progenitors which, therefore, were expected to form black holes, and $k_{\rm push}$ was specifically tuned to reflect that. However, as shown in this paper, this is not necessarily true since 2D, 3D, and also 1D+ simulations indicate that neutrino heating is quite strong for the high-compactness progenitors and likely able to sustain strong explosions for large values of $\xi_{2.0}$.

Based on the discussion in Section~\ref{sec:semi_an_qdot}, it could be expected that a different calibration of $k_{\rm push}$ at large values of $\xi_{2.0}$ would produce similar explosions to the 2D, 3D and 1D+ models. However, as pointed out above, the overall extra heating in PUSH depends on the luminosities and energies of the heavy lepton neutrinos, whereas neutrino-driven convection can be connected to the luminosities and energies of the electron-type neutrinos \citep{Murphy2013_turb_in_CCSNe}, although it is not obvious whether this would cause large discrepancies.

In the stalled-shock phase when neutrino-driven convection is active, the luminosities of all flavors have a very similar time dependence. The neutrino energy, however, is quite different in spherical symmetry between electron flavor neutrinos and heavy lepton neutrinos. The latter is seen to be relatively constant in time in most codes \citep{OConnor2018_comparison}, whereas the electron flavor energies increase in time, as expected by the increase in neutrinosphere temperature (see upper panel of Figure~\ref{fig:Enue_Lnue_vs_time}). Therefore, even with a modified compactness dependence of $k_{\rm push}$ one might still obtain some qualitative differences between PUSH and the 1D+ models discussed here. 

It should be pointed out that, contrary to spherical symmetry, in multi-dimensional simulations, the energy of heavy lepton neutrinos is also increasing in time, most likely due to the multi-dimensional effects, such as PNS convection. A detailed study involving neutrino opacities and transport should be conducted to clarify this, but this goes beyond the scope of the present work.

Another difference between the PUSH models and the 1D+ \citep{Couch2020_STIR,Boccioli2023_explodability} and 2D models \citep{Vartanyan2023_nu_100_2D} discussed in this paper, is the fate of lower mass stars at solar metallicity evolved with the KEPLER stellar evolution code in the mass range \hbox{$12 M_\odot \lesssim M_{\rm ZAMS} \lesssim 15 M_\odot$}. PUSH obtains successful explosions for these progenitors whereas \citep{Couch2020_STIR,Boccioli2023_explodability} and \citep{Vartanyan2023_nu_100_2D} obtain a failed-SN, with the 1D+ models from \citet{Boccioli2023_explodability,Couch2020_STIR} finding even more failed-SN, which might be connected to the discussion in Section~\ref{sec:comparison_multiD}. This discrepancy could be due to a combination of a few factors. 

First, the compactness of these stars is typically $0.1 \lesssim \xi_{2.0} \lesssim 0.3$. Therefore, the extra heating injected by PUSH in these models is sensitive to the low-compactness dependence of $k_{\rm push}$, which mostly depends on the calibration on SN 1987A. The most important quantities to base the calibration on are the explosion energy and the ejected nickel mass. The first shows a degeneracy between $t_{\rm rise}$ and $k_{\rm push}$, i.e. small values of $t_{\rm rise}$ and $k_{\rm push}$ produce similar explosion energies as large values $t_{\rm rise}$ and $k_{\rm push}$. However, small values of $t_{\rm rise}$ also correspond to early explosions (since the maximum of $Q^{+}_{\rm push}$ is reached earlier), which will eject more mass and therefore more nickel. This could in principle be used to break this degeneracy, with the assumption that little to no fallback occurs, which is the assumption adopted for the calibration of PUSH. However, some amount of fallback is to be expected for asymmetric explosions, although the exact amount of this fallback is quite uncertain. Gamma-ray observations of $^{44}$Ti have indeed shown some degree of asymmetry in the case of SN 1987A \citep{Boggs2015_asymmetric_1987A_Ti}, and therefore even if one ejects more mass in the spherically symmetric model in an early explosion, some of that mass likely falls back and is not detected. 

Moreover, the progenitor ZAMS mass of 1987A is not very well constrained and could be anywhere between 18 and 21 $M_\odot$ (or even larger), which for the KEPLER models used in the calibration of PUSH corresponds to values of $\xi_{2.0}$ anywhere between 0.15 and 0.45. Given the uncertainties in the calibration of SN 1987A, one could change the calibration of $k_{\rm push}$ enough to have an overall smaller maximum value of $k_{\rm push}$, located at higher compactnesses which could in principle reproduce the multi-dimensional failed explosions of these KEPLER progenitors with $12 M_\odot \lesssim M_{\rm ZAMS} \lesssim 15 M_\odot$. Of course, this could only be verified by a thorough study, where the PUSH calibration is systematically changed (for example with the goal of reproducing Figure~\ref{fig:Qdot_vs_comp_multiD}). The purpose of this study is however only to highlight the differences among the main 1D explosion models. These differences are fundamentally due to different prescriptions of how the extra-heating of the 1D model depends on the pre-supernova properties (i.e. compactness) as well as on the dynamical properties of the explosion (i.e. neutrino luminosity, average energy, etc...).

\subsection{Ertl model}
\label{sec:Ertl}

The model of \citet{Ertl2016_explodability} and \cite{Sukhbold2016_explodability}, based on the prescription developed by \citet{Ugliano2012}, is another widely adopted 1D method to study explodability. Similar to the PUSH models, it predicts the successful explosion of low-mass KEPLER models with $12 M_\odot \lesssim M_{\rm ZAMS} \lesssim 15 M_\odot$, and a failed explosion of progenitors with large compactness which, for KEPLER progenitors at solar metallicity from both \citet{Sukhbold2016_explodability} and \citet{Sukhbold2018_preSN_KEPLER_bug}, correspond to a mass range $22 M_\odot \lesssim M_{\rm ZAMS} \lesssim 25 M_\odot$ and $M_{\rm ZAMS} \gtrsim 30 M_\odot$. This conflicts with both the 2D results of \citet{Vartanyan2023_nu_100_2D}, who adopt the \citet{Sukhbold2018_preSN_KEPLER_bug} presupernova models, and the 1D+ models of \citet{Boccioli2023_explodability,Couch2020_STIR}, who adopt the \citet{Sukhbold2016_explodability} presupernova models.

In their explosion model, the inner $1.1 M_\odot$ core is excised after bounce, and replaced with a moving inner boundary that sets both neutrino energy and luminosity. The parameters of the model are calibrated on a few different presupernova progenitors that were computed specifically to reproduce the properties of the progenitor of SN 1987A. We refer the reader to \citet{Ugliano2012} for a detailed description of the model and its calibration. Here we summarize the main features relevant to our discussion. 

The electron neutrino luminosity is specified at the inner boundary, and it depends on the gravitational binding energy of the core, and also takes into account the effect of mass accretion onto the inner core (i.e. through the inner boundary). Neutrino transport is then described using the gray treatment of \citet{Scheck2006_MultiD_approx_transport_I}. Relevantly for our discussion, neutrino energies depend on the prescription for the contraction of the PNS. As explained in section 2.3.3 of \citet{Ertl2016_explodability}, in their models the average neutrino energies do not significantly increase in time, as instead seen in multi-dimensional simulations. This is a consequence of their choice of a mild contraction of the 1.1 $M_\odot$ shell, which was motivated by the fact that it poses less stringent time-step constraints compared to a more extreme one. 

In principle, this might not necessarily be an issue, since neutrino heating depends on $L_\nu \langle  \epsilon_\nu^2 \rangle$, and therefore the model can simply be tuned to produce larger luminosities (typically a factor of 2-3 larger than what is seen in self-consistent simulations with accurate neutrino transport), which will compensate for the lower neutrino energies at late times. However, as shown in Section~\ref{sec:semi_an_qdot}, the average energy of neutrinos plays a crucial role in determining the explosion of the high-compactness progenitors, and its time dependence is not only important to recover long-term PNS cooling, but also the short-term shock revival. 

The calibration of the explosion model was performed by choosing an inner boundary contraction and neutrino luminosity that would reproduce the properties of SN 1987A, which resulted in different "engines", depending on what numerical model is chosen to represent SN 1987A. Four engines were calibrated on models with a compactness $\xi_{2.0} \sim 0.2$, whereas the so-called W15 engine was calibrated on a model with a very low compactness $\xi_{2.0} \sim 0.03$. However, as reported by \citet{Ertl2016_explodability}, pre-collapse profiles of this model are not available anymore, and therefore it should be treated as an exception. 

Since neutrino energies experience a very small increase in time, neutrino heating in the gain region is mostly determined by the luminosity. Large mass accretion rates, that positively correlate with compactness, require larger neutrino heating to explode and, therefore, higher luminosities. The different engines are able to explode more stars depending on how strong they are, and specifically they are able to explode stars with low accretion rates, whereas they are not able to explode stars with larger accretion rates. This can be seen, for example, by looking at the bottom panels of Figure 7 from \citet{Ertl2016_explodability}, where it is clear that stars with larger mass accretion rates do not explode.

Therefore, the qualitative difference between the Ertl model and the multi-D and 1D+ results can most likely be explained by the milder increase of the neutrino average energies in time.

\subsection{Muller 2016}
\label{sec:Muller}

Another approach to the explodability is the semi-analytical model of \citet{Muller2016_prog_connection}, which does not require complex simulations at all. It is built upon the theoretical framework developed over the years by \citet{Janka2001_conditions_shk_revival} and \citet{Janka2012_review_CCSNe}, and it expands it and refines it in several way, also thanks to the significant improvement of self-consistent multi-dimensional simulations. The explodability derived by this approach for KEPLER progenitors is relatively similar to the one derived by \citet{Ertl2016_explodability}, but also not too dissimilar to the one derived by \citet{OConnor2011_explodability}. Differently from the 2D and 3D results of \citet{Vartanyan2023_nu_100_2D} and the 1D+ results of \citet{Couch2020_STIR,Boccioli2023_explodability}, progenitors with large compactness tend to yield failed explosions, whereas low-compactness stars are more likely to explode, with a few exceptions for stars around $15 M_\odot$. According to \citet{Muller2016_prog_connection}, most stars in the range $12 M_\odot \lesssim M_{\rm ZAMS} \lesssim 15 M_\odot$ explode, whereas stars in the range $22 M_\odot \lesssim M_{\rm ZAMS} \lesssim 25 M_\odot$ and $M_{\rm ZAMS} \gtrsim 30 M_\odot$ do not. 

Interestingly, some of the stars with compactnesses $0.3 \lesssim \xi_{2.5} \lesssim 0.4$ (which roughly corresponds to $0.5 \lesssim \xi_{2.0} \lesssim 0.7$) experience a successful shock revival that however does not turn into an explosion since the central PNS quickly accretes enough material to form a black hole and therefore the neutrino heating source that is still powering the explosion shuts off, causing the explosion to fail. This has been shown by recent simulations to be indeed a possible scenario \citep{Chan2018_BH_SN_40Msol, Burrows2023_BH_supernova_40Msol, Sykes2024_2D_fallback_SNe, Eggenberger2025_BHSNe_EOS, Burrows2024_BH_formation_3D}. However, despite an early (i.e. $\sim 1.5$ s after bounce) formation of a black hole, one can still achieve relatively moderate to very powerful explosions. Therefore, even though high-compactness progenitors will more quickly form black holes, they can still produce a successful explosion. 

The way in which \citet{Muller2016_prog_connection} include the effects of neutrino-driven convection in their model is by multiplying the shock radius by a coefficient $\alpha_{\rm turb}$ (not to be confused with the coefficient $\alpha_{\rm MLT}$ which sets the efficiency of convection in STIR). As discussed in the section above, this is indeed what happens in multi-D simulations, where the shock does stall at larger radii. However, as can be seen in Figure~\ref{fig:radii_1D_1D+}, in the 1D+ simulation the shock stalls at a large radius for an extended period of time, and the ratio between the shock radius in 1D+ and 1D simulations grows with time. As shown in Section~\ref{sec:semi_an_qdot}, this is crucial to trigger explosions in the high-compactness progenitors, since (as also correctly recognized by \citet{Muller2016_prog_connection}) the late-time accretion onto the PNS will improve the heating conditions due to the increase of the mean energy with neutron star mass. However, this can only lead to successful shock revival if the shock is stalling at large radii and, therefore, the size of the gain region remains large. 

A more refined calibration of $\alpha_{\rm turb}$ that takes this time-variability into account could most likely lead to the explosion of these high-compactness progenitors. It is not as straightforward to analyze the cause of the discrepancy for the explodability of low-mass stars with $12 M_\odot \lesssim M_{\rm ZAMS} \lesssim 15 M_\odot$. Most likely, the source of the discrepancy is caused by the fact that, in 1D+ simulations, the overall impact of neutrino-driven convection is smaller for progenitors with lower compactness, as shown in Figure~\ref{fig:spec_heat}, whereas the effects of convection are taken to be roughly the same for all progenitors in the approach of \citet{Muller2016_prog_connection}.

\subsection{Pejcha \& Thompson 2015}
\label{sec:Pejcha}

This model is a combination of a semi-analytical and a simulation-based approach. The semi-analytical model was developed by \citep{Pejcha2012_antesonic_condition}, who derived the "antesonic condition", which is an alternative approach to deriving the critical luminosity. In their model they solve the steady-state Euler equations where the contribution of neutrinos to the energy is evaluated using heating and cooling functions calculated using a simplified version of the leakage scheme from \cite{Scheck2006_MultiD_approx_transport_I}. Moreover, they also perform simulations using the leakage scheme version of \texttt{GR1D} to provide time-dependent boundary conditions to their steady-state model. For more details, see \citet{Pejcha2012_antesonic_condition,Pejcha2015_explodability}. 

The antesonic condition states that an explosion ensues if $c_{s}^2/v_{\rm esc}^2 \gtrapprox 0.19$ in the accretion flow, where $c_s$ and $v_{\rm esc}$ are the adiabatic sound speed and the escape velocity, respectively. This condition was derived by solving a simple isothermal accretion model, which was shown to be equivalent to a more accurate steady-state model with simplified neutrino physics. It was then shown that the critical luminosity curve $L_\nu^{\rm crit}$, i.e. the maximum luminosity that allows for a steady-state solution, is analogous to the critical sound speed, and that the maximum value of $c_{s}^2/v_{\rm esc}^2$ is always very close to $0.19$, for a wide range of values for the mass accretion rate, mass, and radius of the PNS. Then, in \citet{Pejcha2015_explodability} they provided a parametric explodability criterion:
\begin{equation}
    \frac{L_{\nu_e}}{L_{\nu_e}^{\rm crit}} > p \left(\frac{\dot{M}}{0.4\ M_\odot {\rm s^{-1}}} \right)^q~,
\end{equation}
where $p>0$ and $q>0$. When the luminosity $L_{\nu_e}$ reaches a fraction of the critical luminosity $L_{\nu_e}^{\rm crit}$ given by the RHS of the above equation, one expects an explosion. By varying $p$ and $q$ one finds a different fraction of explosions compared to failed SNe. 

The drawback of this approach is that there is no way to \textit{a priori} determine $p$ and $q$, which are completely free parameters. One has to instead derive observables such as explosion energies, amount of $^{56}$Ni ejected, remnant mass, etc... and use them to derive values of $p$ and $q$.

In general, by varying $p$ and $q$ one finds progressively a larger number of stronger explosions. The explodability found by \cite{Pejcha2015_explodability} for KEPLER progenitors \citep{Woosley2002_KEPLER_models} is similar to the one found by \citet{Ertl2016_explodability}. High-compactness stars (typically $M_\odot > 30$ at all metallicities) do not explode and stars in the mass range $12 M_\odot \lesssim M_{\rm ZAMS} \lesssim 15 M_\odot$ at solar metallicity explode. Instead, stars in the mass range $22 M_\odot \lesssim M_{\rm ZAMS} \lesssim 25 M_\odot$ at solar metallicity do not explode.

An important detail to keep in mind is that convection is not included in this model. The reasoning behind it is that the effects of convection are expected to be proportional to $\dot{M}$, and therefore the authors argue that by varying $p$ and $q$ one would roughly reproduce the effects of convection. While it is true that high-compactness progenitors (i.e. progenitors with higher mass accretion rates) develop stronger convection, as seen in Figure~\ref{fig:spec_heat}, it can also be seen that $\nu$-driven convection grows in time. Therefore, it is not straightforward to estimate how convection would change the critical curve. In our 1D+ model (as well as in the multi-dimensional ones) we showed that the shock stalls at a roughly constant radius for up to $\sim 300$ ms (i.e. solid lines in Figure~\ref{fig:radii_1D_1D+}). During that time, the mass accretion rate significantly drops, whereas the luminosity stays roughly constant. This is not captured by the steady-state model of \cite{Pejcha2012_antesonic_condition}, which instead would find a shock stalling at a radius that is similar to what is seen in 1D simulations (i.e. dashed lines in Figure~\ref{fig:radii_1D_1D+}), due to the lack of convection.

It should be highlighted that, in a steady-state calculation, including convection is not straightforward, since as we discussed above, the strength of convection changes in time. Early attempts were performed by \citet{Yamasaki2006_steady_state_conv}, who however implemented a relatively simple convection model, which only included the effects of the enthalpy flux, without the turbulent dissipation and assuming a complete flattening of the entropy gradient, which is not seen in realistic simulations. A more complete model was developed by \cite{Mabanta2018_MLT_turb}, and more recently a Force Explosion Condition which includes the effects of convection was developed by \citet{Gogilashvili2024_FEC+}, who showed that the effects of convection can reduce the critical curve by $\sim 20\text{--}30 \%$, as seen in multi-dimensional simulations. 

The assumption in the semi-analytical model by \cite{Pejcha2015_explodability} that the effects of convection can be reproduced by adjusting the values of $p$ and $q$ since they are proportional to $\dot{M}$ is too simplistic. In some way, it is a similar assumption to the one made by \cite{Muller2016_prog_connection}, as discussed in Section~\ref{sec:Muller}, who include the effects of convection by multiplying the shock radius by a constant factor. They both fail to capture the fact that the strength of convection grows in time, and causes the shock to stall at a roughly constant radius for a few hundred milliseconds, allowing neutrino heating to grow as well, which is the primary reason behind the explosion of high-compactness progenitors. It is therefore not surprising that both models yield roughly the same explodability.

\section{Summary and Conclusions}
\label{sec:conclusions}
In this paper, we analyzed the dependency of neutrino heating on the compactness of the pre-supernova progenitor, in order to explain the explodability of high-compactness stars. We showed that multi-dimensional simulations consistently yield explosions of high-compactness stars. Moreover, 1D+ models are able to correctly reproduce both the explodability and the neutrino heating found in multi-dimensional simulations, particularly for high-compactness progenitors, whereas they underestimate the amount of neutrino heating in low-compactness models.

By analyzing the differences between 1D+ and 1D models, we showed that the inclusion of neutrino-driven convection allows the shock to stall at larger radii, which causes the volume of the gain region to be roughly constant during the stalled-shock phase. Therefore, the mass of the gain region does not decrease as rapidly as in the 1D case. This allows neutrinos to efficiently heat the material in the gain region, causing net neutrino heating to rapidly increase. In higher-compactness progenitors, characterized by larger mass accretion rates, and therefore a more massive PNS, neutrino energies increase more rapidly, causing neutrino heating to also more rapidly increase.

From simple fits (whose details can be found in Appendix~\ref{sec:appendix_Qdot_comp}) we were able to derive a curve describing how $\dot{Q}_\nu^{\rm max}$ depends on the pre-supernova compactness, and showed that multi-dimensional simulations and 1D+ models closely follow said curve. The rapid increase of $\dot{Q}_\nu^{\rm max}$ as a function of compactness shows that high-compactness progenitors are able to generate very strong neutrino heating, and therefore despite the larger mass accretion rates draining energy from the shock, the shock is always successfully revived. As shown by recent multi-dimensional simulations \citep{Chan2018_BH_SN_40Msol,Powell2021_collapse_PISNe,Burrows2023_BH_supernova_40Msol,Sykes2024_2D_fallback_SNe,Eggenberger2025_BHSNe_EOS,Burrows2024_BH_formation_3D} after shock revival, even if the formation of a black hole is very likely, shock breakout is quite likely, and the ensuing explosion can have energies ranging from 0.01-3.5 $10^{51}$~erg/s. 

We also analyzed the interplay between three different timescales of the problem: $t_{\rm max}$, i.e. the time when $\dot{Q}_\nu^{\rm max}$ is reached; $t_{\rm Si/O}$, i.e. the time when the Si/O interface is accreted through the shock; $t_{\rm expl}$, i.e. the time when the shock crosses 500~km. We showed that, except for a few special cases, $t_{\rm max} \approx t_{\rm Si/O}$. This shows that after the accretion of the Si/O interface neutrino heating significantly drops, and therefore achieving an explosion after that becomes much harder. Therefore, this explains why the accretion of the Si/O interface can be so important in determining the explosion \citep{Boccioli2023_explodability,Wang2022_prog_study_ram_pressure}. However, we also showed that the Si/O interface only influences the explosion for progenitors with $\xi_{2.0} < 0.5$ whereas it has a much smaller impact on high-compactness ones with $\xi_{2.0} > 0.5$, as also recognized by \citet{Boccioli2024_remnant} and \citet{Boccioli2024_FEC+_SiO}.

Finally, we compared 1D+ models to other 1D models widely used in the literature and analyzed the differences in explodability in light of the neutrino-heating dependence on compactness highlighted in this work. In general, other 1D models find that high-compactness progenitors do not explode because: their calibration forces them not to (i.e. PUSH); their neutrino energies do not properly increase in time (\citet{Ertl2016_explodability}); convection is not explicitly included (\citet{Pejcha2015_explodability}) or is included as a constant, time-independent factor (\citet{Muller2016_prog_connection}); it is assumed that neutrino heating efficiency is the same for all progenitors (\citet{OConnor2011_explodability}). An interesting exception to this is the model from \citet{Jost2025_fheat_yields}, where however a very simple strategy is adopted, i.e. artificially increase neutrino heating by a factor of 2-3. This qualitatively produces an explodability compatible with 1D+ and multi-D models, but it also introduces an unphysical behavior of the shock, a potentially incorrect dependence of neutrino heating with the progenitor's compactness, and no clear way to handle neutrino-driven convection. Nonetheless, it is a promising method that still needs more validation and comparisons with more accurate models.

With the advent of important missions such as the Vera C. Rubin Observatory, Cosmic Explorer, and Einstein Telescope, it will become feasible to access observational signatures of high-compactness progenitors, and therefore a better understanding of their explosion is needed. In this paper, we have begun to analyze the early explosion phase and explained why the early shock revival is always successful for high-compactness progenitors. We also showed that, in the future, 1D+ models can be used to carry out similar studies of large sets of simulations, given their agreement with neutrino heating in multi-D models, particularly for high-compactness progenitors. Notice that, however, after shock revival, the continued accretion and the asymmetry in explosion and neutrino emission introduce viewing angle effects and mixing not captured otherwise \citep{Wongwathanarat2015_3D_shk_breakout,Muller2018_stripped_SN_to_breakout,Stockinger2020_3D_low_mass_to_breakout,Sandoval2021_3D_shk_breakout,Vartanyan2025_3D_bounce_to_breakut}. Regarding high-compactness progenitors, there are still some differences among simulations from different groups. Some obtain successful shock revival and late-time shock expansion, yielding successful explosions \citep{Burrows2024_BH_formation_3D}, whereas others obtain either a successful shock revival followed by a very early black-hole formation that stops the explosion, or no shock revival at all \citep{Kuroda2018_BH_formation_GR,Summa2018_rot3D_crit_lum,Walk2020_3DBH_neutrinos}. This further shows that more  work on the later phases of the explosion is needed, and the first efforts have already begun \citep{Stockinger2020_3D_low_mass_to_breakout,Sykes2024_2D_fallback_SNe,Eggenberger2025_BHSNe_EOS,Burrows2024_BH_formation_3D}.

\section*{Acknowledgements}
The authors would like to thank the anonymus referee for significantly contributing to the improvement of this manuscript. They would also like to thank Hans-Thomas Janka and Bernhard M\"uller for their useful comments. L.B. would like to thank Jeremiah Murphy and Mariam Gogilashvili for fruitful discussions. The authors thank Xingzao Li and Oliver Eggenberger Andersen who have simulated many of the 2D FLASH simulations in this work.  L.B. is supported by the U.S. Department of Energy under Grant No. DE-SC0004658 and SciDAC grant, DE-SC0024388. L.B. would like to thank the N3AS center for their hospitality and support. E.O. is supported by the Swedish Research Council (Project No. 2020-00452). D.K. is supported in part by a grant from the Simons Foundation (622817DK). Some of the computations were enabled by resources provided by the National Academic Infrastructure for Supercomputing in Sweden (NAISS), partially funded by the Swedish Research Council through grant agreement no. 2022-06725. This research benefited from support from the Gordon and Betty Moore Foundation through Grant GBMF5076.

\section*{Data Availability}
The data is available upon reasonable request to the corresponding author.



\bibliographystyle{mnras}
\bibliography{References_Books,References_CNO,References_EoS_neutrinos,References_Exp_Obs,References_misc,References_Nucleosynthesis,References_SN,References_Stellar_Models} 

\begin{thebibliography}{}
\makeatletter
\relax
\def\mn@urlcharsother{\let\do\@makeother \do\$\do\&\do\#\do\^\do\_\do\%\do\~}
\def\mn@doi{\begingroup\mn@urlcharsother \@ifnextchar [ {\mn@doi@} {\mn@doi@[]}}
\def\mn@doi@[#1]#2{\def\@tempa{#1}\ifx\@tempa\@empty \href {http://dx.doi.org/#2} {doi:#2}\else \href {http://dx.doi.org/#2} {#1}\fi \endgroup}
\def\mn@eprint#1#2{\mn@eprint@#1:#2::\@nil}
\def\mn@eprint@arXiv#1{\href {http://arxiv.org/abs/#1} {{\tt arXiv:#1}}}
\def\mn@eprint@dblp#1{\href {http://dblp.uni-trier.de/rec/bibtex/#1.xml} {dblp:#1}}
\def\mn@eprint@#1:#2:#3:#4\@nil{\def\@tempa {#1}\def\@tempb {#2}\def\@tempc {#3}\ifx \@tempc \@empty \let \@tempc \@tempb \let \@tempb \@tempa \fi \ifx \@tempb \@empty \def\@tempb {arXiv}\fi \@ifundefined {mn@eprint@\@tempb}{\@tempb:\@tempc}{\expandafter \expandafter \csname mn@eprint@\@tempb\endcsname \expandafter{\@tempc}}}

\bibitem[\protect\citeauthoryear{{Andresen}, {O'Connor}, {Andersen}  \& {Couch}}{{Andresen} et~al.}{2024}]{Andresen2024_GreyM1}
{Andresen} H.,  {O'Connor} E.~P.,  {Andersen} O.~E.,   {Couch} S.~M.,  2024, \mn@doi [\aap] {10.1051/0004-6361/202449776}, \href {https://ui.adsabs.harvard.edu/abs/2024A&A...687A..55A} {687, A55}

\bibitem[\protect\citeauthoryear{{Andrews}, {Weinberg}, {Sch{\"o}nrich}  \& {Johnson}}{{Andrews} et~al.}{2017}]{Andrews2017_GCE_flexCE}
{Andrews} B.~H.,  {Weinberg} D.~H.,  {Sch{\"o}nrich} R.,   {Johnson} J.~A.,  2017, \mn@doi [\apj] {10.3847/1538-4357/835/2/224}, \href {https://ui.adsabs.harvard.edu/abs/2017ApJ...835..224A} {835, 224}

\bibitem[\protect\citeauthoryear{{Andrews} et~al.,}{{Andrews} et~al.}{2024}]{Andrews2024_POSYDON2}
{Andrews} J.~J.,  et~al., 2024, \mn@doi [arXiv e-prints] {10.48550/arXiv.2411.02376}, \href {https://ui.adsabs.harvard.edu/abs/2024arXiv241102376A} {p. arXiv:2411.02376}

\bibitem[\protect\citeauthoryear{{Arnett}}{{Arnett}}{1966}]{Arnett1966}
{Arnett} W.~D.,  1966, \mn@doi [Canadian Journal of Physics] {10.1139/p66-210}, \href {https://ui.adsabs.harvard.edu/abs/1966CaJPh..44.2553A} {44, 2553}

\bibitem[\protect\citeauthoryear{{Beasor}, {Smith}  \& {Jencson}}{{Beasor} et~al.}{2025}]{Beasor2025_RSGP}
{Beasor} E.~R.,  {Smith} N.,   {Jencson} J.~E.,  2025, \mn@doi [\apj] {10.3847/1538-4357/ad8f3f}, \href {https://ui.adsabs.harvard.edu/abs/2025ApJ...979..117B} {979, 117}

\bibitem[\protect\citeauthoryear{{Belczynski}, {Kalogera}, {Rasio}, {Taam}, {Zezas}, {Bulik}, {Maccarone}  \& {Ivanova}}{{Belczynski} et~al.}{2008}]{Belczynski2008_Stratrack_popsynth}
{Belczynski} K.,  {Kalogera} V.,  {Rasio} F.~A.,  {Taam} R.~E.,  {Zezas} A.,  {Bulik} T.,  {Maccarone} T.~J.,   {Ivanova} N.,  2008, \mn@doi [\apjs] {10.1086/521026}, \href {https://ui.adsabs.harvard.edu/abs/2008ApJS..174..223B} {174, 223}

\bibitem[\protect\citeauthoryear{{Bethe} \& {Wilson}}{{Bethe} \& {Wilson}}{1985}]{Bethe_Wilson1985}
{Bethe} H.~A.,  {Wilson} J.~R.,  1985, \mn@doi [\apj] {10.1086/163343}, \href {https://ui.adsabs.harvard.edu/abs/1985ApJ...295...14B} {295, 14}

\bibitem[\protect\citeauthoryear{{Boccioli} \& {Fragione}}{{Boccioli} \& {Fragione}}{2024}]{Boccioli2024_remnant}
{Boccioli} L.,  {Fragione} G.,  2024, \mn@doi [\prd] {10.1103/PhysRevD.110.023007}, \href {https://ui.adsabs.harvard.edu/abs/2024PhRvD.110b3007B} {110, 023007}

\bibitem[\protect\citeauthoryear{{Boccioli}, {Mathews}  \& {O'Connor}}{{Boccioli} et~al.}{2021}]{Boccioli2021_STIR_GR}
{Boccioli} L.,  {Mathews} G.~J.,   {O'Connor} E.~P.,  2021, \mn@doi [\apj] {10.3847/1538-4357/abe767}, \href {https://ui.adsabs.harvard.edu/abs/2021ApJ...912...29B} {912, 29}

\bibitem[\protect\citeauthoryear{{Boccioli}, {Mathews}, {Suh}  \& {O'Connor}}{{Boccioli} et~al.}{2022}]{Boccioli2022_EOS_effect}
{Boccioli} L.,  {Mathews} G.~J.,  {Suh} I.-S.,   {O'Connor} E.~P.,  2022, \mn@doi [\apj] {10.3847/1538-4357/ac4603}, \href {https://ui.adsabs.harvard.edu/abs/2022ApJ...926..147B} {926, 147}

\bibitem[\protect\citeauthoryear{{Boccioli}, {Roberti}, {Limongi}, {Mathews}  \& {Chieffi}}{{Boccioli} et~al.}{2023}]{Boccioli2023_explodability}
{Boccioli} L.,  {Roberti} L.,  {Limongi} M.,  {Mathews} G.~J.,   {Chieffi} A.,  2023, \mn@doi [\apj] {10.3847/1538-4357/acc06a}, \href {https://ui.adsabs.harvard.edu/abs/2023ApJ...949...17B} {949, 17}

\bibitem[\protect\citeauthoryear{{Boccioli}, {Gogilashvili}, {Murphy}  \& {O'Connor}}{{Boccioli} et~al.}{2025}]{Boccioli2024_FEC+_SiO}
{Boccioli} L.,  {Gogilashvili} M.,  {Murphy} J.,   {O'Connor} E.~P.,  2025, \mn@doi [\mnras] {10.1093/mnras/staf066}, \href {https://ui.adsabs.harvard.edu/abs/2025MNRAS.537.1182B} {537, 1182}

\bibitem[\protect\citeauthoryear{{Boggs} et~al.,}{{Boggs} et~al.}{2015}]{Boggs2015_asymmetric_1987A_Ti}
{Boggs} S.~E.,  et~al., 2015, \mn@doi [Science] {10.1126/science.aaa2259}, \href {https://ui.adsabs.harvard.edu/abs/2015Sci...348..670B} {348, 670}

\bibitem[\protect\citeauthoryear{{Bollig}, {Janka}, {Lohs}, {Mart{\'\i}nez-Pinedo}, {Horowitz}  \& {Melson}}{{Bollig} et~al.}{2017}]{Bollig2017_muon_creation}
{Bollig} R.,  {Janka} H.~T.,  {Lohs} A.,  {Mart{\'\i}nez-Pinedo} G.,  {Horowitz} C.~J.,   {Melson} T.,  2017, \mn@doi [\prl] {10.1103/PhysRevLett.119.242702}, \href {https://ui.adsabs.harvard.edu/abs/2017PhRvL.119x2702B} {119, 242702}

\bibitem[\protect\citeauthoryear{{Breivik} et~al.,}{{Breivik} et~al.}{2020}]{Breivik2020_COSMIC_popsynth}
{Breivik} K.,  et~al., 2020, \mn@doi [\apj] {10.3847/1538-4357/ab9d85}, \href {https://ui.adsabs.harvard.edu/abs/2020ApJ...898...71B} {898, 71}

\bibitem[\protect\citeauthoryear{{Bruenn}}{{Bruenn}}{1985}]{Bruenn1985}
{Bruenn} S.~W.,  1985, \mn@doi [\apjs] {10.1086/191056}, \href {https://ui.adsabs.harvard.edu/abs/1985apjsup...58..771B} {58, 771}

\bibitem[\protect\citeauthoryear{{Bruenn} et~al.,}{{Bruenn} et~al.}{2020}]{Bruenn2020_CHIMERA_method}
{Bruenn} S.~W.,  et~al., 2020, \mn@doi [\apjs] {10.3847/1538-4365/ab7aff}, \href {https://ui.adsabs.harvard.edu/abs/2020ApJS..248...11B} {248, 11}

\bibitem[\protect\citeauthoryear{{Buras}, {Janka}, {Keil}, {Raffelt}  \& {Rampp}}{{Buras} et~al.}{2003}]{Buras2003_numutau_pair}
{Buras} R.,  {Janka} H.-T.,  {Keil} M.~T.,  {Raffelt} G.~G.,   {Rampp} M.,  2003, \mn@doi [\apj] {10.1086/368015}, \href {https://ui.adsabs.harvard.edu/abs/2003ApJ...587..320B} {587, 320}

\bibitem[\protect\citeauthoryear{{Burrows}, {Reddy}  \& {Thompson}}{{Burrows} et~al.}{2006}]{BRT2006}
{Burrows} A.,  {Reddy} S.,   {Thompson} T.~A.,  2006, \mn@doi [\nphysa] {10.1016/j.nuclphysa.2004.06.012}, \href {https://ui.adsabs.harvard.edu/abs/2006NuPhA.777..356B} {777, 356}

\bibitem[\protect\citeauthoryear{{Burrows}, {Radice}, {Vartanyan}, {Nagakura}, {Skinner}  \& {Dolence}}{{Burrows} et~al.}{2020}]{Burrows2020_3DFornax}
{Burrows} A.,  {Radice} D.,  {Vartanyan} D.,  {Nagakura} H.,  {Skinner} M.~A.,   {Dolence} J.~C.,  2020, \mn@doi [\mnras] {10.1093/mnras/stz3223}, \href {https://ui.adsabs.harvard.edu/abs/2020MNRAS.491.2715B} {491, 2715}

\bibitem[\protect\citeauthoryear{{Burrows}, {Vartanyan}  \& {Wang}}{{Burrows} et~al.}{2023}]{Burrows2023_BH_supernova_40Msol}
{Burrows} A.,  {Vartanyan} D.,   {Wang} T.,  2023, \mn@doi [\apj] {10.3847/1538-4357/acfc1c}, \href {https://ui.adsabs.harvard.edu/abs/2023ApJ...957...68B} {957, 68}

\bibitem[\protect\citeauthoryear{{Burrows}, {Wang}  \& {Vartanyan}}{{Burrows} et~al.}{2024a}]{Burrows2024_BH_formation_3D}
{Burrows} A.,  {Wang} T.,   {Vartanyan} D.,  2024a, arXiv e-prints, \href {https://ui.adsabs.harvard.edu/abs/2024arXiv241207831B} {p. arXiv:2412.07831}

\bibitem[\protect\citeauthoryear{{Burrows}, {Wang}  \& {Vartanyan}}{{Burrows} et~al.}{2024b}]{Burrows2024_Phys_correlations}
{Burrows} A.,  {Wang} T.,   {Vartanyan} D.,  2024b, \mn@doi [\apjl] {10.3847/2041-8213/ad319e}, \href {https://ui.adsabs.harvard.edu/abs/2024ApJ...964L..16B} {964, L16}

\bibitem[\protect\citeauthoryear{{Cabez{\'o}n}, {Pan}, {Liebend{\"o}rfer}, {Kuroda}, {Ebinger}, {Heinimann}, {Perego}  \& {Thielemann}}{{Cabez{\'o}n} et~al.}{2018}]{Cabezon2018_3Dcomparison}
{Cabez{\'o}n} R.~M.,  {Pan} K.-C.,  {Liebend{\"o}rfer} M.,  {Kuroda} T.,  {Ebinger} K.,  {Heinimann} O.,  {Perego} A.,   {Thielemann} F.-K.,  2018, \mn@doi [\aap] {10.1051/0004-6361/201833705}, \href {https://ui.adsabs.harvard.edu/abs/2018A&A...619A.118C} {619, A118}

\bibitem[\protect\citeauthoryear{{Chan}, {M{\"u}ller}, {Heger}, {Pakmor}  \& {Springel}}{{Chan} et~al.}{2018}]{Chan2018_BH_SN_40Msol}
{Chan} C.,  {M{\"u}ller} B.,  {Heger} A.,  {Pakmor} R.,   {Springel} V.,  2018, \mn@doi [\apjl] {10.3847/2041-8213/aaa28c}, \href {https://ui.adsabs.harvard.edu/abs/2018ApJ...852L..19C} {852, L19}

\bibitem[\protect\citeauthoryear{{Chen}, {Hayden}, {Sharma}, {Bland-Hawthorn}, {Kobayashi}  \& {Karakas}}{{Chen} et~al.}{2023}]{Chen2023_GCE_radial_mixing}
{Chen} B.,  {Hayden} M.~R.,  {Sharma} S.,  {Bland-Hawthorn} J.,  {Kobayashi} C.,   {Karakas} A.~I.,  2023, \mn@doi [\mnras] {10.1093/mnras/stad1568}, \href {https://ui.adsabs.harvard.edu/abs/2023MNRAS.523.3791C} {523, 3791}

\bibitem[\protect\citeauthoryear{{Chieffi} et~al.,}{{Chieffi} et~al.}{2021}]{Chieffi2021_C12_compactness}
{Chieffi} A.,  et~al., 2021, \mn@doi [\apj] {10.3847/1538-4357/ac06ca}, \href {https://ui.adsabs.harvard.edu/abs/2021ApJ...916...79C} {916, 79}

\bibitem[\protect\citeauthoryear{{Colgate} \& {White}}{{Colgate} \& {White}}{1966}]{Colgate_White1966}
{Colgate} S.~A.,  {White} R.~H.,  1966, \mn@doi [\apj] {10.1086/148549}, \href {https://ui.adsabs.harvard.edu/abs/1966ApJ...143..626C} {143, 626}

\bibitem[\protect\citeauthoryear{{C{\^o}t{\'e}}, {O'Shea}, {Ritter}, {Herwig}  \& {Venn}}{{C{\^o}t{\'e}} et~al.}{2017}]{Cote2017_GCE_OMEGA}
{C{\^o}t{\'e}} B.,  {O'Shea} B.~W.,  {Ritter} C.,  {Herwig} F.,   {Venn} K.~A.,  2017, \mn@doi [\apj] {10.3847/1538-4357/835/2/128}, \href {https://ui.adsabs.harvard.edu/abs/2017ApJ...835..128C} {835, 128}

\bibitem[\protect\citeauthoryear{{Couch}, {Warren}  \& {O'Connor}}{{Couch} et~al.}{2020}]{Couch2020_STIR}
{Couch} S.~M.,  {Warren} M.~L.,   {O'Connor} E.~P.,  2020, \mn@doi [\apj] {10.3847/1538-4357/ab609e}, \href {https://ui.adsabs.harvard.edu/abs/2020ApJ...890..127C} {890, 127}

\bibitem[\protect\citeauthoryear{Couch, Carlson, Pajkos, O’Shea, Dubey  \& Klosterman}{Couch et~al.}{2021}]{Couch2021_Spark}
Couch S.~M.,  Carlson J.,  Pajkos M.,  O’Shea B.~W.,  Dubey A.,   Klosterman T.,  2021, \mn@doi [Parallel Computing] {https://doi.org/10.1016/j.parco.2021.102830}, 108, 102830

\bibitem[\protect\citeauthoryear{{Davies} \& {Beasor}}{{Davies} \& {Beasor}}{2018}]{Davies2018_RSG_IIP_ZAMS_mass}
{Davies} B.,  {Beasor} E.~R.,  2018, \mn@doi [\mnras] {10.1093/mnras/stx2734}, \href {https://ui.adsabs.harvard.edu/abs/2018MNRAS.474.2116D} {474, 2116}

\bibitem[\protect\citeauthoryear{{Davies} \& {Beasor}}{{Davies} \& {Beasor}}{2020}]{Davies2020_RSGP_upper_lum_IIP}
{Davies} B.,  {Beasor} E.~R.,  2020, \mn@doi [\mnras] {10.1093/mnras/staa174}, \href {https://ui.adsabs.harvard.edu/abs/2020MNRAS.493..468D} {493, 468}

\bibitem[\protect\citeauthoryear{Diehl \& Prantzos}{Diehl \& Prantzos}{2020}]{Diehl2020_GCE_living_review}
Diehl R.,  Prantzos N.,  2020, Cosmic Radioactivity and Galactic Chemical Evolution.
Springer Nature Singapore, Singapore, pp 1--83, \mn@doi{10.1007/978-981-15-8818-1_107-1}, \url {https://doi.org/10.1007/978-981-15-8818-1_107-1}

\bibitem[\protect\citeauthoryear{{Ebinger}, {Curtis}, {Fr{\"o}hlich}, {Hempel}, {Perego}, {Liebend{\"o}rfer}  \& {Thielemann}}{{Ebinger} et~al.}{2019}]{Ebinger2019_PUSH_II_explodability}
{Ebinger} K.,  {Curtis} S.,  {Fr{\"o}hlich} C.,  {Hempel} M.,  {Perego} A.,  {Liebend{\"o}rfer} M.,   {Thielemann} F.-K.,  2019, \mn@doi [\apj] {10.3847/1538-4357/aae7c9}, \href {https://ui.adsabs.harvard.edu/abs/2019ApJ...870....1E} {870, 1}

\bibitem[\protect\citeauthoryear{{Eggenberger Andersen}, {Zha}, {da Silva Schneider}, {Betranhandy}, {Couch}  \& {O'Connor}}{{Eggenberger Andersen} et~al.}{2021}]{Eggenberger2021_EOS_dependence_GW_2D}
{Eggenberger Andersen} O.,  {Zha} S.,  {da Silva Schneider} A.,  {Betranhandy} A.,  {Couch} S.~M.,   {O'Connor} E.~P.,  2021, \mn@doi [\apj] {10.3847/1538-4357/ac294c}, \href {https://ui.adsabs.harvard.edu/abs/2021ApJ...923..201E} {923, 201}

\bibitem[\protect\citeauthoryear{{Eggenberger Andersen}, {O'Connor}, {Andresen}, {da Silva Schneider}  \& {Couch}}{{Eggenberger Andersen} et~al.}{2025}]{Eggenberger2025_BHSNe_EOS}
{Eggenberger Andersen} O.,  {O'Connor} E.,  {Andresen} H.,  {da Silva Schneider} A.,   {Couch} S.~M.,  2025, \mn@doi [\apj] {10.3847/1538-4357/ada899}, \href {https://ui.adsabs.harvard.edu/abs/2025ApJ...980...53E} {980, 53}

\bibitem[\protect\citeauthoryear{{Eldridge}, {Stanway}, {Xiao}, {McClelland}, {Taylor}, {Ng}, {Greis}  \& {Bray}}{{Eldridge} et~al.}{2017}]{Eldridge2017_BPASS_popsynth}
{Eldridge} J.~J.,  {Stanway} E.~R.,  {Xiao} L.,  {McClelland} L.~A.~S.,  {Taylor} G.,  {Ng} M.,  {Greis} S.~M.~L.,   {Bray} J.~C.,  2017, \mn@doi [\pasa] {10.1017/pasa.2017.51}, \href {https://ui.adsabs.harvard.edu/abs/2017PASA...34...58E} {34, e058}

\bibitem[\protect\citeauthoryear{{Ertl}, {Janka}, {Woosley}, {Sukhbold}  \& {Ugliano}}{{Ertl} et~al.}{2016}]{Ertl2016_explodability}
{Ertl} T.,  {Janka} H.~T.,  {Woosley} S.~E.,  {Sukhbold} T.,   {Ugliano} M.,  2016, \mn@doi [\apj] {10.3847/0004-637X/818/2/124}, \href {https://ui.adsabs.harvard.edu/abs/2016ApJ...818..124E} {818, 124}

\bibitem[\protect\citeauthoryear{{Evans} et~al.,}{{Evans} et~al.}{2021}]{Evans2021_CE_method_paper2}
{Evans} M.,  et~al., 2021, \mn@doi [arXiv e-prints] {10.48550/arXiv.2109.09882}, \href {https://ui.adsabs.harvard.edu/abs/2021arXiv210909882E} {p. arXiv:2109.09882}

\bibitem[\protect\citeauthoryear{{Fischer} et~al.,}{{Fischer} et~al.}{2017}]{Fischer2017_Review_EOS_nu}
{Fischer} T.,  et~al., 2017, \mn@doi [\pasa] {10.1017/pasa.2017.63}, \href {https://ui.adsabs.harvard.edu/abs/2017PASA...34...67F} {34, e067}

\bibitem[\protect\citeauthoryear{{Foglizzo}, {Scheck}  \& {Janka}}{{Foglizzo} et~al.}{2006}]{Foglizzo2006_SASI}
{Foglizzo} T.,  {Scheck} L.,   {Janka} H.~T.,  2006, \mn@doi [\apj] {10.1086/508443}, \href {https://ui.adsabs.harvard.edu/abs/2006ApJ...652.1436F} {652, 1436}

\bibitem[\protect\citeauthoryear{{Fryer} \& {Warren}}{{Fryer} \& {Warren}}{2002}]{Fryer2002_first3D}
{Fryer} C.~L.,  {Warren} M.~S.,  2002, \mn@doi [\apjl] {10.1086/342258}, \href {https://ui.adsabs.harvard.edu/abs/2002ApJ...574L..65F} {574, L65}

\bibitem[\protect\citeauthoryear{{Fryer}, {Belczynski}, {Wiktorowicz}, {Dominik}, {Kalogera}  \& {Holz}}{{Fryer} et~al.}{2012}]{Fryer2012_remnant_popsynth}
{Fryer} C.~L.,  {Belczynski} K.,  {Wiktorowicz} G.,  {Dominik} M.,  {Kalogera} V.,   {Holz} D.~E.,  2012, \mn@doi [\apj] {10.1088/0004-637X/749/1/91}, \href {https://ui.adsabs.harvard.edu/abs/2012ApJ...749...91F} {749, 91}

\bibitem[\protect\citeauthoryear{{Fryer}, {Olejak}  \& {Belczynski}}{{Fryer} et~al.}{2022}]{Fryer2022_nu_conv_remnant_masses}
{Fryer} C.~L.,  {Olejak} A.,   {Belczynski} K.,  2022, \mn@doi [\apj] {10.3847/1538-4357/ac6ac9}, \href {https://ui.adsabs.harvard.edu/abs/2022ApJ...931...94F} {931, 94}

\bibitem[\protect\citeauthoryear{{Gjergo} et~al.,}{{Gjergo} et~al.}{2023}]{Gjergo2023_GCE_GalCEM1}
{Gjergo} E.,  et~al., 2023, \mn@doi [\apjs] {10.3847/1538-4365/aca7c7}, \href {https://ui.adsabs.harvard.edu/abs/2023ApJS..264...44G} {264, 44}

\bibitem[\protect\citeauthoryear{{Glas}, {Just}, {Janka}  \& {Obergaulinger}}{{Glas} et~al.}{2019}]{Glas2019_comparison_nu_transport_AA}
{Glas} R.,  {Just} O.,  {Janka} H.~T.,   {Obergaulinger} M.,  2019, \mn@doi [\apj] {10.3847/1538-4357/ab0423}, \href {https://ui.adsabs.harvard.edu/abs/2019ApJ...873...45G} {873, 45}

\bibitem[\protect\citeauthoryear{{Gogilashvili}, {Murphy}  \& {O'Connor}}{{Gogilashvili} et~al.}{2023}]{Gogilashvili2023_FEC_GR1D}
{Gogilashvili} M.,  {Murphy} J.~W.,   {O'Connor} E.~P.,  2023, \mn@doi [\mnras] {10.1093/mnras/stad2155}, \href {https://ui.adsabs.harvard.edu/abs/2023MNRAS.524.4109G} {524, 4109}

\bibitem[\protect\citeauthoryear{{Gogilashvili}, {Murphy}  \& {Miller}}{{Gogilashvili} et~al.}{2024}]{Gogilashvili2024_FEC+}
{Gogilashvili} M.,  {Murphy} J.~W.,   {Miller} J.~M.,  2024, \mn@doi [\apj] {10.3847/1538-4357/ad1d5e}, \href {https://ui.adsabs.harvard.edu/abs/2024ApJ...962..110G} {962, 110}

\bibitem[\protect\citeauthoryear{{Hannestad} \& {Raffelt}}{{Hannestad} \& {Raffelt}}{1998}]{Hannestad1998_NN_Brem}
{Hannestad} S.,  {Raffelt} G.,  1998, \mn@doi [\apj] {10.1086/306303}, \href {https://ui.adsabs.harvard.edu/abs/1998ApJ...507..339H} {507, 339}

\bibitem[\protect\citeauthoryear{{Hempel} \& {Schaffner-Bielich}}{{Hempel} \& {Schaffner-Bielich}}{2010}]{Hempel2010_HS_RMF}
{Hempel} M.,  {Schaffner-Bielich} J.,  2010, \mn@doi [\nphysa] {10.1016/j.nuclphysa.2010.02.010}, \href {https://ui.adsabs.harvard.edu/abs/2010NuPhA.837..210H} {837, 210}

\bibitem[\protect\citeauthoryear{{Herant}, {Benz}, {Hix}, {Fryer}  \& {Colgate}}{{Herant} et~al.}{1994}]{Herant1994_first2D}
{Herant} M.,  {Benz} W.,  {Hix} W.~R.,  {Fryer} C.~L.,   {Colgate} S.~A.,  1994, \mn@doi [\apj] {10.1086/174817}, \href {https://ui.adsabs.harvard.edu/abs/1994ApJ...435..339H} {435, 339}

\bibitem[\protect\citeauthoryear{{Horowitz}}{{Horowitz}}{2002}]{Horowitz2002}
{Horowitz} C.~J.,  2002, \mn@doi [\prd] {10.1103/PhysRevD.65.043001}, \href {https://ui.adsabs.harvard.edu/abs/2002PhRvD..65d3001H} {65, 043001}

\bibitem[\protect\citeauthoryear{{Horowitz}, {Caballero}, {Lin}, {O'Connor}  \& {Schwenk}}{{Horowitz} et~al.}{2017}]{Horowitz2017_virial}
{Horowitz} C.~J.,  {Caballero} O.~L.,  {Lin} Z.,  {O'Connor} E.,   {Schwenk} A.,  2017, \mn@doi [\prc] {10.1103/PhysRevC.95.025801}, \href {https://ui.adsabs.harvard.edu/abs/2017PhRvC..95b5801H} {95, 025801}

\bibitem[\protect\citeauthoryear{{Iorio} et~al.,}{{Iorio} et~al.}{2023}]{Iorio2023_SEVN_popsynth}
{Iorio} G.,  et~al., 2023, \mn@doi [\mnras] {10.1093/mnras/stad1630}, \href {https://ui.adsabs.harvard.edu/abs/2023MNRAS.524..426I} {524, 426}

\bibitem[\protect\citeauthoryear{{Ivezi{\'c}} et~al.,}{{Ivezi{\'c}} et~al.}{2019}]{Ivezic2019_LSST_method_paper}
{Ivezi{\'c}} {\v{Z}}.,  et~al., 2019, \mn@doi [\apj] {10.3847/1538-4357/ab042c}, \href {https://ui.adsabs.harvard.edu/abs/2019ApJ...873..111I} {873, 111}

\bibitem[\protect\citeauthoryear{{Izzard}, {Tout}, {Karakas}  \& {Pols}}{{Izzard} et~al.}{2004}]{Izzard2004_binaryc_popsynth}
{Izzard} R.~G.,  {Tout} C.~A.,  {Karakas} A.~I.,   {Pols} O.~R.,  2004, \mn@doi [\mnras] {10.1111/j.1365-2966.2004.07446.x}, \href {https://ui.adsabs.harvard.edu/abs/2004MNRAS.350..407I} {350, 407}

\bibitem[\protect\citeauthoryear{{Janka}}{{Janka}}{2001}]{Janka2001_conditions_shk_revival}
{Janka} H.~T.,  2001, \mn@doi [\aap] {10.1051/0004-6361:20010012}, \href {https://ui.adsabs.harvard.edu/abs/2001A&A...368..527J} {368, 527}

\bibitem[\protect\citeauthoryear{{Janka}}{{Janka}}{2012}]{Janka2012_review_CCSNe}
{Janka} H.-T.,  2012, \mn@doi [Annual Review of Nuclear and Particle Science] {10.1146/annurev-nucl-102711-094901}, \href {https://ui.adsabs.harvard.edu/abs/2012ARNPS..62..407J} {62, 407}

\bibitem[\protect\citeauthoryear{Janka \& Bauswein}{Janka \& Bauswein}{2023}]{Janka2023_EOS_and_dynamics}
Janka H.-T.,  Bauswein A.,  2023, Dynamics and Equation of State Dependencies of Relevance for Nucleosynthesis in Supernovae and Neutron Star Mergers.
Springer Nature Singapore, Singapore, pp 4005--4102, \mn@doi{10.1007/978-981-19-6345-2_93}, \url {https://doi.org/10.1007/978-981-19-6345-2_93}

\bibitem[\protect\citeauthoryear{{Johnson} \& {Weinberg}}{{Johnson} \& {Weinberg}}{2020}]{Johnson2020_GCE_VICE}
{Johnson} J.~W.,  {Weinberg} D.~H.,  2020, \mn@doi [\mnras] {10.1093/mnras/staa2431}, \href {https://ui.adsabs.harvard.edu/abs/2020MNRAS.498.1364J} {498, 1364}

\bibitem[\protect\citeauthoryear{{Jost}, {Molero}, {Nav{\'o}}, {Arcones}, {Obergaulinger}  \& {Matteucci}}{{Jost} et~al.}{2025}]{Jost2025_fheat_yields}
{Jost} F.~P.,  {Molero} M.,  {Nav{\'o}} G.,  {Arcones} A.,  {Obergaulinger} M.,   {Matteucci} F.,  2025, \mn@doi [\mnras] {10.1093/mnras/stae2718}, \href {https://ui.adsabs.harvard.edu/abs/2025MNRAS.536.2135J} {536, 2135}

\bibitem[\protect\citeauthoryear{{Just}, {Bollig}, {Janka}, {Obergaulinger}, {Glas}  \& {Nagataki}}{{Just} et~al.}{2018}]{Just2018_1D2D_Vertex_vs_Alcar}
{Just} O.,  {Bollig} R.,  {Janka} H.~T.,  {Obergaulinger} M.,  {Glas} R.,   {Nagataki} S.,  2018, \mn@doi [\mnras] {10.1093/mnras/sty2578}, \href {https://ui.adsabs.harvard.edu/abs/2018MNRAS.481.4786J} {481, 4786}

\bibitem[\protect\citeauthoryear{{Kamlah} et~al.,}{{Kamlah} et~al.}{2022}]{Kamlah2022_BSE-LEVELC_popsynth}
{Kamlah} A.~W.~H.,  et~al., 2022, \mn@doi [\mnras] {10.1093/mnras/stab3748}, \href {https://ui.adsabs.harvard.edu/abs/2022MNRAS.511.4060K} {511, 4060}

\bibitem[\protect\citeauthoryear{{Kubryk}, {Prantzos}  \& {Athanassoula}}{{Kubryk} et~al.}{2015}]{Kubryk2015_GCE_Prantzos_code}
{Kubryk} M.,  {Prantzos} N.,   {Athanassoula} E.,  2015, \mn@doi [\aap] {10.1051/0004-6361/201424171}, \href {https://ui.adsabs.harvard.edu/abs/2015A&A...580A.126K} {580, A126}

\bibitem[\protect\citeauthoryear{{Kuroda}, {Takiwaki}  \& {Kotake}}{{Kuroda} et~al.}{2016}]{Kuroda2016_GR_nu_transport_code}
{Kuroda} T.,  {Takiwaki} T.,   {Kotake} K.,  2016, \mn@doi [\apjs] {10.3847/0067-0049/222/2/20}, \href {https://ui.adsabs.harvard.edu/abs/2016ApJS..222...20K} {222, 20}

\bibitem[\protect\citeauthoryear{{Kuroda}, {Kotake}, {Takiwaki}  \& {Thielemann}}{{Kuroda} et~al.}{2018}]{Kuroda2018_BH_formation_GR}
{Kuroda} T.,  {Kotake} K.,  {Takiwaki} T.,   {Thielemann} F.-K.,  2018, \mn@doi [\mnras] {10.1093/mnrasl/sly059}, \href {https://ui.adsabs.harvard.edu/abs/2018MNRAS.477L..80K} {477, L80}

\bibitem[\protect\citeauthoryear{{Lattimer} \& {Swesty}}{{Lattimer} \& {Swesty}}{1991}]{Lattimer1991_LS}
{Lattimer} J.~M.,  {Swesty} D.~F.,  1991, \mn@doi [\nphysa] {10.1016/0375-9474(91)90452-C}, \href {https://ui.adsabs.harvard.edu/abs/1991NuPhA.535..331L} {535, 331}

\bibitem[\protect\citeauthoryear{{Lentz} et~al.,}{{Lentz} et~al.}{2015}]{Lentz2015_3D}
{Lentz} E.~J.,  et~al., 2015, \mn@doi [\apjl] {10.1088/2041-8205/807/2/L31}, \href {https://ui.adsabs.harvard.edu/abs/2015ApJ...807L..31L} {807, L31}

\bibitem[\protect\citeauthoryear{Li}{Li}{2024}]{Li2024_2D_EOS_GW_thesis}
Li X.,  2024, PhD thesis, Stockholm University, Faculty of Science, Department of Astronomy., \url {https://urn.kb.se/resolve?urn=urn:nbn:se:su:diva-236875}

\bibitem[\protect\citeauthoryear{{Liebend{\"o}rfer}, {Whitehouse}  \& {Fischer}}{{Liebend{\"o}rfer} et~al.}{2009}]{Liebendorfer2009_IDSA}
{Liebend{\"o}rfer} M.,  {Whitehouse} S.~C.,   {Fischer} T.,  2009, \mn@doi [\apj] {10.1088/0004-637X/698/2/1174}, \href {https://ui.adsabs.harvard.edu/abs/2009ApJ...698.1174L} {698, 1174}

\bibitem[\protect\citeauthoryear{{Limongi} \& {Chieffi}}{{Limongi} \& {Chieffi}}{2006}]{Limongi2006_preSN_models}
{Limongi} M.,  {Chieffi} A.,  2006, \mn@doi [\apj] {10.1086/505164}, \href {https://ui.adsabs.harvard.edu/abs/2006ApJ...647..483L} {647, 483}

\bibitem[\protect\citeauthoryear{{Mabanta} \& {Murphy}}{{Mabanta} \& {Murphy}}{2018}]{Mabanta2018_MLT_turb}
{Mabanta} Q.~A.,  {Murphy} J.~W.,  2018, \mn@doi [\apj] {10.3847/1538-4357/aaaec7}, \href {https://ui.adsabs.harvard.edu/abs/2018ApJ...856...22M} {856, 22}

\bibitem[\protect\citeauthoryear{{Maggiore} et~al.,}{{Maggiore} et~al.}{2020}]{Maggiore2020_ET_method_paper2}
{Maggiore} M.,  et~al., 2020, \mn@doi [\jcap] {10.1088/1475-7516/2020/03/050}, \href {https://ui.adsabs.harvard.edu/abs/2020JCAP...03..050M} {2020, 050}

\bibitem[\protect\citeauthoryear{{Marek}, {Dimmelmeier}, {Janka}, {M{\"u}ller}  \& {Buras}}{{Marek} et~al.}{2006}]{Marek2006}
{Marek} A.,  {Dimmelmeier} H.,  {Janka} H.~T.,  {M{\"u}ller} E.,   {Buras} R.,  2006, \mn@doi [\aap] {10.1051/0004-6361:20052840}, \href {https://ui.adsabs.harvard.edu/abs/2006A&A...445..273M} {445, 273}

\bibitem[\protect\citeauthoryear{{Mennekens} \& {Vanbeveren}}{{Mennekens} \& {Vanbeveren}}{2014}]{Mennekens2014_Brussels_popsynth}
{Mennekens} N.,  {Vanbeveren} D.,  2014, \mn@doi [\aap] {10.1051/0004-6361/201322198}, \href {https://ui.adsabs.harvard.edu/abs/2014A&A...564A.134M} {564, A134}

\bibitem[\protect\citeauthoryear{{Miller}, {Wilson}  \& {Mayle}}{{Miller} et~al.}{1993}]{Miller1993_first2D}
{Miller} D.~S.,  {Wilson} J.~R.,   {Mayle} R.~W.,  1993, \mn@doi [\apj] {10.1086/173163}, \href {https://ui.adsabs.harvard.edu/abs/1993ApJ...415..278M} {415, 278}

\bibitem[\protect\citeauthoryear{{Minerbo}}{{Minerbo}}{1978}]{Minerbo1978_closure}
{Minerbo} G.~N.,  1978, \mn@doi [\jqsrt] {10.1016/0022-4073(78)90024-9}, \href {https://ui.adsabs.harvard.edu/abs/1978JQSRT..20..541M} {20, 541}

\bibitem[\protect\citeauthoryear{{M{\"u}ller}}{{M{\"u}ller}}{2019}]{Muller2019_STIR}
{M{\"u}ller} B.,  2019, \mn@doi [\mnras] {10.1093/mnras/stz1594}, \href {https://ui.adsabs.harvard.edu/abs/2019MNRAS.487.5304M} {487, 5304}

\bibitem[\protect\citeauthoryear{{M{\"u}ller}, {Heger}, {Liptai}  \& {Cameron}}{{M{\"u}ller} et~al.}{2016a}]{Muller2016_prog_connection}
{M{\"u}ller} B.,  {Heger} A.,  {Liptai} D.,   {Cameron} J.~B.,  2016a, \mn@doi [\mnras] {10.1093/mnras/stw1083}, \href {https://ui.adsabs.harvard.edu/abs/2016MNRAS.460..742M} {460, 742}

\bibitem[\protect\citeauthoryear{{M{\"u}ller}, {Viallet}, {Heger}  \& {Janka}}{{M{\"u}ller} et~al.}{2016b}]{Muller2016_Oxburning}
{M{\"u}ller} B.,  {Viallet} M.,  {Heger} A.,   {Janka} H.-T.,  2016b, \mn@doi [\apj] {10.3847/1538-4357/833/1/124}, \href {https://ui.adsabs.harvard.edu/abs/2016ApJ...833..124M} {833, 124}

\bibitem[\protect\citeauthoryear{{M{\"u}ller}, {Gay}, {Heger}, {Tauris}  \& {Sim}}{{M{\"u}ller} et~al.}{2018}]{Muller2018_stripped_SN_to_breakout}
{M{\"u}ller} B.,  {Gay} D.~W.,  {Heger} A.,  {Tauris} T.~M.,   {Sim} S.~A.,  2018, \mn@doi [\mnras] {10.1093/mnras/sty1683}, \href {https://ui.adsabs.harvard.edu/abs/2018MNRAS.479.3675M} {479, 3675}

\bibitem[\protect\citeauthoryear{{Murphy} \& {Meakin}}{{Murphy} \& {Meakin}}{2011}]{Murphy2011}
{Murphy} J.~W.,  {Meakin} C.,  2011, \mn@doi [\apj] {10.1088/0004-637X/742/2/74}, \href {https://ui.adsabs.harvard.edu/abs/2011ApJ...742...74M} {742, 74}

\bibitem[\protect\citeauthoryear{{Murphy}, {Dolence}  \& {Burrows}}{{Murphy} et~al.}{2013}]{Murphy2013_turb_in_CCSNe}
{Murphy} J.~W.,  {Dolence} J.~C.,   {Burrows} A.,  2013, \mn@doi [\apj] {10.1088/0004-637X/771/1/52}, \href {https://ui.adsabs.harvard.edu/abs/2013ApJ...771...52M} {771, 52}

\bibitem[\protect\citeauthoryear{{Nagakura}, {Burrows}, {Radice}  \& {Vartanyan}}{{Nagakura} et~al.}{2020}]{Nagakura2020_PNS_convection}
{Nagakura} H.,  {Burrows} A.,  {Radice} D.,   {Vartanyan} D.,  2020, \mn@doi [\mnras] {10.1093/mnras/staa261}, \href {https://ui.adsabs.harvard.edu/abs/2020MNRAS.492.5764N} {492, 5764}

\bibitem[\protect\citeauthoryear{{Nakamura}, {Takiwaki}, {Kuroda}  \& {Kotake}}{{Nakamura} et~al.}{2015}]{Nakamura2015_2D_explodability}
{Nakamura} K.,  {Takiwaki} T.,  {Kuroda} T.,   {Kotake} K.,  2015, \mn@doi [\pasj] {10.1093/pasj/psv073}, \href {https://ui.adsabs.harvard.edu/abs/2015PASJ...67..107N} {67, 107}

\bibitem[\protect\citeauthoryear{{Nakamura}, {Takiwaki}, {Matsumoto}  \& {Kotake}}{{Nakamura} et~al.}{2025}]{Nakamura2025_LS220_3D_suite}
{Nakamura} K.,  {Takiwaki} T.,  {Matsumoto} J.,   {Kotake} K.,  2025, \mn@doi [\mnras] {10.1093/mnras/stae2611}, \href {https://ui.adsabs.harvard.edu/abs/2025MNRAS.536..280N} {536, 280}

\bibitem[\protect\citeauthoryear{{O'Connor}}{{O'Connor}}{2015}]{OConnor2015}
{O'Connor} E.,  2015, \mn@doi [\apjs] {10.1088/0067-0049/219/2/24}, \href {https://ui.adsabs.harvard.edu/abs/2015apjsup..219...24O} {219, 24}

\bibitem[\protect\citeauthoryear{{O'Connor} \& {Couch}}{{O'Connor} \& {Couch}}{2018}]{OConnor2018_2D_M1}
{O'Connor} E.~P.,  {Couch} S.~M.,  2018, \mn@doi [\apj] {10.3847/1538-4357/aaa893}, \href {https://ui.adsabs.harvard.edu/abs/2018ApJ...854...63O} {854, 63}

\bibitem[\protect\citeauthoryear{{O'Connor} \& {Ott}}{{O'Connor} \& {Ott}}{2010}]{OConnor2010}
{O'Connor} E.,  {Ott} C.~D.,  2010, \mn@doi [Classical and Quantum Gravity] {10.1088/0264-9381/27/11/114103}, \href {https://ui.adsabs.harvard.edu/abs/2010CQGra..27k4103O} {27, 114103}

\bibitem[\protect\citeauthoryear{{O'Connor} \& {Ott}}{{O'Connor} \& {Ott}}{2011}]{OConnor2011_explodability}
{O'Connor} E.,  {Ott} C.~D.,  2011, \mn@doi [\apj] {10.1088/0004-637X/730/2/70}, \href {https://ui.adsabs.harvard.edu/abs/2011ApJ...730...70O} {730, 70}

\bibitem[\protect\citeauthoryear{{O'Connor} et~al.,}{{O'Connor} et~al.}{2018}]{OConnor2018_comparison}
{O'Connor} E.,  et~al., 2018, \mn@doi [Journal of Physics G Nuclear Physics] {10.1088/1361-6471/aadeae}, \href {https://ui.adsabs.harvard.edu/abs/2018JPhG...45j4001O} {45, 104001}

\bibitem[\protect\citeauthoryear{{Pan}, {Mattes}, {O'Connor}, {Couch}, {Perego}  \& {Arcones}}{{Pan} et~al.}{2019}]{Pan2019_nu_transport_2D_comparison}
{Pan} K.-C.,  {Mattes} C.,  {O'Connor} E.~P.,  {Couch} S.~M.,  {Perego} A.,   {Arcones} A.,  2019, \mn@doi [Journal of Physics G Nuclear Physics] {10.1088/1361-6471/aaed51}, \href {https://ui.adsabs.harvard.edu/abs/2019JPhG...46a4001P} {46, 014001}

\bibitem[\protect\citeauthoryear{{Pejcha} \& {Thompson}}{{Pejcha} \& {Thompson}}{2012}]{Pejcha2012_antesonic_condition}
{Pejcha} O.,  {Thompson} T.~A.,  2012, \mn@doi [\apj] {10.1088/0004-637X/746/1/106}, \href {https://ui.adsabs.harvard.edu/abs/2012ApJ...746..106P} {746, 106}

\bibitem[\protect\citeauthoryear{{Pejcha} \& {Thompson}}{{Pejcha} \& {Thompson}}{2015}]{Pejcha2015_explodability}
{Pejcha} O.,  {Thompson} T.~A.,  2015, \mn@doi [\apj] {10.1088/0004-637X/801/2/90}, \href {https://ui.adsabs.harvard.edu/abs/2015ApJ...801...90P} {801, 90}

\bibitem[\protect\citeauthoryear{{Perego}, {Hempel}, {Fr{\"o}hlich}, {Ebinger}, {Eichler}, {Casanova}, {Liebend{\"o}rfer}  \& {Thielemann}}{{Perego} et~al.}{2015}]{Perego2015_PUSH1}
{Perego} A.,  {Hempel} M.,  {Fr{\"o}hlich} C.,  {Ebinger} K.,  {Eichler} M.,  {Casanova} J.,  {Liebend{\"o}rfer} M.,   {Thielemann} F.~K.,  2015, \mn@doi [\apj] {10.1088/0004-637X/806/2/275}, \href {https://ui.adsabs.harvard.edu/abs/2015ApJ...806..275P} {806, 275}

\bibitem[\protect\citeauthoryear{{Powell}, {M{\"u}ller}  \& {Heger}}{{Powell} et~al.}{2021}]{Powell2021_collapse_PISNe}
{Powell} J.,  {M{\"u}ller} B.,   {Heger} A.,  2021, \mn@doi [\mnras] {10.1093/mnras/stab614}, \href {https://ui.adsabs.harvard.edu/abs/2021MNRAS.503.2108P} {503, 2108}

\bibitem[\protect\citeauthoryear{{Punturo} et~al.,}{{Punturo} et~al.}{2010}]{Punturo2010_ET_method_paper1}
{Punturo} M.,  et~al., 2010, \mn@doi [Classical and Quantum Gravity] {10.1088/0264-9381/27/19/194002}, \href {https://ui.adsabs.harvard.edu/abs/2010CQGra..27s4002P} {27, 194002}

\bibitem[\protect\citeauthoryear{{Raduta}, {Nacu}  \& {Oertel}}{{Raduta} et~al.}{2021}]{Raduta2021_EOS_review}
{Raduta} A.~R.,  {Nacu} F.,   {Oertel} M.,  2021, \mn@doi [European Physical Journal A] {10.1140/epja/s10050-021-00628-z}, \href {https://ui.adsabs.harvard.edu/abs/2021EPJA...57..329R} {57, 329}

\bibitem[\protect\citeauthoryear{{Reitze} et~al.,}{{Reitze} et~al.}{2019}]{Reitze2019_CE_method_paper1}
{Reitze} D.,  et~al., 2019, in Bulletin of the American Astronomical Society. p.~35 (\mn@eprint {arXiv} {1907.04833}), \mn@doi{10.48550/arXiv.1907.04833}

\bibitem[\protect\citeauthoryear{{Richers}, {Nagakura}, {Ott}, {Dolence}, {Sumiyoshi}  \& {Yamada}}{{Richers} et~al.}{2017}]{Richers2017_nu_transport_comparison}
{Richers} S.,  {Nagakura} H.,  {Ott} C.~D.,  {Dolence} J.,  {Sumiyoshi} K.,   {Yamada} S.,  2017, \mn@doi [\apj] {10.3847/1538-4357/aa8bb2}, \href {https://ui.adsabs.harvard.edu/abs/2017ApJ...847..133R} {847, 133}

\bibitem[\protect\citeauthoryear{{Riley} et~al.,}{{Riley} et~al.}{2022}]{Riley2022_COMPAS_popsynth}
{Riley} J.,  et~al., 2022, \mn@doi [\apjs] {10.3847/1538-4365/ac416c}, \href {https://ui.adsabs.harvard.edu/abs/2022ApJS..258...34R} {258, 34}

\bibitem[\protect\citeauthoryear{{Rybizki}, {Just}  \& {Rix}}{{Rybizki} et~al.}{2017}]{Rybizki2017_GCE_Chempy}
{Rybizki} J.,  {Just} A.,   {Rix} H.-W.,  2017, \mn@doi [\aap] {10.1051/0004-6361/201730522}, \href {https://ui.adsabs.harvard.edu/abs/2017A&A...605A..59R} {605, A59}

\bibitem[\protect\citeauthoryear{{Sandoval}, {Hix}, {Messer}, {Lentz}  \& {Harris}}{{Sandoval} et~al.}{2021}]{Sandoval2021_3D_shk_breakout}
{Sandoval} M.~A.,  {Hix} W.~R.,  {Messer} O.~E.~B.,  {Lentz} E.~J.,   {Harris} J.~A.,  2021, \mn@doi [\apj] {10.3847/1538-4357/ac1d49}, \href {https://ui.adsabs.harvard.edu/abs/2021ApJ...921..113S} {921, 113}

\bibitem[\protect\citeauthoryear{{Sasaki} \& {Takiwaki}}{{Sasaki} \& {Takiwaki}}{2024}]{Sasaki2024_STIR_diffusion}
{Sasaki} S.,  {Takiwaki} T.,  2024, \mn@doi [\mnras] {10.1093/mnras/stad3997}, \href {https://ui.adsabs.harvard.edu/abs/2024MNRAS.528.1158S} {528, 1158}

\bibitem[\protect\citeauthoryear{{Scheck}, {Kifonidis}, {Janka}  \& {M{\"u}ller}}{{Scheck} et~al.}{2006}]{Scheck2006_MultiD_approx_transport_I}
{Scheck} L.,  {Kifonidis} K.,  {Janka} H.~T.,   {M{\"u}ller} E.,  2006, \mn@doi [\aap] {10.1051/0004-6361:20064855}, \href {https://ui.adsabs.harvard.edu/abs/2006A&A...457..963S} {457, 963}

\bibitem[\protect\citeauthoryear{{Schneider}, {Roberts}, {Ott}  \& {O'Connor}}{{Schneider} et~al.}{2019}]{Schneider2019}
{Schneider} A.~S.,  {Roberts} L.~F.,  {Ott} C.~D.,   {O'Connor} E.,  2019, \mn@doi [\prc] {10.1103/PhysRevC.100.055802}, \href {https://ui.adsabs.harvard.edu/abs/2019PhRvC.100e5802S} {100, 055802}

\bibitem[\protect\citeauthoryear{{Schneider}, {O'Connor}, {Granqvist}, {Betranhandy}  \& {Couch}}{{Schneider} et~al.}{2020}]{Schneider2020_EOS_dependence_BH}
{Schneider} A.,  {O'Connor} E.,  {Granqvist} E.,  {Betranhandy} A.,   {Couch} S.~M.,  2020, \mn@doi [\apj] {10.3847/1538-4357/ab8308}, \href {https://ui.adsabs.harvard.edu/abs/2020ApJ...894....4D} {894, 4}

\bibitem[\protect\citeauthoryear{{Segerlund}, {O'Sullivan}  \& {O'Connor}}{{Segerlund} et~al.}{2021}]{Segerlund2021_distance_SN_nns}
{Segerlund} M.,  {O'Sullivan} E.,   {O'Connor} E.,  2021, \mn@doi [arXiv e-prints] {10.48550/arXiv.2101.10624}, \href {https://ui.adsabs.harvard.edu/abs/2021arXiv210110624S} {p. arXiv:2101.10624}

\bibitem[\protect\citeauthoryear{{Skinner}, {Dolence}, {Burrows}, {Radice}  \& {Vartanyan}}{{Skinner} et~al.}{2019}]{Skinner2019_Fornax_methods_paper}
{Skinner} M.~A.,  {Dolence} J.~C.,  {Burrows} A.,  {Radice} D.,   {Vartanyan} D.,  2019, \mn@doi [\apjs] {10.3847/1538-4365/ab007f}, \href {https://ui.adsabs.harvard.edu/abs/2019ApJS..241....7S} {241, 7}

\bibitem[\protect\citeauthoryear{{Smartt}}{{Smartt}}{2015}]{Smartt2015_observations_CCSNe}
{Smartt} S.~J.,  2015, \mn@doi [\pasa] {10.1017/pasa.2015.17}, \href {https://ui.adsabs.harvard.edu/abs/2015PASA...32...16S} {32, e016}

\bibitem[\protect\citeauthoryear{{Smartt}, {Eldridge}, {Crockett}  \& {Maund}}{{Smartt} et~al.}{2009}]{Smartt2009_typeII-P}
{Smartt} S.~J.,  {Eldridge} J.~J.,  {Crockett} R.~M.,   {Maund} J.~R.,  2009, \mn@doi [\mnras] {10.1111/j.1365-2966.2009.14506.x}, \href {https://ui.adsabs.harvard.edu/abs/2009MNRAS.395.1409S} {395, 1409}

\bibitem[\protect\citeauthoryear{{Sonneborn}, {Altner}  \& {Kirshner}}{{Sonneborn} et~al.}{1987}]{Sonneborn1987A}
{Sonneborn} G.,  {Altner} B.,   {Kirshner} R.~P.,  1987, \mn@doi [\apj] {10.1086/185052}, \href {https://ui.adsabs.harvard.edu/abs/1987ApJ...323L..35S} {323, L35}

\bibitem[\protect\citeauthoryear{{Spitoni}, {Verma}, {Silva Aguirre}  \& {Calura}}{{Spitoni} et~al.}{2020}]{Spitoni2020_GCE_Bayesian_MCMC}
{Spitoni} E.,  {Verma} K.,  {Silva Aguirre} V.,   {Calura} F.,  2020, \mn@doi [\aap] {10.1051/0004-6361/201937275}, \href {https://ui.adsabs.harvard.edu/abs/2020A&A...635A..58S} {635, A58}

\bibitem[\protect\citeauthoryear{{Stockinger} et~al.,}{{Stockinger} et~al.}{2020}]{Stockinger2020_3D_low_mass_to_breakout}
{Stockinger} G.,  et~al., 2020, \mn@doi [\mnras] {10.1093/mnras/staa1691}, \href {https://ui.adsabs.harvard.edu/abs/2020MNRAS.496.2039S} {496, 2039}

\bibitem[\protect\citeauthoryear{{Sukhbold}, {Ertl}, {Woosley}, {Brown}  \& {Janka}}{{Sukhbold} et~al.}{2016}]{Sukhbold2016_explodability}
{Sukhbold} T.,  {Ertl} T.,  {Woosley} S.~E.,  {Brown} J.~M.,   {Janka} H.~T.,  2016, \mn@doi [\apj] {10.3847/0004-637X/821/1/38}, \href {https://ui.adsabs.harvard.edu/abs/2016ApJ...821...38S} {821, 38}

\bibitem[\protect\citeauthoryear{{Sukhbold}, {Woosley}  \& {Heger}}{{Sukhbold} et~al.}{2018}]{Sukhbold2018_preSN_KEPLER_bug}
{Sukhbold} T.,  {Woosley} S.~E.,   {Heger} A.,  2018, \mn@doi [\apj] {10.3847/1538-4357/aac2da}, \href {https://ui.adsabs.harvard.edu/abs/2018ApJ...860...93S} {860, 93}

\bibitem[\protect\citeauthoryear{{Summa}, {Hanke}, {Janka}, {Melson}, {Marek}  \& {M{\"u}ller}}{{Summa} et~al.}{2016}]{Summa2016_prog_dependence_vertex}
{Summa} A.,  {Hanke} F.,  {Janka} H.-T.,  {Melson} T.,  {Marek} A.,   {M{\"u}ller} B.,  2016, \mn@doi [\apj] {10.3847/0004-637X/825/1/6}, \href {https://ui.adsabs.harvard.edu/abs/2016ApJ...825....6S} {825, 6}

\bibitem[\protect\citeauthoryear{{Summa}, {Janka}, {Melson}  \& {Marek}}{{Summa} et~al.}{2018}]{Summa2018_rot3D_crit_lum}
{Summa} A.,  {Janka} H.-T.,  {Melson} T.,   {Marek} A.,  2018, \mn@doi [\apj] {10.3847/1538-4357/aa9ce8}, \href {https://ui.adsabs.harvard.edu/abs/2018ApJ...852...28S} {852, 28}

\bibitem[\protect\citeauthoryear{{Sykes} \& {M{\"u}ller}}{{Sykes} \& {M{\"u}ller}}{2025}]{Sykes2024_2D_fallback_SNe}
{Sykes} B.,  {M{\"u}ller} B.,  2025, \mn@doi [\mnras] {10.1093/mnras/staf317}, \href {https://ui.adsabs.harvard.edu/abs/2025MNRAS.538..572S} {538, 572}

\bibitem[\protect\citeauthoryear{{Tsang}, {Vartanyan}  \& {Burrows}}{{Tsang} et~al.}{2022}]{Tsang2022_ML_explodability}
{Tsang} B. T.~H.,  {Vartanyan} D.,   {Burrows} A.,  2022, \mn@doi [\apjl] {10.3847/2041-8213/ac8f4b}, \href {https://ui.adsabs.harvard.edu/abs/2022ApJ...937L..15T} {937, L15}

\bibitem[\protect\citeauthoryear{{Ugliano}, {Janka}, {Marek}  \& {Arcones}}{{Ugliano} et~al.}{2012}]{Ugliano2012}
{Ugliano} M.,  {Janka} H.-T.,  {Marek} A.,   {Arcones} A.,  2012, \mn@doi [\apj] {10.1088/0004-637X/757/1/69}, \href {https://ui.adsabs.harvard.edu/abs/2012ApJ...757...69U} {757, 69}

\bibitem[\protect\citeauthoryear{{Vartanyan} \& {Burrows}}{{Vartanyan} \& {Burrows}}{2023}]{Vartanyan2023_nu_100_2D}
{Vartanyan} D.,  {Burrows} A.,  2023, \mn@doi [\mnras] {10.1093/mnras/stad2887}, \href {https://ui.adsabs.harvard.edu/abs/2023MNRAS.526.5900V} {526, 5900}

\bibitem[\protect\citeauthoryear{{Vartanyan}, {Burrows}, {Radice}, {Skinner}  \& {Dolence}}{{Vartanyan} et~al.}{2018}]{Vartanyan2018_2D_simulations_revival_fittest}
{Vartanyan} D.,  {Burrows} A.,  {Radice} D.,  {Skinner} M.~A.,   {Dolence} J.,  2018, \mn@doi [\mnras] {10.1093/mnras/sty809}, \href {https://ui.adsabs.harvard.edu/abs/2018MNRAS.477.3091V} {477, 3091}

\bibitem[\protect\citeauthoryear{{Vartanyan}, {Tsang}, {Kasen}, {Burrows}, {Wang}  \& {Teryoshin}}{{Vartanyan} et~al.}{2025}]{Vartanyan2025_3D_bounce_to_breakut}
{Vartanyan} D.,  {Tsang} B. T.~H.,  {Kasen} D.,  {Burrows} A.,  {Wang} T.,   {Teryoshin} L.,  2025, \mn@doi [\apj] {10.3847/1538-4357/adb1e4}, \href {https://ui.adsabs.harvard.edu/abs/2025ApJ...982....9V} {982, 9}

\bibitem[\protect\citeauthoryear{{Walk}, {Tamborra}, {Janka}, {Summa}  \& {Kresse}}{{Walk} et~al.}{2020}]{Walk2020_3DBH_neutrinos}
{Walk} L.,  {Tamborra} I.,  {Janka} H.-T.,  {Summa} A.,   {Kresse} D.,  2020, \mn@doi [\prd] {10.1103/PhysRevD.101.123013}, \href {https://ui.adsabs.harvard.edu/abs/2020PhRvD.101l3013W} {101, 123013}

\bibitem[\protect\citeauthoryear{{Wang} \& {Burrows}}{{Wang} \& {Burrows}}{2020}]{Wang2020_Kompaneets_NNS}
{Wang} T.,  {Burrows} A.,  2020, \mn@doi [\prd] {10.1103/PhysRevD.102.023017}, \href {https://ui.adsabs.harvard.edu/abs/2020PhRvD.102b3017W} {102, 023017}

\bibitem[\protect\citeauthoryear{{Wang}, {Vartanyan}, {Burrows}  \& {Coleman}}{{Wang} et~al.}{2022}]{Wang2022_prog_study_ram_pressure}
{Wang} T.,  {Vartanyan} D.,  {Burrows} A.,   {Coleman} M. S.~B.,  2022, \mn@doi [\mnras] {10.1093/mnras/stac2691}, \href {https://ui.adsabs.harvard.edu/abs/2022MNRAS.517..543W} {517, 543}

\bibitem[\protect\citeauthoryear{{Wongwathanarat}, {M{\"u}ller}  \& {Janka}}{{Wongwathanarat} et~al.}{2015}]{Wongwathanarat2015_3D_shk_breakout}
{Wongwathanarat} A.,  {M{\"u}ller} E.,   {Janka} H.~T.,  2015, \mn@doi [\aap] {10.1051/0004-6361/201425025}, \href {https://ui.adsabs.harvard.edu/abs/2015A&A...577A..48W} {577, A48}

\bibitem[\protect\citeauthoryear{{Woosley} \& {Heger}}{{Woosley} \& {Heger}}{2007}]{WH07}
{Woosley} S.~E.,  {Heger} A.,  2007, \mn@doi [\physrep] {10.1016/j.physrep.2007.02.009}, \href {https://ui.adsabs.harvard.edu/abs/2007PhR...442..269W} {442, 269}

\bibitem[\protect\citeauthoryear{{Woosley}, {Heger}  \& {Weaver}}{{Woosley} et~al.}{2002}]{Woosley2002_KEPLER_models}
{Woosley} S.~E.,  {Heger} A.,   {Weaver} T.~A.,  2002, \mn@doi [Reviews of Modern Physics] {10.1103/RevModPhys.74.1015}, \href {https://ui.adsabs.harvard.edu/abs/2002RvMP...74.1015W} {74, 1015}

\bibitem[\protect\citeauthoryear{{Yamasaki} \& {Yamada}}{{Yamasaki} \& {Yamada}}{2006}]{Yamasaki2006_steady_state_conv}
{Yamasaki} T.,  {Yamada} S.,  2006, \mn@doi [\apj] {10.1086/507067}, \href {https://ui.adsabs.harvard.edu/abs/2006ApJ...650..291Y} {650, 291}

\makeatother
\end{thebibliography}




\appendix
\section{Neutrino heating in the gain region}
\label{sec:appendix_diff_Janka}
Here, we show the derivation of Eq.~\eqref{eq:Qdot_semi_an}, and highlight the small differences with the expressions of \citet{Janka2001_conditions_shk_revival} and \citet{Janka2012_review_CCSNe}. We start from the expression for the energy deposition rate in the gain region \citep{Janka2012_review_CCSNe}:
\begin{equation}
    \dot{Q}_{\nu_i}^+ = \int_{\rm gain} q_{\nu_i}^+ \frac{{\rm d} m}{m_u} = \int_{\rm gain} \frac{L_{\nu_i}}{4 \pi r^2} \avg{\sigma_{\rm abs}} \frac{{\rm d} m}{m_u}
\end{equation}
which holds for electron neutrinos and antineutrinos, where the integral is evaluated in the gain region, $m_u = 1.66 \times 10^{-24}$~g is the atomic mass unit, and $\avg{\sigma_{\rm abs}}$ is the average cross section for neutrino absorption in the gain region. Notice that we assumed that the neutrino spectrum is forward peak, so that the integral over all neutrino angles is 1. The cross section for neutrino absorption, which is to a good approximation the same for neutrinos and antineutrinos, is defined in Eq.~19 of \citet{Janka2001_conditions_shk_revival} as:
\begin{equation}
    \avg{\sigma_{\rm abs}} = \frac{3\alpha^2 + 1}{4} \sigma_0 \left(\frac{\avg{\epsilon_{\nu_i}}}{m_e c^2}\right)^2 Y_{\rm n, p}.
\end{equation}
Here, $\sigma_0 = 1.76 \times 10^{-44}$~cm$^2$, $\alpha = -1.26$, $m_e c^2 = 0.511$~MeV is the rest mass of the electron, $\avg{\epsilon_\nu}$ is the average neutrino energy, and $Y_{\rm n}$ and $Y_{\rm p}$ are the neutron and proton fractions, respectively. The former (latter) takes into account the fact that $\nu_e$ ($\bar{\nu}_e$) only undergo charged current reactions with neutrons (protons). In the gain region, to a good approximation one can assume $Y_{\rm n} \approx Y_{\rm p} \approx 0.5$

Putting everything together, using the same numerical factors of \cite{Janka2012_review_CCSNe}, and summing the electron neutrino and antineutrino contributions, the mean value theorem ensures that there exists a radius $R^*$ for which:
\begin{equation}
\label{eq:Qdot+}
\begin{split}
    \dot{Q}_{\nu}^+ &\approx 8.63 \times 10^{51}\ \frac{\rm erg}{\rm s} \left(\frac{R^*}{100\ \rm km}\right)^{-2} \left(\frac{M_{\rm g}}{0.01\ M_\odot}\right)\\
    &\frac{1}{2}\sum_{\nu_i \in \{\nu_e, \bar{\nu}_e \}}\ \frac{\avg{\epsilon^2_{\nu_i}}}{\left(18\ \rm MeV\right)^2} \left(\frac{L_{\nu_i} (R^*)}{3 \times 10^{52}\ \rm erg/s}\right).
\end{split}
\end{equation}
From our simulations, we find that a good choice for $R^*$ is:
\begin{equation}
    \bar{R}_{\rm g} = R_{\rm g} + 0.15 \times (1 + \xi_{2.0}) (R_{\rm s} - R_{\rm g}).
\end{equation}
This is motivated by the fact that progenitors with higher compactnesses have larger densities in the gain region, and therefore neutrino heating extends to larger radii, hence one expects $R^*$ to be slightly larger for higher-compactness progenitors. In the original work from \cite{Janka2012_review_CCSNe}, $R^*$ was taken to be simply the gain radius $R_{\rm g}$. To derive Eq.~\eqref{eq:Qdot_semi_an} we considered that the net neutrino heating $\dot{Q}_\nu$ is to a good approximation some fraction of the total heating $\dot{Q}_{\nu}^+$, and in this work we consider that fraction to be 0.6.

Another difference with Eq.~6 of \citet{Janka2012_review_CCSNe} is that their overall pre-factor is $9.4\times 10^{51}$~erg/s, which is slightly larger than ours. Moreover, we have a factor of $1/2$ in front of the sum. To recover the results from \citet{Janka2001_conditions_shk_revival} and \citet{Janka2012_review_CCSNe}, one should consider that in those works, it is assumed that:
\begin{align}
\label{eq:aprrox1}
\avg{\epsilon^2_\nu} &\approx 21 (k_b T)^2, \\
\label{eq:aprrox2}
L_{\nu_i} (r=\bar{R}_{\rm g}) / \bar{R}_{\rm g}^2 &\approx 1.4-1.5 \times L_{\nu_i} (r=R_{\rm g}) / R_{\rm g}^2, \\
\label{eq:aprrox3}
\avg{\epsilon^2_{\bar{\nu}_e}} &\approx 1.4-1.5 \avg{\epsilon^2_{\nu_e}}, \\
\label{eq:aprrox4}
L_{\bar{\nu}_e} &\approx L_{\nu_e},
\end{align}
where the first expression comes from assuming a blackbody spectrum, and the rest are valid during the accretion phase. Plugging these expressions in Equation~\ref{eq:Qdot+} results in:
\begin{equation}
\begin{split}
    8.63 \times 10^{51} \times &\left(\frac{R^*}{100\ \rm km}\right)^{-2} \left(\frac{M_{\rm g}}{0.01\ M_\odot}\right) \\
    &\frac{1}{2}\sum_{\nu_i}\ \frac{\avg{\epsilon^2_{\nu_i}}}{\left(18\ \rm MeV\right)^2} \left(\frac{L_{\nu_i} (R^*)}{3 \times 10^{52}\ \rm erg/s}\right) \\
     =\ 8.63 \times 10^{51} \times &\ \frac{21}{4.5^2} \times 1.45 \times 1.45 \times \frac{1}{2}\ \left(\frac{R_{\rm g}}{100\ \rm km}\right)^{-2} \left(\frac{M_{\rm g}}{0.01\ M_\odot}\right) \\
    &\left(\frac{T_{\nu_e}}{4\ \rm MeV}\right)^2 \left(\frac{2 L_{\nu_e} (R_{\rm g})}{3 \times 10^{52}\ \rm erg/s}\right) \\
    =\ 9.4 \times 10^{51}\ &\left(\frac{R_{\rm g}}{100\ \rm km}\right)^{-2} \left(\frac{M_{\rm g}}{0.01\ M_\odot}\right) \\
    &\left(\frac{T_{\nu}}{4\ \rm MeV}\right)^2 \left(\frac{L_{\nu} (R_{\rm g})}{3 \times 10^{52}\ \rm erg/s}\right)
\end{split}
\end{equation}
which matches the expression in \citet{Janka2012_review_CCSNe}, with the important detail that, as defined in \citet{Janka2001_conditions_shk_revival}, $T_\nu \equiv T_{\nu_e}$ and $L_{\nu} \equiv 2 L_{\nu_e}$. The reason why we did not directly adopt the expressions from \citet{Janka2001_conditions_shk_revival} and \citet{Janka2012_review_CCSNe} is that the relations in Eq.~\eqref{eq:aprrox1}-\eqref{eq:aprrox4} (in particular Eq.~\eqref{eq:aprrox2}) are only approximately valid, and vary during the accretion phase, hence resulting in a worse agreement with the simulations.

\section{Dependence of the maximum of neutrino heating on compactness}
\label{sec:appendix_Qdot_comp}
To understand the dependence of $\dot{Q}_\nu^{\rm max}$ on $\xi_{2.0}$ shown in Figure~\ref{fig:Qdot_vs_comp_multiD}, one can show that during the accretion phase, typically between $80 - 300$~ms after bounce, the energy increases with time following:
\begin{align}
\label{eq:rmsnue_fit}
\avg{\epsilon^2_{\nu_e}} &= q + m\times t_{\rm pb} \\
\label{eq:ab_1D+}
q^{\rm 1D+} = 127 - 83\, \xi_{2.0}^{1/3} &\qquad m^{\rm 1D+} = 1300\, \xi_{2.0}^{1/3}, \\
\label{eq:ab_1D}
q^{\rm 1D} = 127 - 83\, \xi_{2.0}^{1/3} &\qquad m^{\rm 1D} = 1400\, \xi_{2.0}^{1/3},
\end{align}
where $t_{\rm pb}$ is time post bounce, and the superscripts 1D and 1D+ indicate the values resulting from fits to 1D and 1D+ simulations, respectively. As explained in Section~\ref{sec:semi_an_qdot}, $\avg{\epsilon^2_{\nu_e}}$ is smaller in 1D+ simulations because the shock stalls at larger radii and, therefore, the overall mass accretion onto the PNS is smaller. Notice that, as the mass accretion rate drops, the increase in neutrino energy becomes even slower than the linear trend assumed in the above equation (see Figure~\ref{fig:Enue_Lnue_vs_time}). Therefore, with this simple treatment we expect to overestimate the amount of neutrino heating generated, and we will show later in this section that this is indeed the case.

For 1D simulations, the maximum of $\dot{Q}$ is always reached at roughly $t \approx 100$~ms, and one can show that, at that time, $L_\nu M_{\rm g} / \bar{R}_{\rm g}^2$ is roughly constant. In particular, it can be shown that:
\begin{equation}
    \label{eq:LMR_lowcomp}
    \left(\frac{L_{\nu_e}}{3\times 10^{52}\ {\rm erg/s}}\right) \left(\frac{\bar{R}_{\rm g}}{100\ \rm km}\right)^{-2} \left(\frac{M_{\rm g}}{0.01\ M_\odot}\right) \approx ( 1.75 + 3.84\, \xi_{2.0}^{5/3}).
\end{equation}
Notice that for 1D simulations the above equation is only valid at $t \approx 100$~ms since after that the mass in the gain region drops very quickly (see discussion in Section~\ref{sec:semi_an_qdot}). For 1D+ simulations, however, the above equality holds for a longer time, up to $\sim 300$~ms, because the shock stalls at large radii for longer times, thanks to the support given by $\nu$-driven convection. 

However, for progenitors with large-enough compactness (i.e. $\xi_{2.0} \gtrsim 0.72$), the mass accretion rates are so large that neutrino luminosity continuously increases during the accretion phase, rather than remaining constant (see the highest compactness progenitor in the bottom panel of Figure~\eqref{fig:Enue_Lnue_vs_time}). In that case, one instead finds:
\begin{equation}
    \label{eq:LMR_highcomp}
    \begin{split}
    \left(\frac{L_{\nu_e}}{3\times 10^{52}\ {\rm erg s^{-1}}}\right) &\left(\frac{\bar{R}_{\rm g}}{100\ \rm km}\right)^{-2} \left(\frac{M_{\rm g}}{0.01\ M_\odot}\right) \\
    \approx &3.5 -255\, t + 317\, \xi_{2.0}^{2/3}\, t.
    \end{split}
\end{equation}
Essentially, this is the main reason why $\dot{Q}_\nu^{\rm max}$ rapidly increases for progenitors with large compactnesses $\xi_{2.0} > 0.72$, as shown in Figure~\ref{fig:Qdot_vs_comp_multiD}.

\subsection{1D case}
We can now plug the approximate relations between energies and luminosities of electron neutrino and antineutrinos shown in Eqs.~\eqref{eq:aprrox3} and \eqref{eq:aprrox4}, together with Eqs.\ref{eq:rmsnue_fit}, \eqref{eq:ab_1D}, and \eqref{eq:LMR_lowcomp} into Eq.~\ref{eq:Qdot_semi_an} to obtain:
\begin{equation}
    \label{eq:Qdot_1D}
    \dot{Q}_\nu^{\rm max} = 1.93 \times 10^{49}\ \frac{\rm erg}{\rm s} \left(221 + 99\, \xi_{2.0}^{1/3} + 488\, \xi_{2.0}^{5/3} + 218\, \xi_{2.0}^{2} \right),
\end{equation}
where $\dot{Q}_\nu^{\rm max} = \dot{Q}_\nu(t = t_{\rm max})$, and we have assumed $t_{\rm max} 100$~ms for all progenitors. This is justified by the discussion in Section~\eqref{sec:semi_an_qdot}, where we have shown that, because of the rapid decrease of $M_{\rm g}$ caused by the recession of the shock, neutrino heating decreases immediately after the beginning of the accretion phase.

\subsection{1D+ case}
The same relations used in the 1D case can be applied to the 1D+ case. The first difference is that we have to use Eq.~\eqref{eq:ab_1D+} instead of Eq.~\eqref{eq:ab_1D}, i.e. a smaller slope in time. The reason for the slower increase of the neutrino energy is that in 1D+ simulations the shock is stalling at larger radii, which decreases the mass accretion rate onto the PNS, leading to smaller neutrinosphere temperatures and neutrino energies.

The largest difference with the 1D simulations is however that $t_{\rm max}$ varies for different progenitors. Specifically, as shown in Section~\ref{sec:timescales_correlations}, $t_{\rm max}$ is roughly equal to the time $t_{\rm Si/O}$ when the Si/O interface is accreted through the shock. This is true up to compactnesses of $\sim \xi_{2.0} 0.72$. For progenitors with higher compactnesses than that, $t_{\rm max}$ is reached at $\sim 300$~ms after bounce, i.e. the end of the accretion phase which corresponds to when the shock is revived and mass accretion onto the PNS shuts off. 

Therefore, one can assume $t_{\rm max} \approx t_{\rm Si/O}$ for progenitors with $\xi_{2.0} < 0.72$ and $t_{\rm max} \approx 300$~ms for progenitors with $\xi_{2.0} \geqslant 0.72$. Notice that for the former category of progenitors, we use Eq.~\eqref{eq:LMR_lowcomp}, whereas for the latter we use Eq.~\eqref{eq:LMR_highcomp}, as explained above. Since in both cases $t_{\rm max}$ corresponds to a significant decrease in mass accretion rate, the simple linear increase in time assumed for the neutrino energy in Eq.~\eqref{eq:rmsnue_fit} tends to overestimate the energy by $\sim 20 \%$. This can be seen in the upper panel of Figure~\ref{fig:Enue_Lnue_vs_time}, where towards the end of the accretion phase the increase in energy is not linear anymore. To compensate for that, when we evaluate Eq.~\eqref{eq:rmsnue_fit} at $t=t_{\rm max}$ we multiply it by an extra factor of 0.8.

To express everything in terms of compactness, one can use the fact that the the accretion time of a given layer through the shock is proportional to the free fall time, and as shown by \cite{Boccioli2023_explodability} for the progenitors analyzed in this study, one finds:
\begin{equation}
\begin{split}
    t_{\rm Si/O} &= 0.78\sqrt{\frac{\pi}{4 G \bar{\rho}_{\rm Si/O}}} - 0.13 \\
    &= 0.12\ {\rm s}\ \left(\frac{r_{\rm Si/O}}{100\ {\rm km}}\right)^{3/2}\left(\frac{M_{\rm Si/O}}{M_\odot}\right)^{-1/2} - 0.13\ {\rm s} \\
    &\approx 0.36 \, \xi_{2.0} + 0.07\ {\rm s},
    \end{split}
\end{equation}
where the last equality can be shown to hold for the analyzed progenitors that successfully explode.

Finally, putting everything together, for 1D+ simulations of progenitors with $\xi_{2.0} < 0.72$ one has:
\begin{equation}
    \label{eq:Qdot_STIR_lowxi}
    \dot{Q}_\nu^{\rm max} = 1.54 \times 10^{49}\ \frac{\rm erg}{\rm s} \left(221 + 840\, \xi_{2.0}^{4/3} + 488\, \xi_{2.0}^{5/3} + 1854\, \xi_{2.0}^{3} \right),
\end{equation}
where, as explained above, we have multiplied the overall factor by 0.8 to account for the overestimation of neutrino energies at $t=t_{\rm max}$.

For 1D+ simulations of progenitors with $\xi_{2.0} \geqslant 0.72$ one instead finds:
\begin{equation}
\begin{split}
    \label{eq:Qdot_STIR_highxi}
    \dot{Q}_\nu^{\rm max} = 1.54 \times 10^{49}\ \frac{\rm erg}{\rm s} \times &\left(-9262 - 22352 \, \xi_{2.0}^{1/3}\right.\\
    &\left. + 12087 \, \xi_{2.0}^{2/3} + 29171 \, \xi_{2.0}\right)
\end{split}
\end{equation}

Equations~\eqref{eq:Qdot_1D}, \eqref{eq:Qdot_STIR_lowxi}, and \eqref{eq:Qdot_STIR_highxi} are shown as the black dotted line and the two black dashed lines in Figure~\ref{fig:Qdot_vs_comp_multiD}. They reproduce the trend very well. This serves as further confirmation of what has been shown in Section~\ref{sec:semi_an_qdot} regarding the importance of convection in indirectly influencing neutrino heating. The only difference in $\dot{Q}_\nu^{\rm max}$ between the 1D+ and 1D simulations for progenitors with $\xi_{2.0} < 0.72$ (i.e. between Eqs.~\eqref{eq:Qdot_1D} and \eqref{eq:Qdot_STIR_lowxi}) is $t_{\rm max}$ ($b^{\rm 1D}$ and $b^{\rm 1D+}$ are also different, but the effect is quite small). The fact that $t_{\rm max} > 100$~ms in 1D+ simulations is an indirect effect of convection, which keeps the shock stalling at large radii allowing the neutrino heating to increase during the entire accretion phase, until either a strong Si/O interface is accreted or the explosion develops. 

The change in slope of $\dot{Q}_\nu^{\rm max}$ vs. $\xi_{2.0}$ for progenitors with $\xi_{2.0} \geqslant 0.72$ should instead be attributed to the fact that luminosities for these very high-compactness progenitors rapidly increase during the accretion phase (see bottom panel of Figure~\ref{fig:Enue_Lnue_vs_time} and Eq.~\eqref{eq:LMR_highcomp}) rather than a larger $t_{\rm max}$ which, as explained above, is instead roughly constant (if not slightly decreasing, as shown in Figure~\ref{fig:tmax_vs_taccr_vs_texpl}).

\section{Correlations of Neutrino heating in 1D+ and 2D simulations}
\label{sec:appendix_Qdot_1D_2D}
In this Appendix we analyze some correlations involving neutrino heating to highlight some differences between 1D+ and 2D simulations. We consider the same $341$ 1D+ simulations from \citet{Boccioli2024_remnant} and the 2D F{\sc{ornax}} and FLASH simulations discussed in Section~\ref{sec:comparison_multiD}. We also add to the comparison 88 1D+ simulations of the same progenitors from \citet{Sukhbold2018_preSN_KEPLER_bug} used for the 2D F{\sc{ornax}} simulations. We consider the maximum of neutrino heating, $\dot{Q}_{\nu}^{\rm max}$, and also compute electron neutrino energies ($E_{\nu_e}$), luminosities ($L_{\nu_e}$) and baryonic mass of the PNS ($M_{\rm PNS}$) at the time $t_{\rm max}$ when $\dot{Q}_{\nu}$ reaches its maximum. 

\begin{figure*}
\centering
\includegraphics[width=\textwidth]{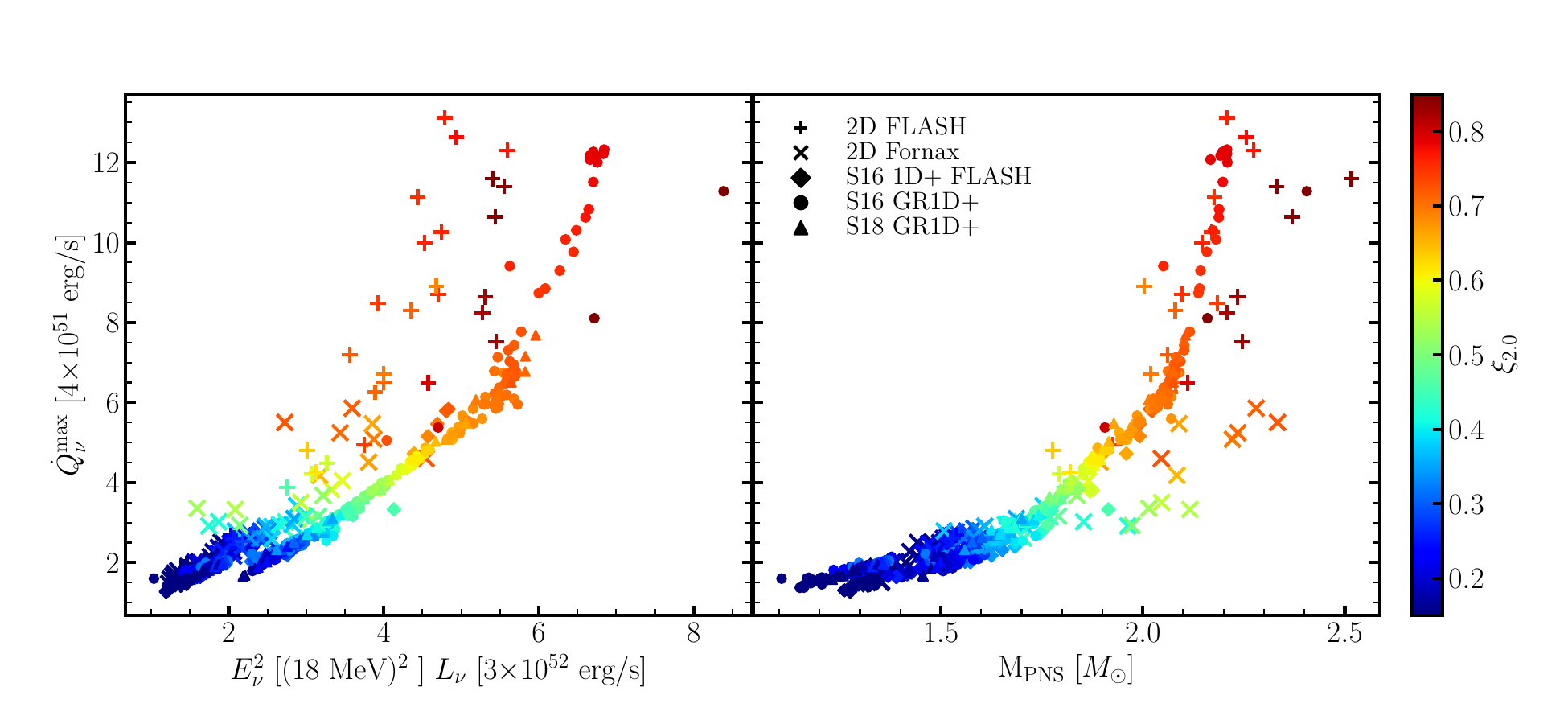}

\caption{Correlations between neutrino heating, neutrino energies and neutrino luminosities calculated at 500 km, and the baryonic mass of the PNS. All quantities are calculated at the time when the maximum of the neutrino heating is reached. Crosses and pluses show the F{\sc{ornax}} and FLASH simulations discussed in Section~\ref{sec:comparison_multiD}, respectively. Diamonds refer to 1D+ simulations from \citet{Couch2020_STIR} for KEPLER progenitors from \citet{Sukhbold2016_explodability}. Circles and triangles refer to 1D+ simulations run with \texttt{GR1D+} for KEPLER progenitors from \citet{Sukhbold2016_explodability} and \citet{Sukhbold2018_preSN_KEPLER_bug}, respectively.}
\label{fig:Qdot_vs_LE2_vs_mbary_multiD}
\end{figure*}

Firstly, from Figure~\ref{fig:Qdot_vs_LE2_vs_mbary_multiD} one can see a clear dependence of $\dot{Q}_{\nu}^{\rm max}$, $E_{\nu_e}$, $L_{\nu_e}$ and $M_{\rm PNS}$ on compactness, as discussed Section~\ref{sec:semi_an_qdot}. In both panels of Figure~\ref{fig:Qdot_vs_LE2_vs_mbary_multiD} there is some scatter among the 2D simulations, whereas the 1D+ simulations are much closer together. This can be attributed to both numerical noise in evaluating neutrino properties (see for example Figure~\ref{fig:compare_1D_2D_3D}), as well as a more complicated non-spherical symmetry compared to the 1D+ simulations which will cause neutrino quantities to deviate from the purely spherical case.

More interestingly, there are two distinct trends in the left panel of Figure~\ref{fig:Qdot_vs_LE2_vs_mbary_multiD}, one followed by the 1D+ simulations and one followed by the 2D ones. On the right panel, however, all simulations follow the same trend of $\dot{Q}_{\nu}^{\rm max}$ as a function of $M_{\rm PNS}$. Since the collapse of matter onto the central object is mostly spherically symmetric, one expects the same mass accretion rates in 1D+ and 2D simulations, at least during the stalled-shock phase before the explosion is launched. Therefore, the time evolution of $M_{\rm PNS}$ is very similar in 1D+ and 2D simulations (again, before the explosion sets in), and therefore from the overlapping 1D+ and 2D trends in the right panel of Figure~\ref{fig:Qdot_vs_LE2_vs_mbary_multiD} one can conclude that the PNS evolution up to the time when $\dot{Q}_{\nu}^{\rm max}$ is reached was very similar in 2D and 1D+, and therefore the time at which the maximum of neutrino heating is reached is also similar in 2D and 1D+ simulations. 

On the contrary, the 1D+ and 2D trends are very different in the left panel of Figure~\ref{fig:Qdot_vs_LE2_vs_mbary_multiD}. This can be traced back to lower neutrino energies in 2D compared to 1D+, whereas the luminosities are roughly the same. As pointed out by \citet{Nagakura2020_PNS_convection}, this is a consequence of PNS convection, which is not properly captured by 1D+ models. Convection causes the PNS to have larger radii, and consequently larger neutrinosphere radii. This would naturally lead to larger luminosities. At the same time, larger neutrinosphere radii also correspond to lower temperatures, and therefore lower energies and luminosities. The net effect is that average energies are smaller in 2D. On the other hand, the luminosities experience two competing effects that, when combined, are quite small, leading to luminosities in 2D that are quite similar to the 1D+ case. Nonetheless, the 1D+ simulations still tend to yield larger luminosities than 2D, at least in the stalled-shock phase as also pointed out by \citet{Nagakura2020_PNS_convection}, although the overall effect is much larger when it comes to neutrino energies.

Despite lower neutrino energies (and, to some extent, lower neutrino luminosities), the maximum of the neutrino heating in 2D is still very similar to the one obtained in 1D+ simulations. This could be due to several factors, for example, asymmetries in the shock and other instabilities such as the standing accretion shock instability (SASI) \citep{Foglizzo2006_SASI} could lead to a larger overall volume in the gain region. More comparisons are needed to understand the reason behind this apparently accidental compensation of the neutrino heating, but this goes beyond the scope of the present work. 

\bsp	
\label{lastpage}
\end{document}